\documentclass[article,12pt]{JHEP3}
\usepackage{graphicx}\usepackage{epsf,shadow,pifont}
\usepackage{amsmath,epsfig}
\usepackage{amssymb,amsfonts}
\usepackage{latexsym}
\usepackage{slashed}
\usepackage{float}
\usepackage{epsfig}
\usepackage{xcolor}
\let\normalcolor\relax
\usepackage[vcentermath,enableskew]{youngtab}

\makeatletter
\newcommand{\globalcolor}[1]{%
  \color{#1}\global\let\default@color\current@color
}
\makeatother

\newcommand{\h}[1]{{\widehat{\rm #1}}}
\newcommand{\vis}[1]{{\rm{#1}}}

\renewcommand{\theequation}{\arabic{section}.\arabic{equation}}

\def\be{\begin{equation}}
\def\ee{\end{equation}}

\newcommand{\bear}{\begin{eqnarray}}
\newcommand{\bea}{\begin{eqnarray}}
\newcommand{\eear}{\end{eqnarray}}
\newcommand{\eea}{\end{eqnarray}}

\def\hri#1#2{\href{http://arxiv.org/abs/#1}{[ArXiv:#1]#2}}
\def\hre#1#2{\href{http://arxiv.org/abs/#1/#2}{[ArXiv:#1/#2]}}

\def\hrj#1#2{\href{www.doi.org/#1}{#2}}
\newbox\pippobox

\def\II{\relax{\rm I\kern-.18em I}}

\def\e{\epsilon}
\def\m{\mu}
\def\n{\nu}
\def\r{\rho}
\def\s{\sigma}
\def\pa{\partial}
\def\t{\theta}
\def\sp{\;\;\;,\;\;\;}

\def\p{\partial}

\def\l{\lambda}

\def\nn{\nonumber}

\def\AA{{\cal A}}
\def\BB{{\cal B}}

\def\DD{{\cal D}}

\def\II{{\cal I}}
\def\JJ{{\cal J}}

\def\LL{{\cal L}}

\def\OO{{\cal O}}
\def\PP{{\cal P}}

\def\WW{{\cal W}}

\def\eqc{{\;\;,}}
\def\eqp{{\;\;.}}

\title{Global symmetries, hidden sectors
and emergent (dark) vector interactions}
\author{P. Betzios$^{\flat}$, E. Kiritsis$^{\flat,\natural}$, V. Niarchos$^{\flat}$, O. Papadoulaki$^*$\\
~\\
$^\flat$ \href{http://hep.physics.uoc.gr}{Crete Center for Theoretical Physics}, Institute for Theoretical and Computational Physics,
Department of Physics,\\
University of Crete, 70013, Heraklion, Greece
~\\
~\\
$^\natural$ \href{http://www.apc.univ-paris7.fr}{Universite de Paris, CNRS, Astroparticule et Cosmologie,  F-75006 Paris, France},\\
~\\
$^*$ \href{https://www.ictp.it/}{International Centre for Theoretical Physics} \\
Strada Costiera 11, Trieste 34151 Italy.
}

\preprint{CCTP-2020-6\\ITCP-IPP-2020/6}

\abstract{

Hidden theories coupled to the SM may provide emergent (dark) vectors, that are composites/bound-states of the hidden fields. This is motivated by paradigms emerging from the AdS/CFT correspondence but it is a more general phenomenon. We explore the general setup and find that UV interactions among currents or charged fields give rise to emergent vectors in the IR. We study the general properties of such vectors and argue that they can be generically different from fundamental dark photons that have been studied so far.

 }

\makeindex
\begin{document}
\newpage

%%%%%%%%%%%%%%%%%%%%%%%%%%%%%%%%%
\section{Introduction}

Modern theories of particle physics generally rely on two types of symmetries. One class of them is global space-time symmetries like Lorentz invariance.
The other class is gauge symmetries and all fundamental interactions of particle physics are based on gauge theories.

Gravity is both similar and different. On one hand, it gauges local space-time symmetries like translations and rotations and it is therefore based again on the gauge principle. On the other, the nature of the gauge symmetry is rather different from gauge theories.

It has long been claimed that both Lorentz invariance as well as gauge symmetry are IR ``accidents", \cite{nielsen}.
The claims on Lorentz invariance have been analyzed on different occasions and different contexts, especially recently, \cite{CG}-\cite{lorentz}. An emerged IR Lorentz invariance remains a possibility, although it seems ``fine tuning" is needed.

On the other hand, a gauge symmetry that emerges in the IR,  is a radical departure from the paradigm that developed in the last 50 years, crowning the success of the Standard Model (SM), where the gauge principle emerged as the fundamental principle. It is believed by most in the field that the gauge principle is central to particle interactions, {\em at all scales}.
However, slowly and sporadically, there were attempts to investigate alternative routes, in which gauge invariance appears or emerges in the IR.

In the condensed matter literature the notion of emergence is less of a taboo, and the issue of emergent gauge invariance has been entertained on various occasions.  An early appearance of emergent gauge fields is at Resonance Valence Bond (RVB) models of antiferromagnetism, \cite{RVB}.
It has been advocated in lattice models that due to string-net condensation emergent gauge invariance and photons can appear, \cite{wen}. And it has been also generalized to emergent gravity, \cite{wen2}.

The notion of emergence from strings is also interesting. Although it is rarely viewed in this light, (fundamental) strings provide emergent gauge bosons and gravitons. The input in NS-R string theory is the two-dimensional theory of scalars and fermions on the world-volume. The output is space-time gauge bosons and gravitons, among others.

In condensed matter another instance of emergence of gauge invariance appeared in the context of fractionalization phenomena and the emergence of fractionalized quasiparticles, \cite{frac}.
This culminated in the so called ``deconfined quantum critical points", \cite{Senthil}.
Such points are second order quantum phase transitions where the critical theory contains an emergent gauge field and ``deconfined degrees of freedom" associated with the fractionalisation of order parameters.
Motivated by the AdS/CFT correspondence, novel condensed matter systems were designed and studied, prtraying the emergence phenomenon,
\cite{Lee1},\cite{Lee2}.

In the high energy community, efforts for describing photons as emergent particles were initially motivated by ideas on superconductivity, transferred to the particle physics realm by Nambu and Jona-Lasinio, \cite{NJL}.
Following the same lead, it was argued that the four-fermion Heisenberg theory with current-current interactions leads to an emergent photon, \cite{Bjorken1,Bjorken2,BB,BZ}.
This theory however, being four-dimensional and non-renormalizable did not allow the study to go far.
There has been however, the same phenomenon, studied in two dimensions in the context of the $CP^{N-1}$ $\s$-model, \cite{cpn,wcpn,Polyakov}.
In such a case, it was established beyond doubt the emergence of a (propagating) vector particle in that theory.

Another form of emergence is at work in the work of Seiberg on the phases of N=1 superQCD, \cite{seiberg}.
In
\cite{Komar} the analogue of the $\rho$-mesons of the strongly coupled IR theory were interpreted as the dual, weakly coupled magnetic gauge bosons of Seiberg, providing another interesting example of (non-abelian) emergence.

  That said, there already existed various models in the context of high energy physics such as the $\mathbb{CP}^N$ model in which a global symmetry becomes effectively gauged at low energies that have a close affinity to the models studied traditionally in condensed matter physics.

The AdS/CFT correspondence, \cite{malda}, and its generalizations (holography), have given an extra boost, and a ``new dimension" to the concept of emergence.
In holographic duals, what is a (gauge-invariant) bound-state of generalized  gluons, becomes in the gravitational description a graviton, and photon, or any other bulk particle. Gauge invariance and diffeomorphism invariance is emergent (similarly to how it happens in string theory).

The emergence of gauge bosons and related constraints was studied  in \cite{Harlow}, where the connection also with the \emph{weak gravity conjecture} was highlighted. In \cite{SMGRAV} a program was outlined, motivated by the holographic correspondence, on how to obtain gravity and other long range interactions from ``hidden" holographic theories, coupled to the Standard Model (SM) weakly in the IR.
The emergence of axions was studied in detail in \cite{axion} and new forms of axionic dynamics has been uncovered.

{The effective action for the emergent vector has many similarities with the actions studied in \cite{KT}. The context studied there is of vector bosons (and gravitons) that are Goldstone bosons of broken Lorentz invariance.}

In composite approaches to gauge interactions, the major difficulties lie in producing a theory with a sensible strong-coupling structure, and appropriate long-distance dynamics.
The popularity of making composite gravitons (and gauge bosons)
has also motivated the well-known Weinberg-Witten (WW) Theorem, \cite{WW} that provides strong constraints on composite gravitons (and composite gauge bosons).

Under a set of assumptions that include Lorentz invariance, well-defined particle states, a conserved covariant energy momentum tensor and a Lorentz covariant and gauge-invariant conserved global currents,  the theorem excludes

$\bullet$ massless particles of spin $s>{1\over 2}$ that can couple to a conserved current, and

$\bullet$ massless particles with spin $s>1$ that can couple to the energy momentum tensor.

Its assumptions however, allow for several loop-holes that help evading the theorem in known cases.
In particular, in does not exclude Yang-Mills theory, as in that case the global conserved currents are not gauge-invariant.
It also does not exclude a ``fundamental" graviton coupled to matter, as  the Lorentz-covariant energy-momentum tensor is not conserved (but only covariantly conserved).

Another counterexample is presented by the massless $\rho$-mesons at the lower-end of the conformal window of N=1 sQCD\footnote{{There is long history on attempts to describe (composite) $\rho$-mesons in terms of a low-energy gauge theory, coming under the name of ``hidden symmetry", \cite{hidden}.
Its proper realization was explained in the context of the AdS/CFT correspondence, \cite{hh}. The situation in sQCD however is distinct, as it has further ingredients that allow the $\rho$-mesons to become light/massless  and weakly interacting.}}, \cite{Komar}.
They evade the WW theorem because of the emergent gauge invariance, associated to them.

In any relativistic quantum field theory, with a global U(1) symmetry, there is at least one state with the quantum number of a U(1) gauge boson. It is the state generated out of the vacuum by the action of the (conserved) current. In weakly-coupled theories this state is unique, while in strongly coupled theories there may be several such states generated by the action of the current. If a theory possesses a  large-N, strong coupling limit, then the width of such states vanishes and there is an infinite number (or a continuum) of them.

In weakly-coupled theories a state generated by the current, is a multi-particle state and therefore its effective interactions are expected to be non-local. In the opposite case, where the interactions are strong, we expect such a state to be tightly bound. If its ``size" is $L$, then we might hope that at distances $\gg L$ the effective interactions of such a state may generate a vector-interaction, plausibly a gauge theory.

In particular, in a theory with an infinite coupling, we expect to have an emergent photon  with a point-like structure.
If the theory is not conformal, and has a gap, then we expect a discrete spectrum of such states, associated to the (generically complex) poles of the two-point function of the U(1) current. In a generic strongly-coupled theory, most such states will be unstable. In YM, to pick a concrete example, such states are generated by a vector composite operator (that is not a conserved current) and are in the trajectory of the $1^{+-}$ glueball.
If we consider QCD instead of YM, then we have conserved currents and novel massive vectors associated with them.
If instead, the strongly-coupled theory is conformal, then the spectrum of vector bound-states forms a continuum.

In the context of holography, the masslessness of the higher-dimensional gauge bosons is explained by the conservation of the associated global current of the dual QFT.
Once this conservation is violated, either explicitly by a source boundary condition, or spontaneously (a vev boundary condition) the gauge field in the bulk obtains a mass, and the associated current an anomalous dimension.

 Therefore the ``masslessness" of the higher dimensional vector is the avatar of the global invariance of the dual QFT and the conservation of the associated boundary current. This does not however imply that the four-dimensional (emergent) vector is massless.
Indeed, in N=4 sYM defined on Minkowski space we obtain a continuum of spin one states starting at zero mass\footnote{This is the reason this theory evades the WW theorem which assumes among other things a isolated bound-state. The WW theorem involves a subtle limit to define the helicity amplitudes
that determine the couplings of massless states to the stress tensor or a
local current.  This limiting procedure is not valid in theories where the states form a continuum.}.
If instead we define the theory on $R\times S^3$, then  the vector spectrum is discrete, but the theory has lost Lorentz invariance.

We therefore learn from the holographic duality that,

$\bullet$ Strong coupling in QFT makes emergent/composite vectors tightly-bound states.

$\bullet$  Large N makes emergent/cosmposite vectors weakly interacting.

Both properties are essential in obtaining a semiclassical and local theory of (composite) vector interactions.

The AdS/CFT intuition therefore suggests that semiclassical effective vector interactions with composite vectors are expected to emerge from holographic quantum field theories, {where the semiclassical nature of the interaction is related to the large-N limit}.
If we are to describe ``dark photons" as emergent vectors coupled to the SM
then we must seek their emergence in a hidden holographic theory.
The simplest way\footnote{There may be more exotic variations on this theme. The SM could be part of the semiclassical holographic theory, and its elementary fields, to be composites of more elementary fields. Or it could be that parts of the SM are composite and others elementary. Crude holographic translations of these possibilities have been considered in the past in the context of the RS realizations of the SM, \cite{csaki}.} is to postulate that, \cite{SMGRAV},

$\bullet$ The whole of physics is described by a four-dimensional QFT.

$\bullet$ The total UV QFT contains a holographic part that is distinct from the (UV limit of the) SM. We shall call it the ``hidden" (holographic) theory.

$\bullet$ This holographic part is coupled in the UV to the standard model via a messenger sector. It consists of fields transforming as bi-fundamentals under the gauge group of the ``hidden" theory, and the gauge group of the SM.
We shall call these fields the messengers. Their mass $M$  shall  be assumed to be much larger than any of the SM scales.

$\bullet$ At energies $\ll M$ we may integrate out the messengers and we end up with an effective theory consisting of the SM coupled to the hidden holographic theory via irrelevant interactions\footnote{There is one possible exception to this statement and it is connected to the gauge hierarchy problem.}.

$\bullet$ Although all operators of the hidden theory are coupled weakly at low energies to the SM, the SM quantum corrections generate ${\cal O}(M)$ masses for all of them with a few notable exceptions that are protected by symmetries: the graviton, the universal axion, \cite{axion}, and exactly conserved global currents, \cite{u1}.

Some of the relevant issues of this rather general setup have been discussed in \cite{SMGRAV}.
In this paper we  undertake a closer look at the emergence of vector interactions. The case of emergent gravity will be treated in a companion paper, \cite{grav}.

\subsection{Results and Outlook}

Our results are as follows
\begin{itemize}

\item We define and establish the dynamics of emergent/composite vectors in a given QFT using an appropriately defined effective action for an abelian global U(1) current that is conserved. We show that the Schwinger functional can be defined so that it is locally gauge invariant. The effective action involves a dynamical vector and has no gauge invariance. It summarizes the propagation and interactions of composites generated by the global U(1) current.

\item In the presence of charged sources the emergent U(1) vector is effectively massive, and in gapped theories one can expand the effective action in a derivative expansion.

\item In the absence of charged sources, the interactions of the emergent vector are non-local. They can be described however using an auxiliary antisymmetric two tensor that turns out to be massive. {The reason for the non-locality is not due to effects of massless modes. Rather it is the effect of an emergent gauge invariance, which produces zero modes and renders the two-point function of the global current non-invertible. To state it simply, in the absence of charged sources, the U(1) current is exactly conserved and its two point-function non-invertible.}

\item When two independent QFTs are coupled via a current current interaction, a linearized analysis indicates that the presence of the coupling induces an vector-mediated interaction of currents in each theory. From the point of view of one of the theories (the ``visible" theory), this interaction is mediated by an emergent vector field and its propagator is the inverse of the current-current correlator of the ``hidden" theory.

    This interaction is always repulsive among identical sources if the theories are unitary. Isolated poles in the two-point function of the current in the hidden theory amount to massive vector exchange while continuous spectra in the correlator provide other non-standard types of behavior for the emergent vector interaction.

\item The linearized analysis can be complemented with a fully non-linear formulation that describes the dynamics of the emergent vector field in the visible theory. There are several distinct cases of coupling the two theories that give rise to distinct symmetries and dynamics that we enumerate below.
    \begin{enumerate}

 \item A hidden theory with a global U(1) symmetry and a current-current coupling to a SM global (non-anomalous) symmetry. An example could be $B-L$ in the SM.
In that case, the result is that the hidden global symmetry will generate a (generically massive) vector boson that will couple to the B-L charges of the SM.  In such a case, the combined theory still has two independent U(1) symmetries, but one of them only is visible in the SM, (the hidden symmetry is not visible from the point of view of the SM).

 \item A hidden theory with a global U(1) symmetry and a coupling between a charged operator in the hidden theory to a charged operator of the SM under a SM  global (non-anomalous) symmetry. Such a coupling breaks the two U(1)'s into a single diagonal U(1). This leftover U(1)  couples to the emergent vector boson.

\item In the two cases above, the U(1) global symmetry of the SM may also be an anomalous global symmetry, like baryon or lepton number.   Although, some of the properties of the new vector interaction remain similar to what was described above, there are new features that are related to the anomaly of the global SM symmetry. In such a case, we expect to have similarities with the anomalous U(1) vector bosons of string theory\footnote{Anomalous U(1) symmetries abound in string theory, \cite{review}. In SM realizations in orientifold vacua, the SM stacks of branes always contain two anomalous U(1) symmetries and generically three, \cite{akt,ADKS}. Their role in the effective theory can be important as (a) they are almost always the lightest of the non-standard model fields, due to the fact that their masses are effectively one-loop effects, \cite{KA,AKR}. A typical such effective theory is analyzed in detail in \cite{CIK}.}.

\item   A hidden theory with a global U(1) symmetry and a current-current coupling to the (gauge-invariant) hypercharge current.
    In this case both the hypercharge gauge field and the emergent vector couple to the hypercharge current. By a (generically non-local) rotation of the two vector fields, a linear combination will become the new hypercharge gauge field while the other will couple to $|H|^2$ where $H$ is the SM Higgs.

\item  A hidden theory with a global U(1) symmetry, whose current is $J_{\m}$  and a coupling with the hypercharge field strength of the SM of the form
\be
S_{int}={1\over m^2}\int d^4 x F^{\m\n}(\pa_{\m}J_{\n}-\pa_{\n}J_{\m})
\ee
where the scale $M$ is of the same order as the messenger mass scale.
By an integration by parts this interaction is equivalent to the previous case, using the equations of motion for hypercharge.

\end{enumerate}

In all of the above we have an emergent U(1) vector boson that plays the role of a dark photon, and in this context the single most dangerous coupling to the standard model is the leading dark photon portal: a kinetic mixing with the hypercharge, \cite{ship}.
We can make estimates of such dangerous couplings using (weak coupling) field theory dynamics, but it is also interesting to make such estimates using dual string theory information at strong coupling. Such a study is underway, \cite{u1}.

\item The case where the hidden theory is a holographic theory (large-N, strong coupling) holds special interest. In such a case, the hidden theory is described by a bulk gravitational theory, that is coupled with the visible theory at a  special bulk scale. The gravitational picture is that of a brane embedded in a holographic bulk, as in earlier brane-world setups. The global U(1) current in the (hidden) holographic theory is now dual to a  bulk gauge field that couples to charged states on the brane. An induced brane localized kinetic term emerges on the brane due to SM quantum corrections.

In the ultimate IR, the effective vector coupling $g_{IR}$ and mass $m_{IR}$ for such a emergent photon are given by
\be
{1\over g_{IR}^2}\simeq {d_2\over d_0^2} {1\over  mg_5^2}+{1\over g_4^2}
\sp {m_{IR}^2\over g_{IR}^2}\simeq {m^2\over d_0(mg_5^2)}+{m_0^2\over g_4^2}
\label{i26}\ee
In the formulae above, the first contribution on the left-hand side is due to the bulk theory and the second contribution is due to the SM.
In particular, $m$ is the dynamical scale of the hidden theory, $g_5$ the bulk gauge coupling constant, $g_4$ is the dimensionless induced gauge coupling constant on the brane, due to the SM quantum corrections.  $m_0$
is a  possible mass generated on the brane, if spontaneous symmetry breaking occurs.
Finally, the coefficients $d_0,d_2$ are dimensionless coefficients appearing in the bulk to bulk propagator of the gauge boson.

\item In the holographic context, the coupling controlling the interactions of the emergent vector is naturally weak, in the large N limit.
    The same applies  to the bulk contributions to its mass, similarly to what happens to gravitons in a similar setting, \cite{self}.

\item  Unlike fundamental dark photons, the emergent photons described here, especially in the holographic context, have propagators that at intermediate energies behave differently from fundamental photons and therefore can have different phenomenology and different constraints from standard elementary dark photons. The same was found recently for , emergent axions in \cite{axion}

\item The most dangerous coupling of emergent vectors to the SM is via the hypercharge portal, \cite{ship}, which is a kinetic mixing with the field strength of the hypercharge. Such a coupling is severely constrained by data. This coupling is naturally suppressed in our context (large N), but a detailed study of this issue will appear in the near future, \cite{u1}. It turns out that when the hidden theory is at  strong coupling, there is additional suppressions of the hypercharge mixing term.

\end{itemize}

The structure of this paper is as follows. In section \ref{Generalsetup}
we describe our general setup concerning the required properties of the hidden theory and its coupling to the SM. In section \ref{effe} we describe the definition and properties of the effective action for a global U(1) current in a QFT. In section \ref{JJinteractionlinearised} we describe the coupling of a hidden and a visible theory, and the induced vector interaction in the visible theory in the linearized approximation.
In section \ref{couplednonlinear} the non-linear theory for the emergent photon is formulated. In section \ref{Hologaxion} the holographic emergence theory is treated. In section \ref{singleNL} we analyze the non-local effective action for the emergent vector in the absence of sources. Finally section \ref{Dis} contains a discussion of the results.

In appendix \ref{Examples} we present the current two point function in free bosonic and fermionic theory.
Appendix \ref{Spectralrepresentation} contains a review of the K\"allen-Lehmann
representation of the current two-point function. In appendix \ref{effectivestaticpotential} we survey the long-distance behavior of the static potential mediated by the emergent vector. In appendix \ref{multiplefields} we analyze the effective action of the current vev in the presence of multiple charged sources. In appendix \ref{high} we analyze the structure of the higher derivative terms in the effective action and the issue of emergent gauge invariance. In appendix \ref{complete} we derive the full effective action by Legendre-transforming also the charged sources. Finally in appendix \ref{gau12} we derive the vector bulk propagators relevant for the holographic case.

\section{The general setup}\label{Generalsetup}

{

The starting point of our analysis is described in \cite{SMGRAV}, namely a UV-complete theory that in the IR splits into two weakly interacting IR sectors\footnote{There are several similarities between these models and what have been called ``hidden valley" models in the literature, \cite{stras}.} . One of these sectors, will be identified with the ``visible" theory  which for all practical purposes is the Standard Model and some of its direct extensions.
The other, that we call the ``hidden sector", and we shall denote as $\h{QFT}$, is a theory that in the IR is weakly interacting with the SM.

The complete theory  is defined on a flat (non-dynamical) space-time background $g_{\m\n}\equiv \eta_{\m\n}$.
From the IR point of view, the two IR sectors are connected by irrelevant interactions. These interactions have a characteristic scale $M$, that si assumed to be well above all scales of the SM and the IR $\h{QFT}$.

A way to think about the origin of such a coupling, is to think of two distinct gauge theories that are coupled together with a set of bi-fundamental fields (we call them the messengers) with mass $M$ in such a way that the total theory is UV-complete.
This amounts to the fact that the UV limit of the total theory is well defined and is given by a four-dimensional CFT.
The messenger mass $M$ controls the strength of the interactions between the two IR sectors, the SM and the hidden theory, $\h{QFT}$.

At energy scales much smaller than the messenger mass, $M$, the messengers can be integrated-out leaving the hidden $\h{QFT}$ interacting with the visible one via a series of non-renormalizable interactions. In this paper, we  focus in particular in studying the effective induced interactions that are related to global $U(1)$ symmetries. The study of other types of symmetries is undertaken in \cite{grav,axion} and results in effective theories of emergent gravity and axions. We now give a few more details and review the precise setup originally described in~\cite{SMGRAV,grav}.

\subsection{The assumptions}

Having described the idea, we now make explicit our assumptions on the class of theories we consider.

}

Our starting point is a local relativistic quantum field theory. We assume that this quantum field theory has the following features:
\begin{itemize}

\item[$(a)$] It possesses a large scale $M$ and all the other characteristic mass scales $m_i \ll M$.

\item[$(b)$] At energies $E\gg M$ the dynamics is described by a well-defined ultraviolet theory. For example this could be a UV fixed point described by a four-dimensional conformal field theory.\footnote{We could also envisage a more exotic UV behavior involving higher dimensional QFTs or some form of string theory.}

\item[$(c)$] At energies $E\ll M$ there is an effective description of the low energy dynamics in terms of two separate sets of distinct quantum field theories communicating to each other via irrelevant interactions. We shall  call the first quantum field theory the {\it visible} QFT and shall denote all quantities associated with that theory with normal font notation. We  call the second quantum field theory the {\it hidden} $\h{QFT}$ and shall denote all its quantities with a hat notation. Schematically,  we have the following low energy description in terms of an effective action
\be
\label{setupaa}
S_{IR} = S_{visible}(\Phi) + S_{hidden}(\h{\Phi}) + S_{int}(\Phi,\h{\Phi})
~,
\ee
where $\Phi$ are collectively the fields of the visible QFT and $\h{\Phi}$ the fields of the hidden QFT. The interaction term $S_{int}$ can be formally described by a sum of irrelevant interactions of increasing scaling dimension
\be
\label{setupab}
S_{int} = \sum_i \int d^4 x\, \lambda_i \, \vis{O}_i(x) \h{O}_i (x)
~,
\ee
where $\vis{O}_i$ are general operators of the visible QFT and $\h{O}_i$ are general operators of the hidden QFT. $S_{int}$ arises by integrating out massive messenger degrees of freedom of the UV QFT with characteristic mass scale $M$. This scale then defines a natural UV cutoff of the effective description. It is hence a physical scale determining the point in energy where the theory splits into two sectors, weakly interacting with each other at low energies.

\item[$(d)$] If we further assume that the hidden $\h{QFT}$ is a theory with mass gap $m$, at energies $E$ in the range $m\ll E \ll M$ we can employ the description \eqref{setupaa} to describe a general process involving both visible and hidden degrees of freedom. For energies $E\ll m \ll M$ on the other hand, it is more natural to integrate out the hidden degrees of freedom and obtain an effective field theory in terms of the visible degrees of freedom only.

\end{itemize}

Our main focus then will be the low energy $(E\ll M)$ behaviour of observables {\it defined exclusively in terms of elementary or composite fields in the visible QFT}, relevant for observers who have only access to visible QFT fields. In addition we  focus in an effective description of global $U(1)$ symmetries and the possibility that these symmetries have to appear as gauged symmetries in the low energy effective description.

More explicitly, we  consider the generating functional of correlation functions (Schwinger functional) for the visible QFT defined as
\be
\label{setupac}
e^{- W(\JJ)} = \int [D\Phi] [D\h{\Phi}] \, e^{-S_{visible}(\Phi,\JJ) - S_{hidden}(\h{\Phi}) - S_{int}(\vis{O}_i, \h{O}_i)}
~.
\ee
We use a Euclidean signature convention (that can be rotated to Lorentzian) where $\JJ$ is collective notation that denotes the addition of arbitrary sources in the visible QFT. This path integral is a Wilsonian effective action below the UV cutoff scale $M$. By integrating the hidden sector fields $\h{\Phi}$ we obtain
\be
\label{setupad}
e^{- W(\JJ)} = \int [D\Phi] \,  e^{-S_{visible}(\Phi,\JJ) - \WW (\vis{O}_i) }
~,
\ee
where $\WW$ is the generating functional in the hidden QFT,
\be
 e^{-{\cal W}(\h{J})}\equiv \int  [D\h{\Phi}] \, e^{- S_{hidden}(\h{\Phi}) - \int \h{O}\h{J}}
~.
\ee
We first observe that from the point of view of the hidden QFT, the visible operators $\vis{O}_i$ appearing in the interaction term $S_{int}$ in \eqref{setupaa} and \eqref{setupab} are dynamical sources.

 {In \eqref{setupad}, an observer in the visible sector registers a formal series of increasingly irrelevant interactions. We would like to understand when it is possible to reformulate these interactions by integrating-in a set of (semi-)classical fields. We  focus in the case of $U(1)$ symmetries and we shall try to understand under which conditions, the effective action for these fields is a sensible $U(1)$ vector  theory.}

\section{The effective action for a global conserved current\label{effe}}

In this section we consider a simpler framework for the emergence of gauge invariance. We  consider a single QFT with an exact U(1) global symmetry. Our analysis is general but we also present an explicit example by expanding the Schwinger functional in an IR derivative expansion in section \ref{Quadraticexample}. Other global internal symmetries can be treated in a similar fashion. Anomalies in global symmetries can be treated also but we shall not do this here. Some elements in this direction were given in \cite{lorentz}. In appendix~\ref{Effectiveactionexamples} we present more general effective actions, for example in the case of multiple fields or in the presence of higher derivative corrections.

\subsection{The effective action for a U(1) symmetric theory}\label{effectiveactionsingletheory}

Consider a theory with a global U(1) symmetry, and an associated conserved current, $J_{\m}$. We  define and study the Schwinger source functional for the current as well as charged operators in that theory.
  To simplify matters, we include a single charged scalar operator $O(x)$ with charge one, as it is enough to indicate the relevant issues.

Let $O$ be a complex operator charged (with charge 1) under the global symmetry current $J_{\m}$. We construct the extended source functional by adding the appropriate sources to the action of the theory
\be
 S(\phi,A,\Phi)=S(\phi)+\int d^4x\left[ J^{\m}(x)A_{\m}(x)+\Phi(x) O^*(x)+\Phi^*(x) O(x)\right]+\cdots
\label{A18p}\ee
where $\phi$ denote collectively the quantum fields of the theory.
The Schwinger source functional is then defined as
\be
e^{-W(A,\Phi)}\equiv \int  {\cal D}\phi~e^{-S(\phi,A,\Phi)}
\label{A19p}\ee

It is well known that the Schwinger functional of a source gauge field $A_{\m}$ coupled to a conserved global symmetry current $J_{\m}$ has  a local gauge invariance if defined properly, (see \cite{lorentz} for example). This remains true if the global symmetry has the usual triangle anomalies, \cite{lorentz}.
The ellipsis in the formula (\ref{A18p}) contains possible terms that have to be added to restore local gauge invariance.

The functional $W(A,\Phi)$ is therefore locally gauge-invariant under the U(1) gauge transformations
\be
A_{\m}\to A_{\m}+\pa_{\m}\e(x) \sp \Phi\to \Phi~e^{-i\e(x)}
\label{A1}\ee
This is equivalent to the standard Ward identity
\be
\pa_{\m}{\delta W\over \delta A_{\mu}}+i\left(\Phi{\delta W\over \delta \Phi}-\Phi^*{\delta W\over \delta \Phi^*}\right)=0
\label{A20p}\ee

The gauge invariant completion of the Schwinger functional is not unique.
Like in the case of translational symmetries, the conserved current can be modified  by adding topological\footnote{By ``topological" we mean currents that are identically conserved independent of the dynamics of the theory. Standard topological currents that are associated with topological invariants, as for example the Chern-Simons current, are special cases of our definition here. For example, for any antisymmetric local gauge-invariant operator $O_{\m\n}$, $J_{\n}\equiv \pa^{\m}O_{\m\n}$ is conserved identically.
This example is of course the tip of the iceberg. For example the current $J^{\m}={\delta W\over \delta A_{\m}}$ with $W(A)$ an arbitrary local gauge-invariant functional, without minimally charged sources, is identically conserved.} currents (currents that are conserved identically without using the equations of motion).
This is equivalent to adding local gauge invariant terms to $W$. As an example, the following addition
\be
W'=W+{1\over 4}\int d^4x ~Z(|\Phi|^2)F_{\m\n}F^{\m\n}
\label{a2}\ee
modifies the current by
\be
J'_{\m}\equiv {\delta W'\over \delta A^{\m}}=J_{\m}-\delta J_{\m}\sp \delta J_{\m}= \partial^{\n}\left[Z(|\Phi|^2)F_{\n\m}\right]
\label{a3}\ee
As is obvious $\delta J_{\m}$ is identically conserved,
\be
\partial^{\m}\delta J_{\m}=0\;.
\ee
without using the equations of motion.

To fix this ambiguity we define the gauge-invariant extension of the Schwinger functional by minimal substitution, which in the U(1) case is a unique prescription as covariant derivatives commute. Generically however, if we define the U(1) current in the presence of arbitrary gauge fields, this fixes the enormous  scheme ambiguity of the Schwinger functional.

It should be also mentioned that the gauge invariance of the Schwinger functional is explained by the fact that the operator $\partial^{\m}J_{\m}$ is zero on-shell, and is therefore a redundant operator of the theory. Perturbing the theory with it, has no effect.

It should be stressed that the gauge invariance of the Schwinger functional is always underlying a U(1) global symmetry. However, in the presence of arbitrary non-trivial charged sources, the U(1) current obtained from the Schwinger functional is not anymore conserved. Its divergence is a linear combination of charged sources and ``equations of motion".
 This is explicitly visible in the Ward identity (\ref{A20p}).
 In the special case that the Schwinger functional is extremal with respect to the charged sources, ($\Phi$ in our case), then the current is conserved.
This amounts to setting ${\delta W\over \delta \Phi}={\delta W\over \delta \Phi^*}=0$ in (\ref{A20p}). Extremality of the functional with respect to the charged sources is equivalent to the absence of charged expectation values, as expected.

Having fixed this ambiguity, we should remember that, as usual in QFT, $W(A)$ is UV divergent and requires regularization and renormalization. We shall  not delve a lot in this direction as it has been studied for decades, but suffice to say that this can be done even for chiral non-anomalous symmetries. That may be done without breaking the global symmetry  in the regularized theory.  Even when the global symmetry is broken at finite cutoff, techniques exist to provide a finite renormalized Schwinger functional that is gauge-invariant, and has a few extra parameters associated with standard scheme dependence because of  renormalization.

 We now assume that we have in hand a concrete renormalized Schwinger functional with U(1) gauge invariance.
We shall define the effective action for the current by Legendre-transforming $W$, only with respect to the gauge field. Since the fields $\Phi$ could also be sources with no dynamical equations of motion, we therefore need an appropriate definition that holds both for dynamical and non-dynamical $\Phi$. The Legendre transform should also respect the original gauge invariance exemplified by~\ref{A1} and~\ref{A20p}.

We now define the Legendre transform with respect to an arbitrary original background $\mathbf{A}_\mu$. $\mathbf{A}_\mu$ could be also trivial. The Legendre transform is
\be
\Gamma(\tilde V,\Phi,\mathbf{A}_\mu)=\int d^4 x\left[-\tilde V^{\mu}(A_{\mu}-\mathbf{A}_\mu) \right] + W(A,\Phi) \, .
\label{B22p}\ee
The expectation value of the current is
\be
\langle \tilde V_{\m} \rangle \equiv {\delta W\over \delta A^{\mu}} \Bigg|_{A_\mu = \mathbf{A}_\mu} \, .
\label{B21p}\ee
The conservation law (\ref{A20p}) hence becomes
\be
\pa^{\mu} \langle \tilde V_{\m} \rangle + i\left(\Phi{\delta W\over \delta \Phi}-\Phi^*{\delta W\over \delta \Phi^*}\right)  \Bigg|_{A_\mu = \mathbf{A}_\mu} = 0
\label{aa3}\ee
With these definitions, we maintain gauge invariance explicitly, since both the background and the gauge field transform in the same way, while $\tilde{V}$ is invariant.

The effective action $\Gamma$ is constructed to have the property that it is extremal with respect to $\tilde V_{\m}$.
To show this we first obtain by direct functional differentiation
\be
{\delta \Gamma\over \delta \tilde V^{\m}} = (\mathbf{A}_\mu - A_\mu)   -{\delta A^{\n}\over \delta \tilde V^{\m}}\tilde V_{\n}+{\delta W\over \delta \tilde V^{\m}}
\label{a4}\ee
and by using the chain rule
\be
{\delta W\over \delta \tilde V^{\m}}={\delta W\over \delta A^{\n}}{\delta A^{\n}\over  \delta \tilde V^{\m}}=\tilde V_{\n}{\delta A^{\n}\over  \delta \tilde V^{\m}} \, ,
\label{a5}\ee
the variation with respect to the emerging vector field $\tilde{V}_\mu$ is
\be
\frac{\delta \Gamma (\tilde V,\Phi,\mathbf{A}_\mu)}{\delta \tilde{V}_\mu} = (\mathbf{A}_\mu - A_\mu)  \, .
\ee
We therefore find that it is extremal on the background solution
\be
\frac{\delta \Gamma (\tilde V,\Phi,\mathbf{A}_\mu)}{\delta \tilde{V}_\mu} \Bigg|_{A_\mu = \mathbf{A}_\mu} = 0
\ee
In addition with this definition we find
\be
\frac{\delta \Gamma (\tilde V,\Phi,\mathbf{A}_\mu)}{\delta \Phi} = (\mathbf{A}_\mu - A_\mu) \frac{\delta \tilde{V}_\mu}{\delta \Phi} + \frac{\delta W}{\delta \Phi}
\ee
So on the background, the result is
\be\label{B20}
\frac{\delta \Gamma (\tilde V,\Phi,\mathbf{A}_\mu)}{\delta \Phi} \Bigg|_{A_\mu = \mathbf{A}_\mu} = \frac{\delta W}{\delta \Phi} \Bigg|_{A_\mu = \mathbf{A}_\mu} = 0
\ee
where in the last equality we used the on-shell condition on $W$ from the EOM's of $\Phi$. In case $\Phi$ is a non-dynamical source,  equation (\ref{B20}) determines the expectation values of the charged sources. The original conservation law~\ref{aa3} is now written as
\be
\pa_{\mu}\tilde V^{\mu}+i\left(\Phi{\delta \Gamma \over \delta \Phi}-\Phi^*{\delta \Gamma \over \delta \Phi^*}\right)\Bigg|_{A_\mu = \mathbf{A}_\mu}=0
\ee
We therefore find that the two formulations of the problem are equivalent on the background $\mathbf{A}_\mu$, regardless on whether the fields $\Phi$ are dynamical or not.

The effective action $\Gamma(V_{\m})$, describes the complete quantum dynamics of the state generated out of the vacuum by the global U(1) current $J^{\m}$.
The poles of the current two-point function,  become by construction the zeros of the quadratic part of the effective action, $\Gamma(V_{\m})$, and determine therefore the mass and coupling constant of the emergent vector.

\subsection{An explicit example}\label{Quadraticexample}

We would like now to investigate what is the structure of the vector theory described by the action $\Gamma$ and whether it is a gauge theory in disguise.
To analyze this  in  detail, we  use a large-distance expansion parametrization of the Schwinger functional, valid in theories with a mass gap\footnote{The mass gap may not be explicit, but generated by appropriate non-trivial sources.}. We assume, beyond the gauge field source $A_{\m}$, the presence of the source $\Phi$, coupled to an operator with non-trivial U(1) global charge. In the Schwinger functional, $\Phi$ is minimally charged under the gauge field $A_{\m}$.
We expand the Schwinger functional in a long-distance (derivative) expansion,
\be
W(A,\Phi)=\int d^4x\left(W_0(|\Phi|^2)+{W_1(|\Phi^2|)\over 4}F_A^2+{W_2(|\Phi|^2)\over 2}|D\Phi|^2+{\cal O}(\pa^4)\right)
\label{A24p}\ee
where $F_A=d A$,
\be
D_{\m}\Phi=(\pa_{\m}+iA_{\m})\Phi\sp D_{\m}\Phi^*=(\pa_{\m}-iA_{\m})\Phi^*\;.
\label{A25p}\ee

From \eqref{A24p} we can compute the current as
\be
\tilde V_{\nu}\equiv {\delta W\over \delta A^{\n}}=-\partial^{\m}(W_1F^A_{\m\n})-i{W_2\over 2}(\Phi^*\pa_{\n}\Phi-\Phi\pa_{\n}\Phi^*)+W_2A_{\n}|\Phi|^2\,+\,{\cal O}(\pa^3)
\label{A26p}\ee

We  now invert the previous expression and compute $A_{\m}$ as a function of $\tilde V_{\m}$ in a derivative expansion. In particular, we obtain an expansion of the form $A_\mu = \sum_{i=0}^\infty A_\mu^{(i)}$, where the $A_\mu^{(i)}$ contains $(i)$-derivatives.
The result for the first few terms is
\be\label{A31p}
A_\n^{(0)} = \hat{V}_{\n} \, ,
\ee
\be
A_\n^{(1)} = \frac{i}{2}\pa_{\n}\log{\Phi\over \Phi^*} \, ,
\ee
\be
A_\n^{(2)} = \frac{1}{W_2\,|\Phi|^2}\,\partial^\m\left(W_1\,F^{\hat V}_{\m\n}\right)\;,
\ee
where
\be
\hat{V}_\m\,\equiv\,{\tilde V_{\m}\over W_2|\Phi|^2}
\label{a6}\ee
and $F^{\hat{V}}_{\m\n}=\pa_{\m} \hat{V}_{\n}-\pa_{\n} \hat{V}_{\m}$.

We shall  truncate our expansion up to two derivatives since our original functional~\ref{A24p} was also valid up to two derivative terms.
Note, that from (\ref{A31p},~\ref{a6}) that $\hat{V}_{\m}$ is gauge-invariant under the original gauge transformation~\ref{A1}.
We may rewrite the equation in (\ref{A26p}) on the background as
\be\label{B21}
\frac{1}{W_2\,|\Phi|^2}\partial^\m\left(W_1\,F^{\hat V}_{\m\n}\right)+~\left(\hat V_{\n} +\frac{i}{2}\pa_{\n}\log{\Phi\over \Phi^*}\right)+\cdots = {\bf A}_\nu
\ee
Interestingly, this equation is gauge-invariant under a different gauge transformation as well:
\be
\hat V_{\m}\to \hat V_{\m}+\pa_{\m}\l \sp \Phi\to \Phi~e^{ i\l}  \, .
\ee
This is however  an artifact of the first orders in the derivative expansion as shown in appendix \ref{high}.

We now proceed to derive explicit expressions for the functionals in a derivative expansion.
Using the definition~\ref{B22p} and the original functional~\ref{A24p}, we compute the effective action $\Gamma$ first in terms of $A_\mu$
\bea
&\Gamma(A_\mu,\Phi,\mathbf{A}_\mu) = \int d^4x\left(W_0(|\Phi|^2)+{W_1(|\Phi^2|)\over 4}F_A^2+{W_2(|\Phi|^2)\over 2}|D\Phi|^2+\cdots\right) + \nn \\
&+ \int d^4 x (\mathbf{A}^\nu - A^\nu) \left(-\partial^{\m}(W_1F^A_{\m\n})-i{W_2\over 2}(\Phi^*\pa_{\n}\Phi-\Phi\pa_{\n}\Phi^*)+W_2A_{\n}|\Phi|^2\,+\,\dots \right)  \nn \\
\eea
and by using \ref{B21} and keeping terms up to two derivatives we find
\bea\label{B22}
\Gamma(\hat{V}_\mu,\Phi,\mathbf{A}_\mu) &=& \int d^4x\left[W_0 - {1\over 4}W_1 (F^{\hat V})^2+{W_2\over 2}|\partial\Phi|^2-{W_2\over 2}|\Phi|^2\left(\hat{V}_{\m}+\frac{i}{2}\pa_{\m}\log{\Phi\over \Phi^*}\right)^2 \right] +
  \nn \\
  &+& \int d^4x \left[ W_2 |\Phi|^2  \mathbf{A}^\mu  \hat{V}_\mu +\cdots\right]
\eea
or equivalently
\bea\label{B22b}
\Gamma(\hat{V}_\mu,\Phi,\mathbf{A}_\mu) &=& \int d^4x\left[W_0 - {1\over 4}W_1 (F^{\hat V})^2+{W_2\over 2}\left(\partial|\Phi| \right)^2-{W_2\over 2}|\Phi|^2 \hat{V}_{\m} \hat{V}^{\m}\right]+
  \nn \\
  &+& \int d^4x \left[ W_2 |\Phi|^2   \hat{V}_\mu \left( \mathbf{A}^\mu - \frac{i}{2}\pa_{\m}\log{\Phi\over \Phi^*} \right) +\cdots\right]\;.
\eea

Splitting into radial and phase components $\Phi = R e^{- i \Theta}$
\bea\label{B22c}
\Gamma(\hat{V}_\mu,R, \Theta,\mathbf{A}_\mu) &=& \int d^4x\left[W_0 - {1\over 4}W_1 (F^{\hat V})^2+{W_2\over 2}\left(\partial R \right)^2-{W_2\over 2} R^2 \hat{V}_{\m} \hat{V}^{\m}  \right]+
  \nn \\
  &+& \int d^4x \left[ W_2 R^2   \hat{V}_\mu \left( \mathbf{A}^\mu -\partial_\mu \Theta \right) +\cdots\right]
\eea

Some remarks are in order, regarding the EOM's and gauge invariance. In particular the EOM's are given now from (\ref{B21}) where the gauge field on the right-hand side must be replaced by the background gauge field ${\bf A}_{\n}$.

The equations of motion in (\ref{B21})  (as well as the functionals~\ref{B22},~\ref{B22b},~\ref{B22c}) are gauge-invariant with respect to the original gauge symmetry
\be
A_\mu \rightarrow A_\mu + \partial_\mu \epsilon \, , \qquad \Phi \rightarrow \Phi e^{-i \epsilon} \, .
\label{g1}\ee
 We can improve on~\ref{B22c} by absorbing the non-dynamical term $\partial_\mu \Theta$ into the background by shifting $\mathbf{A}_\mu \rightarrow \mathbf{A}_\mu + \partial_\mu \Theta$ to obtain
\bea\label{B24}
\Gamma(\hat{V}_\mu,R, \Theta,\mathbf{A}_\mu) &=& \int d^4x\left[W_0 - {1\over 4}W_1 (F^{\hat V})^2+{W_2\over 2}\left(\partial R \right)^2-{W_2\over 2} R^2  \hat{V}_{\m}  \hat{V}^{\m}   + \right]
  \nn \\
  &+& \int d^4x \left[{W_2\over 2} R^2 \hat{V}_{\m}   \mathbf{A}^\mu  +\cdots\right]
\eea

We end up with a dynamical theory for the vector $V_{\m}$ that has the structure of a vector theory without gauge invariance. The gauge-degree of freedom of the Schwinger functional has disappeared. In the single field case, studied here,  this degree of freedom  corresponds to the phase of $\Phi$. In the multi-charged field case that is worked out in appendix  \ref{multiplefields}, what is removed is the overall gauge degree of freedom.
The end result is that  the effective action involves the vector $V_{\m}$ and gauge-invariant (ie. chargeless) combinations of the charged fields.

The structure of the action in (\ref{B24}) is that of a single vector with a standard kinetic terms, but coupled to a real (uncharged) field $R=|\Phi|$.
 It was shown by Coleman that a massive U(1) theory without gauge invariance has a sensible quantum theory, \cite{Coleman}. The vector here has a mass term that is a function of $R$. This is generic in the presence of charged sources. The uncharged case will be analyzed later in section \ref{singleNL}.

The class of theories we are considering here are generalizations of the above.
Among others, their dynamics contains a general effective potential for the emergent vector.

\section{Emergent coupled U(1)'s: the linearized theory}\label{JJinteractionlinearised}

So far our discussion concerned  the case of a single theory. We shall now move our discussion to the case of a system of coupled QFT's, focusing only on the current operators in the two theories. The effective description of the various interactions, discussed in~\ref{Generalsetup} much below the messenger mass scale $M$, is in terms of an effective current-current interaction $\lambda J \hat{J}$. Later we  also discuss the case where the interaction between the two sectors is mediated by charged operators with a term $\lambda \mathcal{O} \widehat{\mathcal{O}}$. This more general possibility is discussed in section~\ref{couplednonlinear}.

In particular, we  now study the case of two QFTs coupled by the $J \hat{J}$ interaction via (we now work directly in mostly plus Lorentzian signature)
\be
\label{linearaac}
S_{int} = \lambda \, \int d^d x\,  J^\mu(x) {\hat J}_\mu(x)
~,\ee
where $J^\mu$ and ${\hat J}^\mu$ are conserved abelian U(1) currents for the visible and hidden QFTs. The parameter
scales as $\lambda \sim 1/M^{2}$.
For all dimensions above two, this is an irrelevant interaction. In two dimensions it is marginal top leading order and has been studied widely in the past.
A similar setup of this deformation, similar to how we currently treat the $TT$ interactions was advanced in \cite{gk} in the marginal case, and more recently in \cite{sf} for the relevant case.

In the presence of \eqref{linearaac}, the generating functional of correlation functions in the visible QFT is
\begin{align}
\begin{split}
\label{linearab}
e^{i W(\JJ)} &= \int [D\Phi] [D\h{\Phi}] \, e^{i S_{vis}(\Phi,\JJ) + i S_{hid}(\h{\Phi}) + i \lambda \int d^4 x\,  J^\mu(x) {\hat J}_\mu(x)}
\\
&=
\int [D\Phi] [D\h{\Phi}] \, e^{i S_{vis}(\Phi,\JJ) + i  S_{hid}(\h{\Phi})}
\bigg[ 1 + i \lambda \int d^4 x\,   J^\mu(x) {\hat J}_\mu(x)
\\
&\hspace{1.5cm} - \frac{1}{2} \lambda^2 \int d^4 x_1 d^4 x_2 \,  J^\mu(x_1) {\hat J}_\mu(x_1)  J^\nu(x_2) {\hat J}_\nu(x_2)  +\OO(\lambda^3) \bigg]
~,
\end{split}
\end{align}
where in the second equality we expanded the path integral perturbatively in $\lambda$ up to second order.
The second term on the second line involves the one-point function of the current ${\hat J}_\mu$ in the undeformed hidden theory and the term in the third line its two-point function. We  also assume that in the absence of the interaction \eqref{linearaac}, ${\hat J}_\mu$ is the conserved current of a Lorentz-invariant QFT. This means that the one point function of the current operators in the vacuum is taken to be zero.

Recall now, the standard derivation of the Ward identities associated with the global U(1) symmetry in the hidden QFT
\bea
\label{genpertuae}
0 = \int \DD {\hat \Phi} \, e^{i \int d^4 x {\hat \LL}} \bigg\{ -i \int d^4 x \, \p_\mu \theta(x)
\bigg[ {\hat J}^\mu(x) {\hat J}^\nu(y) \bigg]
+ \delta_\theta {\hat J}^\nu (y) \bigg\}
~.\eea
Using
$$\delta_\theta {\hat J}^\nu = -i \p^\rho \theta \, {\hat {\mathfrak J}}^{\nu}~\hskip-5pt_{\r}$$
 and dividing by the partition function $Z$ we obtain the Ward identity
\be
\label{genpertuaf}
\langle \p_\mu {\hat J}^\mu (x) {\hat J}^\nu (y) \rangle = - \p^\rho \left(\delta(x-y) \langle {\hat {\mathfrak J}}^{\nu}~\hskip-5pt_{\r} \rangle \right)
~.
\ee
Integrating both sides with $\int d^d x \, e^{-ik x}$ converts to the momentum space expression
\be
\label{genpertuag}
k_\mu \langle {\hat J}^\mu (k) {\hat J}^\nu (-k) \rangle = - k^\rho \langle {\hat {\mathfrak J}}^{\nu}~\hskip-5pt_{\r}\rangle
~.
\ee
The 1-point function on the RHS of this equation does not necessarily vanish. Typically, the operator ${\hat {\mathfrak J}}^{\nu}~\hskip-5pt_{\r}$ is non-zero and its 1-point function will not vanish if the operator is mixing with the identity. Such mixing is possible in theories with intrinsic scales, for example if the hidden theory has a mass gap $m$. From Lorentz invariance we therefore expect
\be
\label{genpertuai}
\langle {\hat {\mathfrak J}}^{\nu}~\hskip-5pt_{\r}(y) \rangle = i \AA\, {\delta^\nu}_\rho
\ee
where $\AA$ is a dimensionfull constant. Its dimension arises from the mass scale $m$ of the hidden QFT. Then,
\be
\label{genpertuaj}
k_\mu \langle {\hat J}^\mu (k) {\hat J}^\nu (-k) \rangle = - i \AA ~k^\nu
~.
\ee
As explicit examples, in appendix~\ref{Examples}, we consider the case of free massive bosons and fermions. For free massive bosons $\varphi$ we have
\be
{\hat {\mathfrak J}}^{\nu}~\hskip-5pt_{\m} \propto \varphi \varphi^* \, {\delta^{\nu}}_\mu\;,
\ee
and the vev $\langle (\varphi \varphi^*)(x) \rangle$ is indeed non-vanishing at non-vanishing mass $m$ (as a simple perturbative computation in $m$ reveals). We have verified this mixing by a straightforward computation of the 2-point function $\langle {\hat J}^\mu (k) {\hat J}^\nu (-k) \rangle$ in the appendix \ref{Examples}. On the other hand, for free massive fermions, ${\hat {\mathfrak J}}^{\mu}~\hskip-5pt_{\n}=0$ identically. As a result, in this case we do not expect a contact term violation of the classical Ward identity as can be again verified by explicit computation.

Since we have coupled the hidden theory to the visible sector, we  now use the upper-index $(0)$ and the lower-index $hid$ to denote that such expectation values are to be computed in the undeformed hidden theory. In particular the undeformed one and two-point functions that we shall use are
\be
\label{genpertuai2}
\langle {\hat {\mathfrak J}}^{\nu}~\hskip-5pt_{\r} \rangle^{(0)}_{hid} = i \AA\, {\delta^\nu}_\rho \, ,
\ee
\be
\label{genpertuad}
i {\hat G}_{\mu\nu}(k) = \langle {\hat J}_\mu (k) {\hat J}_\nu(-k) \rangle^{(0)}_{hid}
~,\ee
where ${\hat G}_{\mu\nu}(k)$ is the momentum space propagator.
A spectral representation of the two point function for a general hidden theory can be found in appendix~\ref{Spectralrepresentation}.

Finally, denoting the partition function of the undeformed hidden theory as $e^{i W^{(0)}_{hid}}$ and assuming that the currents are conserved in the corresponding undeformed theories --- so that we can use the results of the Ward identity --- we can recast \eqref{linearab} up to $\OO(\lambda^3)$ as
\begin{align}
\begin{split}
\label{linearad}
e^{i W(\JJ)} & = e^{i W^{(0)}_{hid}} \int [ D \Phi ]\, e^{i  S_{vis}(\Phi, \JJ)}
\bigg[  1 -
\frac{i}{2}\lambda^2
\int d^4 x_1 d^4 x_2 \, J^\m (x_1)\, J^\n (x_2) {\hat G}_{\mu\nu}(x_1 - x_2)  \bigg]
~.
\end{split}
\end{align}
In this expression, the one-point function of the hidden current is taken to be zero in the Lorentz invariant vacuum.
In addition, this expression reveals that from the point of view of the visible theory, the interaction \eqref{linearaac} with the hidden theory has induced effective interactions for the visible current. Working up to quadratic order in $\lambda$, we can exponentiate these interactions in an effective action of the form
\be
\label{linearaf}
\delta S_{vis}  =
- \frac{1}{2} \lambda^2 \int d^4 x_1 d^4 x_2 \, J^\m (x_1)\, J^\n (x_2)\,{\hat G}^{c}_{\mu\nu}(x_1 - x_2)
~.
\ee
In this last equation we have also used an upper script $c$ to denote the connected part of this two point function, since it is this connected part that appears in the exponent and obeys the Ward-identity \eqref{genpertuaf}. We observe the emergence of a quadratic visible current-current interaction. In the most general case, we can use the spectral representation of the current two point function (analysed in appendix~\ref{Spectralrepresentation})
\be\label{spectralpositionmain}
 {\hat G}^{c}_{\mu\nu}(x_1 - x_2) = - \int_0^\infty d \mu^2 \int \frac{d^d k}{(2\pi)^4} \frac{e^{- i k (x_1 - x_2)}}{k^2 + \mu^2 - i \epsilon} \rho_{\m \n}(k, \, \mu) \, ,
\ee
where the spectral weight, $\r$, is split into longitudinal and transverse parts
\be
\rho_{\m \n}(k, \, \mu) = \left( \eta_{\m \n}   -  \frac{k_\m k_\n}{\mu^2}  \right){\cal B}( \mu) + \eta_{\m\nu}  \, {\cal A}( \mu)\;.
\label{splitmain}\ee
The current-current interaction can also be expressed in momentum space as
\be
\label{linearafa}
\delta S^{JJ}_{vis} \equiv - \frac{\lambda^2}{2}  \int \frac{d^4 k }{(2\pi)^4}  \, J^\m(-k)\, J^\n(k)\, {\hat G}^{c}_{\mu\nu}(k)
~.
\ee
This part can be reformulated as an interaction with a classical spin-1 field $A_{\mu}$. At quadratic order, the effective action of an emergent vector field $A_\mu$ reads
\be
\label{genpertuab}
S_{eff} = \int d^d k \bigg[ A_\mu (-k) J^\mu (k) - \frac{1}{2} \PP^{\mu\nu}(k) A_\mu(-k) A_\nu(k) \bigg]
~.\ee
In this action, the tensor $\PP^{\mu\nu}$ is proportional to the inverse of the hidden current-current 2-point function
\be
\label{genpertuac}
\left( \PP^{-1} \right)_{\mu\nu} (k) =  -  \lambda^2 {\hat G}^{(c)}_{\mu\nu}(k)
~,\ee
The 2-point function is evaluated in the undeformed, $\lambda=0$, theory.

The Ward identity \eqref{genpertuaj} (together with \eqref{spectralpositionmain}) implies that
\be
\label{genpertuak}
{{\hat G}_{(c)}}^{\mu\nu}(k) =  - \AA \, \eta^{\mu\nu} +  \BB(k^2) \left( k^\mu k^\nu - k^2 \eta^{\mu\nu} \right)
~,\ee
so that
\be
\label{genpertual}
\left( \PP^{-1} \right)^{\mu\nu}(k) =  \lambda^2 \left( \AA \, \eta^{\mu\nu} - \BB(k^2) \left( k^\mu k^\nu - k^2 \eta^{\mu\nu} \right) \right)
~.\ee

There are now two distinct possibilities. If the constant $\AA \neq 0$, then the inversion is straightforward and gives
\be\label{genpertuam}
\PP^{\mu\nu}  ={1 \over \lambda^2 \AA}\left(\eta^{\m\n}+{\BB\over  \AA+ \BB k^2}(k^{\m}k^{\n}-k^2 \eta^{\m\n})\right) \, .
\ee
We therefore find up to quadratic order in the momentum expansion
\be
\label{genpertuam2}
\PP^{\mu\nu} =  \lambda^{-2} \AA^{-1} \left( \eta^{\mu\nu} + \frac{\BB(0)}{\AA} (k^\mu k^\nu - k^2 \eta^{\mu\nu} ) \right) + \OO(k^4)
\ee
The effective action \eqref{genpertuab} takes the form (in real space)
\be
\label{genpertuan}
S_{eff} = \int d^d x \bigg[
A_\mu J^\mu - \frac{1}{2}  \lambda^{-2} \AA^{-1} A_\mu A^\mu - \frac{1}{4} \lambda^{-2} \AA^{-2} \BB(0) F_{\mu\nu}F^{\mu\nu} + \OO(\p^4) \bigg]
~.
\ee
This is an example of a massive-photon action, with its mass arising from a Higgs effect due to the non-vanishing vacuum expectation value $\AA \neq 0$.

On the other hand, if the constant $\AA=0$ (as in the case of free fermions) then the inversion of $\PP^{-1}$ can be performed in two different ways. One possibility is to form a \emph{non-local} effective action \eqref{genpertuab} (taking into account properly gauge fixing conditions etc.).This will be explored further in section \ref{singleNL}. The other possibility is to add and subtract a contact term in \eqref{genpertuak}, so that the inversion is possible. Of course this second possibility is ambiguous and reflects the fact that the true effective interaction between visible sector currents is contained in the action \eqref{linearafa} and not in the IR expansion of \eqref{genpertuab}. Using therefore \eqref{genpertuan} as an effective action can sometimes be misleading, since it can truncate important degrees of freedom\footnote{By expanding in momenta a propagator and its inverse we miss-estimate the mass term that is relevant for the interaction and it can obtain admixtures from contact terms that are irrelevant.}  . We conclude that the effective interaction (\ref{linearafa}) for the visible theory current is unambiguous, whereas the resolved dynamical action in   (\ref{genpertuan}) is scheme dependent.

Some specific examples of the effective action in the cases that the hidden sector fields are free bosons or fermions are given in appendix~\ref{Examples}. In this case the spectral weight $\rho_{\m \n}(\mu)$ has a mass gap $m$ above which there is a continuum of states. For a more general theory, we can use the general spectral representation for the hidden theory current correlator provided in~\ref{Spectralrepresentation}. In particular, for a strongly coupled hidden theory with a discrete spectrum, we expect the appearance of poles in the spectral function. Near such poles one finds a massive photon state as shown in appendix~\ref{Spectralrepresentation}. In particular the effective current interaction on the visible sector takes the following form near such poles
\be
\label{linearafapoles}
\delta S^{JJ}_{vis} \equiv  \frac{\lambda^2}{2}  \int \frac{d^4 k }{(2\pi)^4}  \, J^\m(-k)\, J^\n(k)\,  \frac{R(m_i)}{k^2 + m_i^2} \left(\eta_{\m \n} - \frac{k_\m k_\n}{m_i^2} \right)
~,
\ee
where $R(m_i)>0$ is the positive spectral weight residue near the pole. This interaction can be resolved with a standard Proca field. The effective action is similar to \eqref{genpertuan}, with the difference that the mass term is now governed by the location of the pole $m_i$, and the coupling to the emergent vector field by the residue $R(m_i)$. The interaction between two charged sources of equal sign is repulsive as expected.

In appendix \ref{effectivestaticpotential}  we survey the long -distance behavior of the emergent vector interaction, as a function of the structure of the spectral density of the vector two-point function.
In the case of the discrete spectrum with zero widths and masses $m_i$ we obtain for the static potential a sum of Yukawa interactions.

\be\label{s9}
\Phi(r) \sim \frac{1}{4 \pi r} \sum_i  e^{- m_i r} \, .
\ee
The force between two equal charges is then a repulsive force as expected from the exchange of a massive vector boson. The sign is fixed due to the positivity of the spectral weight. If the isolated pole is at zero momentum, we obtain a long range potential due to an exchange of a massless photon-like state
\be
\Phi(r) \sim \frac{1}{4 \pi r} \, .
\ee
{This case is not obviously excluded by the WW theorem, \cite{WW}, as one of the assumptions is that the massless pole should correspond to a charged state under the current whereas the states generated by a U(1) currets are chargeless.}

A continuum spectral density starting above a mass $M$ as $\rho^{(1)}(\m)\sim (\m^2-M^2)^a$ gives a static potential that at large distances behaves as
\be
\Phi(r) \sim {e^{-Mr}\over r^{a+2}}
 \label{s10}\ee
If the continuum starts at $M=0$ then  (\ref{s10}) is modified to
\be
\Phi(r) \sim {1\over r^{2a+3}}
\label{s11} \ee
It is clear that in both cases, (\ref{s10}) and (\ref{s11}), as the spectral density must be integrable, the exponent in the denominator, is allowed to approach 1, but cannot reach it as in that limit a logarithmic divergence appears in the density of states.

Finally for a conserved current in a CFT, we obtain
\be
\Phi(r) \sim {1\over r^{5}}
\ee
in agreement with (\ref{i22}).

We conclude by mentioning that the form of the IR expansion \eqref{genpertuan} is \emph{universal} for all the possible choices of a hidden theory, and is dictated solely by the $U(1)$ symmetry and the associated Ward identity that leads to \eqref{genpertuaj} and \eqref{genpertuak}.

\section{The non-linear theory of the coupled system}\label{couplednonlinear}

In this section we present, in general terms, the non-linear extension of the mechanism explained in the previous section in the two-theory case. In particular, we shall make use of the global symmetries of the system, to advocate for the emergence of a dynamical vector field and its consistent dynamics.

We first assume that both the visible and hidden theory, when uncoupled, have an independent U(1) global invariance, therefore they have  a total $U(1) \times \widehat{U(1)}$ symmetry. We start by defining the generating functional of the correlation functions of the visible theory. We write down this Schwinger functional in terms of external vector potentials $A_\m, \hat{A}_\m$ that can couple to the visible/hidden sector respectively. It is straightforward to generalise this functional with the addition of scalar and other types of sources, but we refrain from doing so, in order to keep our equations as transparent and compact as possible. The full theory is also defined on a flat and not dynamical geometric background $g_{\m\n}\equiv \eta_{\m\n}$. When we wish to set the external sources to their background values, we will use a bold notation i.e  ${\bf A}_\m, {\bf \hat{A}}_\m$. Normally these are taken to be zero in a Lorentz invariant vacuum.

The Schwinger functional is therefore given by
\begin{equation}
e^{-W(A, \hat{A})}\,=\,\int\,\left[D \Phi\right] [D \h \Phi]\,e^{-S_{visible}\left(\Phi,A\right)-S_{hidden}\left(\h \Phi,\hat{A}\right)\,-\,S_{int}\left(\mathcal{O}_i,\h {\mathcal{O}}_i \right)} \label{fun}
\end{equation}
where $\Phi^i$ and $\h \Phi^i$ are respectively the fields of the visible QFT and the hidden $\h{QFT}$ and the interacting part is defined as:
\begin{equation}
S_{int}\,=\,\int\,d^4x\,\sum_i\,\lambda_i\,\mathcal{O}_i(x)\,\h{\mathcal{O}}_i(x)
\label{12}\end{equation}
where $\mathcal{O}_i$ are operators of the visible QFT,  $\h{\mathcal{O}}_i$ operators of the hidden $\h{QFT}$ and the $\lambda_i$ are generic couplings.

There are now, two different possibilities. The first is that the operators appearing in \eqref{12} are uncharged
under the independent global symmetries and the second is that some of them are charged. This first possibility is also the one analysed at the linearised level in section~\ref{JJinteractionlinearised}.

For the second possibility, the operators in \eqref{12} are chosen to be charged under the visible and hidden U(1) as follows\footnote{It is clear that this case can be easily generalized to more complicated cases but we shall  refrain from doing so, here.}:
\begin{equation}
\mathcal{Q}\left(\mathcal{O}_i\right)\,=\,1\,,\quad \h{\mathcal{Q}}(\h{\mathcal{O}}_i)\,=\,-1\,.
\end{equation}
This means that the two independent U(1) global symmetries are broken into the diagonal subgroup
\begin{equation}
U(1)\,\times\,\h{U(1)}\,\,\rightarrow\,\,U(1)_{diag}\label{sym}
\end{equation}
which corresponds to the U(1) invariance of the total functional defined in \eqref{fun}. Had we chosen the first possibility, the functional would simply have retained the two separate global symmetries. The difference can be summarised in the following statement: In the presence of only a single global symmetry, we may  identify $A_\m \equiv \hat{A}_\m$, since there is a common background field for the single $U(1)_{diag}$ and this is the field for which we expect a gauge invariance.

We now remark that the theory, as written in \eqref{fun}, has a natural cutoff represented by the mass $M$ of the messenger fields. For all energies below the cutoff scale $M$, as we are interested in the visible theory observables,  we shall integrate out the hidden $\h{QFT}$ to obtain
\begin{align}
e^{-W(A, \hat{A})} \, &= \,\int\,[D \Phi] [D\h \Phi]\,e^{-S_{visible}\left(\Phi,A\right)-S_{hidden}\left(\h \Phi,\hat A\right)\,-\,S_{int}}\,\nonumber\\&=\,\int\,[D \Phi] \,e^{-S_{visible}\left(\Phi,A\right)\,-\,\mathcal{W}\left(\mathcal{O}_i, \hat{A}\right)}
\label{6}
\end{align}
where $\mathcal{W}\left(\mathcal{O}_i,\hat{A} \right)$ is the generating functional for the hidden theory with external sources given by the operators $\mathcal{O}_i$ of the visible theory.\\
The low energy dynamics of the visible theory is now described by the total action:
\begin{equation}
S_{total}=S_{visible}+\mathcal{W} \, .
\label{3}
\end{equation}

We define the expectation value of the current of the hidden theory (including the interaction)
\begin{equation}
\tilde{V}_{\mu}\,\equiv\,\frac{\delta \mathcal{W}\left(\mathcal{O}_i,\hat{A}\right)}{\delta \hat A^{\mu}}\,=\,\langle \h {J_{\mu}}\rangle\label{defJ2}
\end{equation}
with the idea that such an object could act as an emergent vector field for the visible theory. More precisely, the functional derivative appearing in \eqref{defJ2} must be computed at $\hat A^\mu={\bf \hat A^\mu}$.

In the case of an uncharged coupling both currents are independently conserved
\be
\partial^{\m} \, \tilde{V}_\mu  \, = \, 0 \, \qquad \partial^\mu \, J_\mu^{visible} \, =\,0
\label{v3}\ee
In this case, we defined the current of the visible theory as:
\begin{equation}
J_{\mu}^{visible}\,\equiv\,\frac{\delta S_{visible}\left(\Phi,A\right)}{\delta A^{\mu}}\Big|_{A={\bf \hat A}}\label{vis}\;.
\end{equation}

In the case of a single $U(1)$ invariance of the full theory \eqref{sym} defined when $A_{\mu} = \hat{A}_{\mu}$, only the total current is conserved:
\begin{equation}
\partial^{\m}\left(\tilde{V}_\mu\,+\,J_\mu^{visible}\right)\,=\,0\label{conserv}
\end{equation}

At this point,  we  invert equation \eqref{defJ2}:
\begin{equation}
\hat A^\mu\,=\, \hat A^\mu(\tilde{V}^\mu,\mathcal{O}_i) \, .
\end{equation}
To proceed, we define the Legendre-transformed functional with arbitrary background sources ${\bf A}, {\bf \hat{A}}$ as
\begin{align}
\Gamma\left( \tilde{V},\mathcal{O}_i , {\bf A} , {\bf \hat{A}}\right)\,\equiv \,\int d^4x\,\tilde{V}_\m\, \left( \hat{A}^\mu(\tilde{V}^\mu,\mathcal{O}_i) - {\bf \hat{A}}^\mu \right)\,-\,\mathcal{W}\left(\mathcal{O}_i,\hat{A}^\mu(\tilde{V}^\mu,\mathcal{O}_i)\right)\label{leg}
\end{align}

Using this we define an ``effective action'' $S_{eff}(\tilde{V},\Phi)$ which contains the same information as the original functional, but acts as an action for the visible sector's fields coupled to an induced dynamical vector field $\tilde{V}_\mu$
\begin{equation}\label{leg2}
S_{eff}(\Phi, \tilde{V}, {\bf A} , {\bf \hat{A}} )\,=\,S_{visible}(\Phi, {\bf A})\,-\,\Gamma\left(\tilde{V},\mathcal{O}_i, {\bf A} , {\bf \hat{A}} \right)=
\end{equation}
$$
=\,S_{visible}(\Phi, {\bf A})\,-\,\int d^4x\,\,\tilde{V}_\m\, \left( \hat A^\mu(\tilde{V}^\mu,\mathcal{O}_i) - {\bf \hat{A}}^\mu \right) \,+\,\mathcal{W}\left(\mathcal{O}_i,\hat A^\mu(\tilde{V}^\mu,\mathcal{O}_i)\right)
$$

We shall prove that the effective action defined in \eqref{leg2}, once extremised with respect to the emergent vector  field $\tilde{V}^\mu$, has the following desired feature
\begin{equation}
S_{eff}(\tilde{V}^\star,\Phi)\,=\,S_{visible}(\Phi,A= {\bf A})\,+\,\mathcal{W}(\mathcal{O}_i,\hat{A} = {\bf \hat A})\,\equiv\,S_{total}\vert_{{\bf A}, {\bf \hat{A} }}
\end{equation}
where $\tilde{V}=\tilde{V}^\star$ is the solution that extremises the effective action $S_{eff}(\tilde{V},\Phi)$.\\

In order to achieve this task, we start by computing the variation of the Legendre transformed functional \eqref{leg} with respect to the vector field $\tilde{V}^\m$:
\begin{equation}
\frac{\delta \Gamma\left(\tilde{V},\mathcal{O}_i, {\bf A} , {\bf \hat{A}}\right)}{\delta \tilde{V}_\m}\,= \hat A_\m \, - {\bf \hat{A}}_\m \, ,
\end{equation}
where the visible current is defined in \eqref{vis} and we used the definition of the emergent vector field \eqref{defJ2}. {By setting the source gauge field $\hat A^\m$ to its background value, we obtain the simple expression
\begin{equation}
\frac{\delta \Gamma\left(\tilde{V},\mathcal{O}_i\right)}{\delta \tilde{V}_\m}\Big|_{\hat A = {\bf \hat{A}}}\,=\,0 \, .\label{dynm}
\end{equation}
Therefore,  the Legendre transformed functional as we defined it, is extremal on the background.
Using \eqref{dynm} we also conclude that $S_{eff}$ is extremal with respect to $\tilde{V}^\m$ on the hidden source background $\hat{A}_\mu = {\bf \hat{A}}_\m$.}
\\

What is left to show is that the effective action $S_{eff}$ once evaluated on the solution of the equation \eqref{defJ2} reduces to the original induced action for the original theory \eqref{3}. Let us denote the solution of the equation \eqref{defJ2} $\tilde{V}_\m^\star$. By construction it corresponds to the vev of the current of the hidden theory
\begin{equation}
\tilde{V}_{\mu}^\star\,=\,\frac{\delta \mathcal{W}\left(\mathcal{O}_i,\hat A\right)}{\delta \hat A^{\mu}}\Big|_{\hat A= {\bf \hat{A}}}\,=\,\langle \h {J_{\mu}}\rangle
\end{equation}
From \eqref{leg} we can evaluate the effective action at $\tilde{V}_\m=\tilde{V}_\m^\star$ (which coincides with $\hat A_\m= {\bf \hat{A}}_\m$) and we obtain indeed the already advertised result
\begin{equation}
S_{eff}(\tilde{V}^\star,\Phi)\,=\,S_{visible}(\Phi,A= {\bf A})\,+\,\mathcal{W}(\mathcal{O}_i,\hat{A} = {\bf \hat A})\,\equiv\,S_{total}\vert_{{\bf A}, {\bf \hat{A} }} \, .
\end{equation}
This is the description in the case of a $U(1) \times \widehat{U(1)}$ global invariance that is translated into a
$U(1)_{global} \times \widehat{U(1)}_{local}$ invariance of the effective action. The whole procedure can be repeated with almost no differences in the case of a single $U(1)_{diag}$: one simply replaces ${\bf A} = {\hat{\bf A}}$ in the formulae above and the final effective action has only a single $U(1)_{diag}$ local invariance.

To conclude, we have shown that the imprints of the hidden theory on the visible
theory can quite generically be reformulated as the visible theory being coupled to an emergent dynamical
vector field (denoted by $\tilde{V}^\m$). The dynamics of this emergent vector field encode the effects that the hidden sector has to the visible one.
 In the next subsection, we shall perform a low energy derivative expansion to the various functionals, in order to make the features of the induced interactions more explicit.

\subsection{The low-energy U(1) dynamics}

In this section, we shall employ the generic procedure described above in a simple choice of the functional \eqref{6} dictated by symmetry and an IR derivative expansion. This case hence assumes the presence of a mass gap for the hidden theory, so that the derivative expansion is organised in inverse powers of the mass gap. For simplicity, we also focus in the case ${\bf A}_\m = {\bf \hat{A}}_\m = 0$ with a single $U(1)$ symmetry for the total system and a Lorentz invariant vacuum.

We hence assume the low-energy dynamics of the hidden theory plus interactions to be described by the effective Schwinger functional
\begin{equation}
\mathcal{W}\left(A^\mu\right)\,=\,\int d^4x\,\left(Z_0(|\Phi|^2)-\frac{Z_1(|\Phi|^2)}{4}F^2\,+\,\frac{Z_{ij}(|\Phi|^2)}{2}(D_\mu \Phi^i)(D^\mu \Phi^j)^*\,+\,\dots\right)\label{effC}
\end{equation}
We also consider the presence of a set of complex operators $\Phi^i$ of equal charge, that belong to the visible QFT\footnote{In general the theory contains many charged operators. The relevant analysis in this more general setup is treated  in appendix~\ref{multiplefields}.}. $\Phi_i$ denote here what we called ${\mathcal O}_i$ in the previous section.
The notation $|\Phi|^2$ in the potential functions that appear in (\ref{effC}) is sketchy and stands for gauge-invariant combinations of the scalar sources without derivatives.
Furthermore, the covariant derivative is defined as
\begin{equation}
D_\mu\,\equiv\,\partial_\mu\,+\,i\,q\,A_\mu\;,
\end{equation}
where we keep the charge $q$ generic and the field strength is given by $F=dA$.\\
The ellipsis in (\ref{effC}) is there to remind us that the effective action \eqref{effC} is a low-energy description where higher energy terms, \textit{i.e.} terms with more than two derivatives, are neglected.\\

Following the generic procedure explained in the previous section, we define the emergent vector field:
\begin{equation}
\tilde{V}^\mu\,\equiv\,\frac{\delta \mathcal{W}\left(\Phi_i,A^\mu\right)}{\delta A_\mu}\equiv \langle \h{J}^\m \rangle\label{defj3}
\end{equation}
namely the expectation value of the U(1) current of the Schwinger functional (that includes cross interactions) of the hidden $\h{QFT}$.\\
More explicitly we obtain:
\begin{equation}
\tilde{V}^\mu\,=\,\partial_\n \left(Z_1\,F^{\m\n}\right)\,+\,q^2\,A^\mu\,Z_{ij}\,\Phi^i\,{\Phi^j}^*\,-\,\frac{i\,q}{2}\,Z_{ij}\left(\Phi^i \partial^\mu {\Phi^j}^*\,-\,{\Phi^i}^*\,\partial^\mu \Phi^j\right) +\cdots \label{lala}
\end{equation}
At this stage we want to invert the previous expression \eqref{lala}:
\begin{equation}
A^\mu=A^\mu\left(\tilde{V}^\mu,\Phi\right)\label{inv}
\end{equation}
and identify $\tilde{V}^\mu$ as the new emergent and dynamical degrees of freedom. We  perform such a task within a perturbative derivative expansion.\\
At zeroth order in derivatives, we find:
\begin{equation}
A^\m\,=\,\frac{\tilde{V}^\mu}{q^2\,Z_{ij}\,\Phi^i\,{\Phi^j}^*}\equiv \,V^\mu
\end{equation}
where for simplicity, we have rescaled the original current $\tilde{V}^\mu$ appearing in \eqref{defj3}.\\
Up to two derivatives we find the result:
\begin{equation}
A^\mu \,=\,V^\mu\,+\,\frac{i\,Z_{ij}}{2\,q\,Z_{kl}\,\Phi^k\,{\Phi^l}^*}\,\left(\Phi^i \partial^\mu {\Phi^j}^*\,-\,{\Phi^i}^*\,\partial^\mu \Phi^j\right)\,-\,\frac{1}{q^2\,Z_{ij}\,\Phi^i\,{\Phi^j}^*}\,\partial_\n\left(Z_1\,F_V^{\m\n}\right)\,+\,\mathcal{O}(\partial^3)\label{inva}
\end{equation}
where $F_V=d V$ is the field strength of the emerged vector field.

We can manipulate this expression to bring it to the following form:
\begin{equation}
q^2\,Z_{ij}\,\Phi^i\,{\Phi^j}^*\,A^\mu\,=\,q^2\,Z_{ij}\,\Phi^i\,{\Phi^j}^*\,V^\mu\,+\,Z_{ij}\,(J_\Phi^{ij})^\mu-\,\partial_\n\left(Z_1\,F_V^{\m\n}\right)+\cdots
\label{pp}
\end{equation}
where we have defined the current associated to the field $\Phi$ as
\be
(J_\Phi^{ij})^\m \equiv \frac{i\,q}{2}\left(\Phi^i \partial^\mu {\Phi^j}^*\,-\,{\Phi^i}^*\,\partial^\mu \Phi^j\right)
\;.
\ee
We can now set the external source to zero ${\bf A}^\mu=0$ and rewrite our result eqn.\eqref{pp} as a Maxwell equation for the vector field $\tilde{V}^\m$:
\begin{equation}
{\partial_\n\left(Z_1\,F_V^{\m\n}\right)\,=\,q^2\,Z_{ij}\,\Phi^i\,{\Phi^j}^*\,V^\mu\,+\,Z_{ij}\,(J_\Phi^{ij})^\m +\cdots \,}\label{max1}
\end{equation}
This equation exhibits a ``dual" gauge invariance
\be
 V_{\m}\to  V_{\m}+\pa_{\m}\l \sp \Phi_i \to \Phi_i ~e^{- i q \l}  \, .
\label{dual}\ee
As in the case of a single theory, in section \ref{effe}, it is an illusion of the low order in the derivative expansion, as explained in appendix \ref{high}.

Defining the original current of the visible theory as:
\begin{equation}
J_{visible}^\mu\,=\,\frac{\delta S_{visible}(A^\m)}{\delta A^\m}\Big|_{{\bf A}=0}\,,
\end{equation}
two conservation laws follow (the one of which is redundant and reflects our specific definition of variable). The first physical conservation law is represented by the Ward identity for the total action:
\begin{equation}
S_{total}\,=\,S_{visible}\,+\,\mathcal{W}
\end{equation}
with respect to the diagonal preserved U(1) global group. This takes the form of the conservation of the total current:
\begin{equation}
\partial_\m\,\left(\frac{\delta S_{total}(A^\mu,\Phi,\dots)}{\delta A_\mu}\right)\,=\,0 \quad \rightarrow \quad\partial_\m\,\left[\tilde{V}^\mu \,+\,  J^\mu_{visible}\right]\,=\,0\label{cons1}
\end{equation}

The second ``conservation law" comes directly from the specific definition of the emergent vector field that led to eqn. \eqref{max1}. It implies (on the background)
\begin{equation}
\partial_\m\,\left[q^2\,Z_{ij}\,\Phi^i\,{\Phi^j}^*\,V^\mu\,+\,Z_{ij}\,(J_\Phi^{ij})^\m \right]\,=\,0\label{cons2}
\end{equation}

The two conservation laws \eqref{cons1} and \eqref{cons2} can also be combined into
\begin{equation}\label{cons2n}
\partial_\m\,\left[J_{visible}^\mu\, -\,Z_{ij}\,( J_\Phi^{ij})^\m \right]\,=\,0
\end{equation}

Equations \eqref{max1} together with \eqref{cons2n} represent the main result of this section. The dynamical equation \eqref{max1}  indicates that the low-energy effects of the hidden sector on the visible one, can be captured by a Maxwell equation of a dynamical emergent vector field $V_\m$ . The total current is conserved, as in \eqref{cons2n},  but it is split into the visible sector piece labelled by $J_{visible}^\mu$ and a piece of the hidden sector including interactions $Z_{ij}\,( J_\Phi^{ij})^\m$.

As we discussed in section~\ref{effectiveactionsingletheory}, we may add to the various functionals ``improvement terms" that shift the definition of the various currents by identically conserved quantities. A similar ambiguity is also reflected in the splitting $S_{total}\,=\,S_{visible}\,+\,\mathcal{W}$ ---it is a form of scheme dependence. Nevertheless, once a particular scheme is chosen, the procedure described above follows consistently and one obtains unambiguous results.

\section{The holographic emergent photon}\label{Hologaxion}

We  now investigate the special case where the hidden theory $\widehat{QFT}$ is a large-$N$ holographic theory. In this case $\widehat{QFT}$ has a gravity dual that we shall assume to be five-dimensional.

The general action can be written as
\be
S=\hat S+S_{int}+S_{SM}
\label{b19}\ee
where the interaction term $S_{int}$ has been defined in (\ref{linearaac}), $\hat S$ is the action of the holographic theory, and $S_{SM}$ the action of the SM.
Applying the holographic correspondence, we can write\footnote{{For a conserved U(1) vector $\hat J_{\m}$  of dimension $\Delta=3$, dual to a U(1) gauge field $A_M(x,z)$, the asymptotic behaviour near the boundary is
$A_{\mu}(x,z)=B_{\m}(x)$ while the radial component can be gauged away.
$B_{\m}(x)$ is the source that couples to $\hat J_{\m}$ in the $\h{QFT}$ action. In our example $B_{\m}=J_{\m}$ the SM current.
It is should be stressed, that there are, in general, many other couplings of hidden theory operators to SM Operators. These will generate further couplings between the bulk gravitational theory and the SM. We neglected them here, but they can be
readily included.}}
\be
\langle e^{iS_{int}}\rangle_{\widehat{QFT}}=\int_{\lim_{z\to 0}A_{\m}(x,z)= J_{\m}(x)} {\cal D}A_{\m}~e^{iS_{\rm bulk}[A_{\m}]}
\label{e1}\ee
where $A_{\m}$ is a bulk (five-dimensional) gauge field dual to the global current $\hat J_{\m}$ of $\widehat{QFT}$.
 $S_{\rm bulk}[A_{\m}]$ is the bulk gravity action, $z$ is the holographic coordinate, and the gravitational path integral has boundary conditions for $A_{\m}$ to asymptote to the operator $ J_{\m}$ near the AdS boundary. We have also neglected the other bulk fields.

By inserting a functional $\delta$-function  we may rewrite (\ref{e1}) as
\be
\langle e^{iS_{int}}\rangle=\int_{\lim_{z\to 0}A_{\m}(x,z)= B_{\m}(x)} {\cal D}A_{\m}(x,z){\cal D}B_{\m}(x){\cal D}C_{\m}(x)~e^{iS_{\rm bulk}[A_{\m}]+i\int C^{\m}(x)(B_{\m}(x)-J_{\m}(x))}
\label{e2}\ee
If we now integrate $B_{\m}(x)$ first in the path integral transform, we obtain the Legendre transform of the Schwinger functional of the bulk gauge field, which becomes the bulk effective action. This corresponds in holography to switching boundary conditions at the AdS boundary from Dirichlet to Neumann, and where $C_{\m}(x)$ is the expectation value of the operator $\hat J_{\m}$.
We finally obtain
\be
\langle e^{iS_{int}}\rangle=\int_{\lim_{z\to 0}A_{\m}(x,z)=A^{(0)}(x)+z^2 C_{\m}+\cdots} {\cal D}a(x,z){\cal D}k(x)~e^{iS_N[A_{\m}]-i\int C_{\m}(x)J^{\m}(x)}
\label{e22}\ee
This analysis is valid, with the SM action coupled to the holographic theory at the UV  (the shifted boundary). When however the coupling is at a cutoff scale, the SM must be positioned as a brane in the appropriate radial position giving rise to the  brane-world coupling.

We therefore imagine the SM action as coupled at the radial scale $z_0\sim 1/M$ to the bulk action.
Following holographic renormalization \cite{Bianchi:2001kw, Bianchi:2001de}, we may then rewrite the full bulk+brane action of the emergent vector field  as
 \be
S_{total}=S_{bulk}+S_{brane}
\label{e3}\ee
\be
S_{bulk}=M_P^3\int d^5x\sqrt{g}\left[Z~F^2+{\cal O}(F^4)\right]
\label{e4}\ee
\be
S_{brane}=\delta(z-z_0)\int d^4x\sqrt{\gamma}\left[M^2(\hat F)^2+\hat A_{\m}J^{\m}+\cdots\right]
\label{b5}\ee
where $\hat F_{\m\n}(x)\equiv F_{\m\n}(z_0,x)$ is the induced gauge field  on the brane and we are working in the axial gauge $A_5=0$.
As we shall  be interested at energies $E\ll M$, we can ignore higher derivative terms like $\hat F^4$ on the brane.
Here the U(1) gauge invariance is intact on the brane as the induced gauge field on the brane transforms under bulk gauge transformations induced on the brane.

 In the boundary action (\ref{b5}) $\gamma$ is the induced four-dimensional metric. We have suppressed the metric and other bulk fields. The kinetic coefficient $Z$ depends in general on scalar bulk fields.
On the brane, we have suppressed the standard model fields some of which are charged under the gauge field $A$. There are also localized terms for other bulk fields that we have suppressed. All the localized kinetic terms of the bulk fields on the brane, like the $\hat F^2$ term are due to the quantum corrections of the SM fields.
The graviton also couples to the SM action and provides emergent gravity, \cite{grav}.

Importantly, the gauge symmetry on the bulk is unbroken and so is on the brane, as long as charged fields have no vev. However as we shall see, and in agreement with our earlier analysis, the dark-photon exchange on the brane is not massless.

Finally, the boundary conditions for the bulk  action are Neumann.
It should be noted that what we have here is a close analogue of the DGP mechanism, \cite{DGP}, with two differences: here we have a vector field and also the bulk data are non-trivial as in the setup of \cite{self}.

The main difference in the physics of an emergent vector field  originating in a holographic theory is that,  due to the strong coupling effects, there is an infinity of vector-like  resonances coupled to the SM charged fields. They correspond to the poles of the two-point function of the global current $\hat J^{\m}$, of the ``hidden" holographic theory.

If the holographic theory is gapless, then there is a continuum of modes and, as mentioned earlier, in such a case the induced vector interaction  is non-local.
If the theory has a gap and a discrete spectrum (like QCD) then there is a tower of nearly stable states at large N that are essentially the vector meson trajectories, and act as the KK modes of the bulk gauge field.

To investigate these interactions we should  analyze the propagator of the gauge field on the SM brane.

For this we  introduce a $\delta$-function source for the vector on the brane and we  solve the bulk+brane equations in the linearized approximation, assuming a trivial profile for the bulk gauge field\footnote{ This will be the case where  the hidden QFT is at zero (global) charge density.}  while the metric and other scalars have the holographic RG flow profile of a Lorentz-invariant QFT, namely
 \be
 ds^2=dz^2+e^{2A(z)}dx_{\m}dx^{\m}\sp Z(\Phi_i(z))
 \label{b6}\ee
 In a transverse gauge, the bulk fluctuation equation is given by the bulk Laplacian, plus corrections that come from the brane couplings. We  factor out the space-time index dependence (this is taken into account in  appendix \ref{gau12}).
 The equations read
\be
M_P^3Z\left[\pa_z^2+\left({Z'\over Z}+4A'\right)\pa_z +e^{-2A}\square_4\right]G(x,z)+
\label{b7}\ee
$$
+\delta(z-z_0)G_b~G(x,z)=\delta(z-z_0)\delta^{(4)}(x)
$$
where, $G(x,z)$ is the bulk to bulk gauge field propagator with Neumann boundary conditions and $G_b$ is the two-point operator of the brane current . We work in Euclidean 4d space along the brane and primes stand for derivatives with respect to $z$.
$m^2$ is a potential mass term for the gauge field on the brane,  in the case there are non-trivial charged  vevs on the brane.
The two terms on the brane originate in  the IR expansion of the two-point function of the brane current, that couples to the bulk gauge field.

We Fourier transform along the four space-time dimensions to obtain
\be
M_P^3Z\left[\pa_z^2+\left({Z'\over Z}+4A'\right)\pa_z -e^{-2A}p^2\right]G(p,z)=\delta(z-z_0)-
\label{b8}\ee
$$
-\delta(z-z_0)G_b(p)~G(p,z)
$$where $p^2=p^ip^i$ is the (Euclidean) momentum squared. Later on we  also use $p=\sqrt{p^2}$.
We have also substituted the low-energy expansion of the current two-point function on the brane\footnote{The presence of the $p^2\log p^2$ terms is associated with the logarithmic RG running of the coefficient of the $F^2$ term in four-dimensions. It is the first non-analytic term in the two-point function of currents.}
\be
G_b(p)=M_4^2(p^2+m_0^2)+{\cal O}(p^4)
\label{b8a}\ee
and we also added a brane mass $m_0$, in case symmetry breaking on the brane generates one.
One can also add the logarithmic running of the brane coupling constant, due to the brane quantum corrections, in which case equation (\ref{b8a}) is modified to
\be
G_b(p)=M_4^2\left(p^2+b_0 p^2\log{p^2\over m_{e}^2}+m_0^2\right)+{\cal O}(p^4)
\label{b8ab}\ee

 To solve (\ref{b8}), we must first solve this equation for $z>z_0$ and for $z<z_0$
 obtaining two branches of the bulk propagator, $G_{IR}(p,z)$ and $G_{UV}(p,z)$ respectively. The IR part, $G_{IR}(p,z)$ depends on a single multiplicative integration constant as the regularity constraints in the interior of the bulk holographic geometry fix the extra integration constant.
$G_{UV}(p,z)$ is defined with Neumann boundary conditions at the AdS boundary and depends on two integration constants. In the absence of sources and fluctuations on the SM brane, the propagator is continuous with a discontinuous $z$-derivative at the SM brane\footnote{For Randall-Sundrum branes this condition is replaced by $G_{UV}(p,z-z_0)=G_{IR}(p,z_0-z)$, which identifies the UV side with the IR side. This corresponds to a cutoff holographic QFT in the bulk.}
\be
G_{UV}(p,z_0;z_0)=G_{IR}(p,z_0;z_0)\sp \partial_zG_{IR}(p,z_0;z_0)-\partial_zG_{UV}(p,z_0;z_0)={1\over Z~M_P^3}
\label{b9}\ee
where $M_P$ is the five-dimensional Planck scale in (\ref{e4}).
In this case there is a single multiplicative integration constant left and the standard AdS/CFT procedure extracts from this solution the two-point function of the global current of the hidden QFT. We denote the bulk gauge field propagator in the absence of the brane as $G_0(p,z;z_0)$ that satisfies
\be
M_P^3Z\left[\pa_z^2+\left({Z'\over Z}+4A'\right)\pa_z -e^{-2A}p^2\right]G_0(p,z;z_0)=\delta(z-z_0)
\label{b10}\ee

In our case the presence of an induced action on the SM brane changes the matching conditions to
\be
G_{UV}(p,z_0)=G_{IR}(p,z_0)
\label{b11}\ee
\be
\label{b11a} \partial_zG_{IR}(p,z_0)-\partial_zG_{UV}(p,z_0)={1+G_b(p)~G_{IR}(p,z_0)\over Z~M_P^3}
\ee
The general solution can be written in terms of the bulk propagator $G_0$ with Neumann boundary conditions at the boundary as follows\footnote{Recall that $G(p,z;z_0)$ and $G_0(p,z;z_0)$ are bulk propagators in coordinate space in the radial/holographic direction $z$ and in Fourier space $p^\mu$ for the remaining directions $x^\mu$.}
\cite{self}
\be
G(p,z;z_0)=-{G_0(p,z;z_0)\over 1+G_b(p)~G_{0}(p,z_0;z_0)}
\label{b12}\ee
 The propagator on the brane is obtained by setting $z=z_0$ and becomes
\be
G(p,z_0;z_0)=-{1\over {G_0(p,z_0;z_0)}^{-1}+G_b(p)}={G_0(p,z_0;z_0)\over 1+G_b(p)~G_{0}(p,z_0;z_0)}
\label{b13}\ee
The general structure of the bulk propagator $G_0$ is derived in appendix (\ref{gau12}) and follows a similar structure of the scalar bulk propagator,  \cite{self}.
Similar to our manipulations in the appendix, we perform a scale transformation in order to bring the induced metric on the brane to be $\eta_{\m\n}$. Then $G_0$ in (\ref{b13}) is replaced by $\bar G_0$ given in (\ref{i8}) and (\ref{i9a}). The brane current correlator $G_b$ is also computed in the metric $\eta_{\m\n}$.

We assume that the bulk holographic QFT has a single dynamical scale\footnote{The case where the bulk theory has several such scales can be treated in a similar manner, albeit having more regimes in the energy scale to analyse.}, that  we shall denote by $m$.
Another scale in the problem is the position of the brane, $z_0$. In cases where this is determined dynamically as in \cite{self}, this is of the same order as $m$. But there can be also cases where it is hierarchically different, \cite{self}.
Assuming that $z_0\sim R_0$, we obtain
\be
\bar G_0(p,z_0;z_0)={1\over 2ZM_P^3}
\left\{ \begin{array}{lll}
\displaystyle {1\over ~p}, &\phantom{aa} &p\gg m\\ \\
\displaystyle {1\over m}\left[d_0-\left(d_2+d_2'\log\left({p^2\over m^2}\right)\right){p^2\over m^2}+{\cal O}(p^4)\right],&\phantom{aa}& p\ll m.
\end{array}\right.
 \label{b14}
\ee
The IR expansion above is valid for all holographic RG flows. It starts having non-analytic terms starting at $p^2\log p^2$. The expansion coefficients can be determined either analytically or numerically from the bulk holographic RG flow solution. Analytic formulae for them in terms of the bulk solution were given in \cite{self}.
The dimensionless coefficients $d_i$ above are functions of $mz_0$. Their size is typically of order one  unless $z_0$ is very different from $m$.
The UV expansion in (\ref{b14}) is given, as expected,  by the flat space result.

 Using (\ref{b14}), we now investigate the interaction induced by the vector on the SM brane from (\ref{b13}). It is known that $\bar G_0(p,z_0;z_0)$ is monotonic as a function of $p$, vanishes at large $p$ and attains its maximum at $p=0$ compatible with (\ref{b14}). On the other hand, $G_b(p)$, that captures the two point function of the brane  current, is diverging at large $p$ as $p^2\log(p)^2$ and asymptotes to a constant in the IR. Therefore the function
$G_b(p)\bar G_0(p,z_0,z_0)$ that appears in the denominator of (\ref{b13}) starts from zero or constant in the IR, and asymptotes to +$\infty$ in the UV.

We  denote by the $E_t$ the transition scale at which it  reaches the value one:
\be
G_b(E_t)\bar G_0(E_t,z_0,z_0)\equiv 1
\label{b20}\ee
This is the analogue of the DGP scale, \cite{DGP} for the vector field.
We may therefore write
\be
\bar G(p,z_0;z_0)=
\left\{ \begin{array}{lll}
\displaystyle G_0(p,z_0,z_0), &\phantom{aa} &p\ll E_t\\ \\
\displaystyle {1\over G_b(p)},&\phantom{aa}& p\gg E_t.
\end{array}\right.
 \label{b21}
\ee
When $E_{t}\gg m$ then
\be
\bar G(p,z_0;z_0)=
\left\{ \begin{array}{lll}
\displaystyle {1\over  2ZM_P^3m}\left[d_0-\left(d_2+d_2'\log\left({p^2\over m^2}\right)\right){p^2\over m^2}+{\cal O}(p^4)\right], &\phantom{aa} &p\ll m\\ \\
\displaystyle {1\over 2ZM_P^3}{1\over ~p}+{\cal O}(p),&\phantom{aa}& m\ll  p\ll E_t\;,
\\\\
\displaystyle {1\over G_b}= {1\over M_4^2}{1\over p^2}+{\cal O}(p^4) &\phantom{aa}&  p\gg E_t.
\end{array}\right.
\label{b22}
\ee

Up to now we used the five-dimensional definitions for the gauge field that is dimensionless. Now we  pass to four-dimensional QFT language by $A_{\m}\to {A_{\m}\over M_P}$ and define
\be
{g_5^2}={1\over 2ZM_P}\sp g_4^2={M_P^2\over M_4^2}
\label{b23}\ee
where $g_5^2$ has dimension of length and $g_4$ is dimensionless.
With these normalizations, the vector interaction on the brane in (\ref{b22}) becomes
\be
\bar G(p,z_0;z_0)={1\over M_P^2}
\left\{ \begin{array}{lll}
\displaystyle {g_{IR}^2\over m_{IR}^2+\left(1+{d_2'\over d_2}\log{p^2\over m^2}\right)p^2}+{\cal O}(p^4), &\phantom{aa} &p\ll m\\ \\
\displaystyle {g_5^2\over ~p}+{\cal O}(p),&\phantom{aa}& m\ll  p\ll E_t\;,
\\\\
\displaystyle {1\over G_b}= {g_4^2\over p^2}+{\cal O}(p^4) &\phantom{aa}&  p\gg E_t.
\end{array}\right.
\label{b24}
\ee
with
\be
g_{IR}^2\equiv (m g_{5}^2){d_0^2\over d_2}
\sp
m_{IR}^2={d_0\over d_2}m^2
\label{b25}\ee

Therefore at short enough distances, $p\to\infty$, the interaction mediated by the vector is four-dimensional, and is controlled by the dimensionless coupling constant $g_4$.
At intermediate distances, $m\ll  p\ll E_t$, the induced interaction becomes five-dimensional due to the coupling of the KK modes and the respective five dimensional coupling constant $g_5^2$ has dimension of length.
At large enough distances, $p\ll m$, the interaction is determined by the bulk dynamics and is that of  a massive photon with (dimensionless) coupling constant $g_{IR}$ and mass $m_{IR}$.

We therefore find the emergent (dark) photon is always massive (like the graviton in \cite{self}), and its coupling constant is small,   $g_{IR}\sim {1\over N}$.
More precisely, the emergent photon is a resonance of the associated two point function, that is produced in the interplay between the bulk theory  and the brane dynamics.
Moreover, a similar analysis as in \cite{self} indicates that $m_{IR}\over M_{p}$, where $M_{p}$ is the emergent four-dimensional Planck scale, scales as $N^{-{1\over 3}}$ and can be made arbitrarily small at large enough $N$.

Turning on a vector mass $m_0$ on the brane and keeping all IR contributions, the IR parameters in (\ref{b25}) become
\be
{1\over g_{IR}^2}\simeq {1\over g_4^2}+{d_2\over d_0^2} {1\over  mg_5^2}
\sp {m_{IR}^2\over g_{IR}^2}\simeq {m^2\over d_0(mg_5^2)}+{m_0^2\over g_4^2}
\label{b26}\ee
The effective IR coupling constant for the vector interactions $g_{IR}$ received contributions from the bulk and the brane. The weakest of the two interactions dominates and determines $g_{IR}$. Typically, this can be the bulk interaction as it behaves as $g_{5}\sim {1\over N}$. At hierarchically large $N$, it will dominate the vector interactions on the brane also and the associated coupling can be arbitrarily small.

A similar argument indicates that for a generic bulk holographic theory the vector boson mass, $m_{IR}$ is determined by the bulk physics and is of order ${\cal O}(m)$. For special bulk theories, like theories where the coupling runs slowly at intermediate scales (walking theories\footnote{Holographic walking theories have been discussed in \cite{Nu}-\cite{KJ}.}), the ratio $d_0/d_2$ can become hierarchically small and the vector mass can be $\ll ~m$.

   There are other parameter ranges in (\ref{b26}) that offer more phenomenologically interesting windows but we shall  not pursue this analysis here.

Depending on the parameters of the bulk and the brane theory we could have a differing ordering of scales ie. $E_t\ll m$.  In such a case, there is no intermediate five-dimensional regime for the vector-mediated interaction.

We conclude this section, by observing that when the hidden theory is holographic, the setup generates an emergent dark photon coupled to the visible theory. It is always massive, but both its coupling as well as its mass can be made arbitrarily small, by taking $N$ to be sufficiently large.

\section{A non-local emergent gauge theory}\label{singleNL}

In section \ref{effe} we have analyzed the effective action of a global current in the presence of charged sources and have shown how this can describe an emergent ``photon" that is essentially massive because of the presence of the charged sources. In the presence of a mass gap that theory admits a local IR expansion.

In this section, we analyze a similar effective action in the absence of charged sources\footnote{The treatement of the effective action in theories wirth massless degrees of freedom has pitfalls that are catalogue on page 7 of \cite{IS}. Although seberal issues mentioned there are not problems here, one should keep them alwys  in mind.}.

We start with the simplest setup possible. Consider a theory with a global U(1) symmetry, and an associated conserved current, $\pa_{\m}J^{\mu}=0$.
The variation of the action of the theory under a space-time dependent U(1) parameter $\e(x)$ is
\be
\delta S=\int d^4 x \, \pa_{\mu}\e \, J^{\mu}
\label{C1}\ee
Integrating by parts and demanding invariance when $\e=constant$, \textit{i.e.} a global transformation, gives the conservation of the current advertised before.

Consider now coupling the current to a background gauge field source and improving this so that the coupling is fully (locally) invariant
\be
S\to S(A)=S+\int d^4 x \, J^{\m}A_{\m}+\cdots \, .
\label{A2}\ee
where the ellipsis denotes subleading terms that may be important for gauge invariance, namely $S(A+d\e)=S(A)$ provided all charged fields are appropriately transformed.
We now consider the Schwinger functional
\be
e^{-W(A)}\equiv \int {\cal D}\phi~e^{-S(\phi,A)}
\label{A3}\ee
where $\phi$ denotes collectively the quantum fields of the theory.
The functional $W(A)$ is locally gauge invariant.
This is also equivalent to the standard Ward identity
\be
\pa_{\m}{\delta W\over \delta A_{\mu}}=0
\label{A4}\ee
We now Legendre-transform
\be
\Gamma(V, {\bf A})=\int d^4 x\left[V^{\mu} (A_{\mu} - {\bf A}_\mu) -W(A)\right]
\label{A6}\ee
by defining the current vev on the background ${\bf A}_\mu$ as usual
\be
V_{\m}\equiv {\delta W\over \delta A_{\mu}} \Big|_{A_\mu = {\bf A}_\mu}
\label{A5}\ee
To vary $\Gamma(V, {\bf A})$ with respect to $V$ we must be careful as $V_{\m}$ are not independent variables but satisfy the constraint $\pa^{\m}V_{\m}=0$. In the presence of charged sources analysed in section \ref{effe}, the inversion procedure was local and straightforward. In the present  case we introduce a Lagrange multiplier function $\varphi$ and consider the modified effective action
\be
\Gamma_{\varphi}(V, {\bf A})=\Gamma(V, {\bf A})-\int d^4 x ~\varphi\pa_{\m}V^{\mu}=
\Gamma(V)+\int d^4 x ~\pa_{\m}\varphi V^{\mu} \, .
\label{A7}\ee
A consistency check of the inversion procedure, is that the degree of freedom corresponding to the Lagrange multiplier should decouple and that the constraint $\pa^{\m}V_{\m}=0$ will be automatically satisfied in the effective action for $V_\mu$ that we derive.

We now have to vary $\Gamma_{\varphi}(V, {\bf A})$ both with respect to $V_{\m}$ and $\varphi$.
We find
\be
{\delta \Gamma \over \delta V_{\m}}= A^{\mu} - {\bf A}^\m +V^{\nu}{\delta A_{\n}\over \delta V_{\m}}-{\delta W\over \delta V_{\m}}
\label{A8}\ee
Using
\be
{\delta W\over \delta V_{\m}}={\delta W\over \delta A^{\n}}{\delta A_{\n}\over \delta V_{\m}}=V^{\n}{\delta A_{\n}\over \delta V_{\m}}
\label{A9}\ee
and substituting above we obtain
\be
{\delta \Gamma_{\varphi} \over \delta V_{\m}}=A^{\mu}- {\bf A}^\m+\pa^{\m}\varphi\sp
{\delta \Gamma_{\varphi} \over \delta \varphi}=\pa^{\m}V_{\m}=0
\label{A10}\ee
These are the two equations that fully describe the dynamics of the current vev, $V_{\m}$.  The Lagrange multiplier has become a gauge parameter. The original theory defined on the background $A_{\m}= {\bf A}_\m$ is also equivalent to the one on the gauge transformed background $A_{\m}={\bf A}_\m - \pa_{\m}\e$ because in this case $\e$ couples to the  redundant operator $\partial^\mu V_\mu$.

Consider now a low-energy expansion for  the functional $W(A)$,
\be
W=W_0+\int d^4 x \left[{W_1\over 4}~F^2+{W_3\over 8}(F^2)^2+\right.
\label{A11}\ee
$$
\left.+{W_4\over 8}F_{\m\n}F^{\n\rho}F_{\rho\sigma}F^{\sigma\mu}+{W_5\over 4}F_{\m\n}\square F^{\m\n}+{\cal O}(\pa^6)\right]
$$
with $W_i$ constants\footnote{More generally, $W_i$ may depend on neutral sources.}.

Using the definition \eqref{A5} we obtain
\be
V_{\n}=-W_1\pa^{\m}F_{\m\n}-W_3\pa^{\m}(F^2 F_{\m\n})-W_4\pa^{\m}(F^3)_{\n\mu}-W_5\square \pa^{\m}F_{\m\n} +{\cal O}(\pa^6)
\label{A12}\ee
where\footnote{The seemingly independent term $
(\pa^{\rho}F_{\rho\m})(\pa^{\sigma}{F_{\sigma}}^{\m})$ upon integration by parts is equal to twice $F_{\m\n}\square F^{\m\n}$.}
\be
F^2_{\m\n}\equiv F_{\m\rho} {F^{\rho}}_{\nu}\sp F^{3}_{\m\n}= F_{\m\rho} F^{\rho\sigma}F_{\sigma\nu}
\label{A13}\ee
We have the following identities
\be
\pa^{\m}\pa^{\n}F_{\m\n}=0\sp \pa^{\m}\pa^{\n}(F^2 F_{\m\n})=0\sp
\pa^{\m}\pa^{\n}F^3_{\m\n}=0
\label{A14}\ee
Notice that $V_{\m}$ defined in \eqref{A12} automatically satisfies $\pa^{\mu}V_{\m}=0$.

From (\ref{A12}) we can calculate the two-point function of the currents as
\be
\langle J^{\m}(x)J^{\n}(y)\rangle={\delta V^{\m}(x)\over \delta A_{\n}(y)}\Big|_{A=0}=-W_1\left[\eta^{\m\n}\square -\pa^{\m}\pa^{\n}\right]\delta^{(4)}(x-y)-
\label{A17}\ee
$$
-W_5\square\left[\eta^{\m\n}\square -\pa^{\m}\pa^{\n}\right]\delta^{(4)}(x-y)+{\cal O}(\pa^6)
$$
It is clear that in the derivative expansion the most general function of conserved currents contains only contact terms and is of the form
\be
\langle J^{\m}(x)J^{\n}(y)\rangle=f(\square)\left[\eta^{\m\n}\square -\pa^{\m}\pa^{\n}\right]\delta^{(4)}(x-y)
\label{A18}\ee
where $f(x)$ is an arbitrary function with a regular expansion around $x=0$.
Note also that the two-point function of a conserved current does not have an inverse, as it is annihilated by $\pa_{\m}$ and this is a priori why the quadratic terms in the effective action (that are typically given by the inverse two point function) are ill-defined.

We can also calculate
\be
{\cal F}_{\m\n}=\pa_{\m}V_{\n}-\pa_{\n}V_{\m}=-W_1~\square F_{\m\n}-W_5~\square^2 F_{\m\n}-
\label{A15}\ee
$$
-W_3\left(F^2\square F_{\m\n}+(\pa^{\rho}F^2)\pa_{\rho}F_{\m\n}+\pa_{\m}F^2\pa^{\r}F_{\rho\n}-
\pa_{\n}F^2\pa^{\r}F_{\rho\m}+\pa_{\m}\pa^{\rho}F^2 F_{\rho\nu}-\pa_{\n}\pa^{\rho}F^2 F_{\rho\m}\right)+
$$
$$
+W_4\left(\pa_{\m}\pa^{\rho}(F^3_{\r\n})-\pa_{\n}\pa^{\rho}(F^3_{\r\m})\right)+{\cal O}(\pa^6)
$$

where we have also used the Bianchi identity.
This can be inverted to
\be
F_{\m\n}=-{1\over W_1}\square^{-1}{\cal F}_{\m\n}+{W_5\over W_1^2}{\cal F}_{\m\n}+{\cal O}(\pa^4)
\label{A19}\ee
Notice that, as anticipated, the inversion is non-local.
We  define the action of the inverse Laplacian as
\be
\square^{-1}_x f(x^{\m})\equiv \int d^4x'~G(x;x')f(x')
\label{A20}\ee
where $G(x,x')$ is the appropriate Green's function of the Laplacian, $\square_x G=\delta(x-x')$.
With this definition, ``integration by parts" works trivially on $\square^{-1}$.
Of course there are  boundary conditions implicit in the Green's function that should be correlated with the absence of zero modes.
Here, we shall be cavalier about these issues.

Using the previous result we can now compute
\be
\Gamma_{\varphi}=\int d^4x\left[V_{\m} (A^{\m} - {\bf A}^\m)-W\right]+\int d^4 x ~\pa_{\m}\varphi V^{\mu}=
\label{A16}\ee
$$=-W_0+{W_1\over 4}\int d^4 x F^2+
{W_5\over 4}\int d^4 x ~F\square F+
\int d^4 x ~(\pa_{\m} - {\bf A}_\m)\varphi V^{\mu}+{\cal O}(\pa^6)=
$$
 $$
=   -W_0+{1\over 4W_1}\int d^4 x \left[{\cal F}_{\m\n}\square^{-2}{\cal F}^{\m\n}-{W_5\over W_1}{\cal F}_{\m\n}\square^{-1}{\cal F}^{\m\n}\right]+\int d^4 x ~(\pa_{\m}\varphi - {\bf A}_\m) V^{\mu}+{\cal O}({\cal F}^3)
$$
where we kept only quadratic terms in the field strength for simplicity.

We now vary this effective action to verify that we obtain (\ref{A10})
\be
-{1\over W_1}\square^{-2}\pa^{\m}{\cal F}_{\m\n}+{W_5\over W_1^2}\square^{-1}\pa^{\m}{\cal F}_{\m\n}+\pa_{\n}\varphi+{\cal O}(\pa^4)=  {\bf A}_\n \sp \pa_{\m}V^{\m}=0
\label{aa10}
\ee
Notice that the previous equation automatically implies
\be
\Box\,\varphi=\partial^\n {\bf A}_\n\;.
\ee
There is a ``massless" degree of freedom that decouples and can be subsumed in the background ${\bf A}_\n$.

Using (\ref{A15}), (\ref{aa10})  can be translated to
\be
\square^{-1}\pa^{\m}F_{\m\n}+\pa_{\n}\varphi - {\bf A}_\n +{\cal O}(\pa^4)= A_{\n}+\pa_{\n}\left[\varphi-\square^{-1}\pa^{\m}A_{\m}\right]+{\cal O}(\pa^4)=0
\label{A21}\ee

 (\ref{aa10}) implies that, modulo the subtleties of defining $\square^{-1}$, the effective equation  is gauge-invariant under $V_{\m}$ gauge transformations. On the other hand it is non-local.

The invariance of this action can be shown beyond the derivative expansion. Our assumptions imply, that since there are no minimally-charged fields, the dependence of $W(A_{\m})$ on $A_{\m}$ is via $F_{\m\n}$ only, and this includes all possible multipole couplings. We parametrize
\be
W(A_{\m})=\int d^4x~{\cal L}(F_{\m\n})
\label{A22}\ee
where we dropped the dependence on other fields.
Then the current is given by
\be
V_{\m}={\delta W\over \delta A^{\m}}=-4\pa^{\n}{\delta {\cal L}\over \delta F^{\n\m}}\sp \pa^{\m}V_{\m}\equiv 0
\label{def1}\ee
The right-hand side is a functional of the field strength of $A_{\m}$ which implies that $F_{\m\n}$ is a non-local functional of ${\cal F}_{\m\n}$ proving the emergent gauge invariance.

In the following, we propose a different set of dynamical variables that  renders the effective description, quasi-local. We start again with the functional $\Gamma[V]$ defined in \eqref{A6}
\be
\Gamma(V, {\bf A})=\int d^4 x\left[V^{\mu}(A_{\mu}- {\bf A}^\mu) -W(A)\right]
\label{A23}\ee

 and we change variables to a two form, $B_{\m\n}$
\be
V_{\m}={1\over 2}\e_{\m\n\r\s}\pa^{\n}B^{\rho\sigma}={1\over 3!}\e_{\m\n\r\s}H^{\n\r\s}\sp H_{\m\n\r}\equiv \pa_{\m}B_{\n\r}+\pa_{\n}B_{\rho\m}+\pa_{\r}B_{\m\n}
\label{A244}\ee
so that the constraint
\be
 \pa^{\mu}V_{\m}=0
\label{A255}\ee
is obeyed identically.

This dual substitution solves the constraint explicitly, but introduces a dual gauge invariance in terms of the new unconstrained variable, $\Lambda_{\m}$
\be
B'_{\m\n}=B_{\m\n}+\pa_{\m}\Lambda_{\n}-\pa_{\n}\Lambda_{\m}
\label{A266}\ee

We compute
\be
{\cal F}_{\m\n}={1\over 3!}(\e_{\n\r\s\t}\pa_{\m}H^{\r\s\t}-\e_{\m\r\s\t}\pa_{\n}H^{\r\s\t})
\label{A277}\ee
or equivalently in form notation
\be
{\cal F}=d^*dB
\label{A288}\ee
We have that
\be
(^*{\cal F})_{\m\n}=\pa^{\m}H_{\m\n\rho}
\label{29}\ee

Integrating by parts (\ref{A16}) we can simplify it as
\be
\int d^4 x ~{\cal F}_{\m\n}{\cal F}^{\m\n}={1\over 3}\int d^4x~H_{\m\n\r}\square H^{\m\n\r}
\label{30}\ee
The action (\ref{A16}) can be  therefore  written as
\be
\Gamma[V, {\bf A}]=\int d^4x\left[
 -W_0 - {1\over 6}\e_{\m\n\r\s}H^{\n\r\s} {\bf A}^\m +{1\over 12W_1}\left[H_{\m\n\r}\square^{-1} H^{\m\n\r}-{W_5\over W_1}H_{\m\n\r}H^{\m\n\r}\right]+{\cal O}(\pa^4)\right]
\label{A311}\ee
The coupling to the background term can also be written as the standard anomaly term
\be
  {1\over 6}\int \e_{\m\n\r\s}H^{\n\r\s} {\bf A}^\m= {1\over 6}\int \e_{\m\n\r\s}B^{\r\s} F_{\m\n}({\bf A})
 \label{A322}\ee
From now on we set the background field ${\bf A}_\m$ to zero, since it only acts as an external source for the dynamics of the emergent degree of freedom described by $B_{\m \n}$.

To compute variations we need,
\be
\delta{\cal F}_{\m\n}={1\over 2}(\e_{\n\r\s\t}\pa_{\m}\pa^{\r}\delta B^{\s\t}-\e_{\m\r\s\t}\pa_{\n}\pa^{\r}\delta B^{\s\t})
\label{A333}\ee
Therefore, after some integrations by parts we obtain,
\be
\delta\Gamma[V, {\bf A}=0]=\int d^4x\left[{1\over 2W_1}\left[\delta{\cal F}_{\m\n}\square^{-2}{\cal F}^{\m\n}-{W_5\over W_1}\delta{\cal F}_{\m\n}\square^{-1}{\cal F}^{\m\n}\right]+{\cal O}(\pa^4)\right]=
\label{A344}\ee
$$
=\e_{\n\r\s\t}\int d^4x~{\delta B^{\s\t}\over 2W_1}\left[ \pa^{\r}\left(\square^{-2}\pa_{\m}{\cal F}^{\m\n}-{W_5\over W_1}\square^{-1}\pa_{\m}{\cal F}^{\m\n}\right)+{\cal O}(\pa^4)\right]
$$
We now use the Bianchi identity for ${\cal F}$
\be
\e_{\n\r\s\t}\pa^{\rho}\pa_{\m}{\cal F}^{\m\n}=\e_{\n\r\s\t}\square {\cal F}^{\rho\nu}
\label{A355}\ee
to rewrite
\be
\delta\Gamma[V, {\bf A}=0]=\e_{\n\r\s\t}\int d^4x~\left[{\delta B^{\s\t}\over 2W_1}\left(
\square^{-1}{\cal F}^{\n\r}-{W_5\over W_1}{\cal F}^{\n\r}\right)+{\cal O}(\pa^4)\right]
\label{A366}\ee
We also have
\be
\e_{\n\r\s\t}{\cal F}^{\n\r}=-2\pa_{\m}H^{\m\s\t}
\label{A377}\ee
so that the equations of motion are
\be
\left[\square^{-1}-{W_5\over W_1}\right]\pa_{\m}H^{\m\s\t}+{\cal O}(\pa^2)=0
\label{A388}\ee
and they appear to be non-local.
Nevertheless, if we fix the analogue of the  ``Lorentz gauge" $\pa^{\m}B_{\m\n}=0$ then the equation above becomes local
\be
\left[\square-{W_1\over W_5}\right]B^{\s\t}+{\cal O}(\pa^4)=0
\label{A399}\ee
and is indeed a free field equation for a massive two form.
This gives three propagating degrees of freedom, which is the correct number carried by a conserved vector $V_{\m}$ satisfying $\partial^\m V_\m=0$.

In the Lorentz gauge, equation (\ref{A377}) can be written as
\be
\square^{-1}{\cal F}=-{1\over 3!}~^* B
\label{Bia}\ee
which indicates that in this gauge the theory is local.
Indeed, the Bianchi identity for ${\cal F}$ is via equivalently, via (\ref{Bia}), the Lorentz gauge condition for $B$.

In conclusion, this theory is an interacting theory of a massive two-form in four-dimensions. The two-form theory , including the coupling to the external source in  (\ref{A322}), is reminiscent of the tensor theory in \cite{QT} that was inspired by the ideas in \cite{JT}.

\subsection{A different non-local case}

In this section, we  consider a separate case. More specifically we assume the presence of a single global U(1) symmetry within the hidden $\h{QFT}$. On the contrary, the visible one will not have any U(1) global symmetry and it will inherit it from the hidden sector through the procedure we shall explain in what follows. Moreover, we assume the absence of any charged interaction between the visible QFT and the hidden $\h{QFT}$. As a consequence of that choice the emergent theory will become non local and will have to be treated in a different way.

To proceed we make the two theories to interact via a coupling involving a set of uncharged fields. In such a way the original global U(1) symmetry of the hidden sector is preserved and the symmetry pattern  can be summarized as:
\begin{equation}
\h{U(1)}\,\times\,\bullet \quad \rightarrow \quad U(1)
\label{A400}\end{equation}
where $\bullet$ stands for ``no global symmetry''.
In particular, no operator of the visible theory is charged under the emergent vector field, and quantum corrections coupling the two theories can only generate multiple couplings between the emergent vector field and the visible fields.

In order to describe  such a setup we assume the Schwinger functional to take the form
\begin{equation}
W\left(A^\mu,\chi^I\right)=\int d^4x\Big[Y^{(0)}(\chi^I)-\frac{Y^{(1)}(\chi^I)}{4}F^2\,\,+\,\frac{Y^{(2)}_{IJ}}{2}\,\partial_\mu \chi^I\,\partial^\m \chi^J\,+\,\mathcal{L}_{int}+{\cal O}(\pa^4)\,\Big]\label{act2}
\end{equation}
where the $\chi^I$ are a collection of neutral scalar fields and the leading interaction term takes the ``dipole'' form:
\begin{equation}
\mathcal{L}_{int}\,=\,\frac{Y^{(int)}_{IJ}(\chi^I)}{2}\,F^{\mu\nu}\,\partial_{[\mu} \chi^I\,\partial_{\nu]} \chi^J
\label{A43}\end{equation}
The ellipsis stands for higher derivative terms, namely terms with more than three derivatives.

From the low energy functional \eqref{act2} the emergent vector field defined via the relation
\begin{equation}
\mathcal{V}^\m\,\equiv\,\frac{\delta W(A,\chi^i)}{\delta A_\m}\Big|_{A= {\bf A}}\;,
\label{A44}\end{equation}
takes the form
\begin{equation}
\mathcal{V}^\mu\,=\,\partial_\n \left(Y^{(1)}\,F^{\n\m}\,-\,Y^{(int)}_{IJ}\,\partial^{[\nu} \chi^I\,\partial^{\mu]} \chi^J\right)+{\cal O}(\pa^4)
\label{j2}
\end{equation}
where the brackets indicate the antisymmetrized part.
  In the absence of minimally charged sources, the constraint $\partial_\m \mathcal{V}^\m=0$ is trivially satisfied.

Once now we try to invert the previous expression as:
\begin{equation}
A^\mu\,=\,A^\mu(\mathcal{V}^\mu,\chi^I)
\label{A45}\end{equation}
the same issue as before appears. In particular, it is clear that because of the structure of
\eqref{j2}, the inversion cannot be performed in a local way. As a consequence the effective action for the emergent vector field $\mathcal{V}^\mu$ has a non local formulation. This example is the non-linear manifestation of the same problem we encounter in the single theory setup in the absence of charged operators.

The first step is to construct the field strength of the vector field $\mathcal{V}^\m$ as
\begin{equation}
\mathcal{F}^{\m\n}\,\equiv\,\partial^\m \mathcal{V}^\n\,-\,\partial^\n \mathcal{V}^\m\label{F1}
\end{equation}
From equation \eqref{j2} we can compute such a quantity and we obtain
\begin{equation}
\mathcal{F}^{\m\n}\,=\,\Box\,\left(\bar{F}^{\m\n}\,-\,\mathcal{K}^{\m\n}\right)\,+{\cal O}(\pa^5)\label{NL1}
\end{equation}
where we have defined the antisymmetric two form
\begin{equation}
\mathcal{K}^{\m\n}\,\equiv\,2\,Y^{(int)}_{IJ}\,\partial^{[\m}\chi^I \partial^{\n]}\chi^J\label{twoF}
\end{equation}
and
\begin{equation}
\bar{F}^{\m\n}\,\equiv\,2\,Y^{(1)}\,F^{\m\n}
\label{def}\end{equation}
The previous equation can be inverted into the non local form
\begin{equation}
\bar{F}^{\m\n}\,=\,\Box^{-1}\left(\mathcal{F}^{\m\n}\right)\,+\,\mathcal{K}^{\m\n}+{\cal O}(\pa^3)\;.
\label{A46}\end{equation}
 We can then compute the gauge-invariant density
\begin{equation}
 \frac
{1}{4 \,{Y^{(1)}}^2}\,\bar{F}_{\m\n}\bar{F}^{\m\n}\,=\,\frac{1}{4 \,{Y^{(1)}}^2} \,\Box^{-1}\, \mathcal{F}_{\m\n}\,\Box^{-1}\,\mathcal{F}^{\m\n}\,+\,\frac{1}{4 \,{Y^{(1)}}^2}\,\mathcal{K}_{\m\n}\,\mathcal{K}^{\m\n} \, +
\label{A47}\end{equation}
$$
+\, \frac{1}{2 \,{Y^{(1)}}^2}\,\mathcal{K}_{\m\n}\,\Box^{-1}\,\mathcal{F}^{\m\n}+{\cal O}(\pa^5)
$$
in terms of the field strength of the emerging vector field $\mathcal{F}$
defined in \eqref{F1} and the antisymmetric two-form $\mathcal{K}$ \eqref{twoF}.

Furthermore, the interacting part becomes
\begin{equation}
\mathcal{L}_{int}\,=\,\mathcal{K}_{\m\n}\,\frac{\bar{F}^{\m\n}}{8 \,Y^{(1)}}\,=\,\frac{1}{8 \,Y^{(1)}}\,\mathcal{K}_{\m\n}\,\Box^{-1}\mathcal{F}^{\m\n}\,+\,\frac{\mathcal{K}^2}{8 \,Y^{(1)}}+{\cal O}(\pa^5)
\label{A48}\end{equation}
where we defined $\mathcal{K}^2\equiv \mathcal{K}_{\m\n}\mathcal{K}^{\m\n}$. It is easy to see then that the Schwinger functional is rewritten in the new variables as
\begin{equation}
W = \int d^4x\Big[Y^{(0)} \, -\frac{1}{16 Y^{(1)}} \,\Box^{-1}\mathcal{F}_{\m\n}\,\Box^{-1}\,\mathcal{F}^{\m\n}\, + \frac{1}{16 Y^{(1)}} \mathcal{K}^2 \, + \frac{Y_{IJ}^{(2)}}{2} \partial_\mu \chi^I \partial^\mu \chi^J   \, +{\cal O}(\pa^5)\,\Big]
\label{A49}\end{equation}
and the effective action via the Legendre transform becomes
\be
\Gamma(\mathcal{V}_\m,\chi_I,\mathbf{A}_\mu) =\int d^4 x\left[-\mathcal{V}^{\mu}(A_{\mu}-\mathbf{A}_\mu) \right] \, + \, W \,
\label{A50}\ee
$$
=  \int d^4 x~\frac{1}{8 Y^{(1)}}  \left[ \Box^{-1}\mathcal{F}_{\m\n}\,\Box^{-1}\,\mathcal{F}^{\m\n}+\mathcal{F}_{\m\n}\,\Box^{-1}\mathcal{K}^{\m\n}\right]\,+\mathcal{V}^{\mu}\mathbf{A}_\mu+ \, W\nn
$$
$$
=\int d^4x\Big[Y^{(0)} \, +\frac{1}{16 Y^{(1)}}\left( \Box^{-1}\mathcal{F}_{\m\n}+ \mathcal{K}_{\m\n}\right)\left(\Box^{-1}\mathcal{F}^{\m\n} +\mathcal{K}^{\m\n}\right) + \frac{Y_{IJ}^{(2)}}{2} \partial_\mu \chi^I \partial^\mu \chi^J   \, +{\cal O}(\pa^5)\,\Big]
$$

We observe, that as in the previous section, and unlike the case were there are minimally charged fields, here the action for $\mathcal{V}_{\m}$ is invariant under
\be
\mathcal{V}_{\m}\to \mathcal{V}_{\m}+\pa_{\m}\e
\label{A51}\ee
and this invariance can be shown to exist to all orders.

To summarize, we obtained a gauge-invariant effective description in terms of the emergent gauge field $\mathcal{V}^\m$ which appears to be non-local due to the presence of the inverse Laplacian in the effective action. This result is analogous of what we already obtained in the single theory case in section \ref{singleNL}.\\

One way to proceed is to follow the same method presented in section \ref{singleNL}.\\
To see however the structure of the next order terms  we have to supplement the effective functional \eqref{act2} with the leading higher-derivative corrections, namely the terms containing four derivatives,
\begin{equation}
W_4(A^\mu,\chi^I)=\int d^4x\,\Big(\frac{Y^{(3)}}{8}(F^2)^2\,+\,\frac{Y^{(4)}}{8}F_{\m\n}F^{\n\rho}F_{\rho\sigma}F^{\sigma \mu}\,+\,\frac{Y^{(5)}}{4}F_{\m\n}\Box F^{\m\n}\,+\nonumber
\label{A52}\end{equation}
\begin{equation}
+\,\frac{Y^{(6)}_{IJKL}}{4}\,\partial_\m \chi^I \partial^\m \chi^J\,\partial_\n \chi^K \partial^\n \chi^L\,+ \frac{Y^{(8)}_{IJKL}}{4} \chi^I \Box \chi^J\, \chi^K \Box \chi^L\,+  \frac{Y^{(10)}_{IJKL}}{4} \chi^I \Box \chi^J\, \partial_\n \chi^K \partial^\n \chi^L\,+\mathcal{O}(\partial^5)\Big)\label{higher1}
\end{equation}
along with
\begin{equation}
\mathcal{L}_{int}^{4}(A^\mu,\chi^I)=\int d^4x\,\left(\frac{Y_{IJ}^{(7)}}{4}F^2\,\partial_\m \chi^I \partial^\m \chi^I\,+\frac{Y_{IJ}^{(9)}}{4}F^{\mu\rho}\,\partial_\m \chi^I F_{\rho\nu} \partial^\nu \chi^I\,+\,\mathcal{O}(\partial^5)\right)\label{higher2}
\end{equation}

For simplicity, we  assume in the following that all the $Y^{(n)}$ couplings are constant, independent of the neutral fields $\chi^I$. Furthermore from \eqref{higher1}, \eqref{higher2} we  consider only corrections which are at most quadratic in the field strength $F^{\m\n}$ and in the derivative of the neutral scalars $\partial \chi$ as well as only linear terms, neglecting the terms proportional to $Y^{(3)},Y^{(4)},Y^{(8)}, Y^{(6)},Y^{(9)},Y^{(10)}$. Such a simplified setup shall be enough for the scope of this section, which is to demonstrate the propagating degrees of freedom.

Under the previous assumptions we have

\be \label{W4}
W(A^\mu,\chi^I)=\int d^4x\,\Big(Y^{(0)}(\chi^I)-\frac{Y^{(1)}(\chi^I)}{4}F^2\,\,+\,\frac{Y^{(2)}_{IJ}}{2}\,\partial_\mu \chi^I\,\partial^\m \chi^J\,+\,\frac{Y^{(5)}}{4}F_{\m\n}\Box F^{\m\n}\,+\,\mathcal{L}_{int}\Big)
\ee
where
\be
\mathcal{L}_{int} = \frac{\mathcal{K}_{\mu\nu}F^{\mu\nu}}{4}+\frac{Y_{IJ}^{(7)}}{4}F^2\,\partial_\m \chi^I \partial^\m \chi^I
\label{A53}\ee

The higher derivative corrections we have introduced add new contributions to the definition of the induced vector \eqref{j2}; in particular
\begin{equation}
\mathcal{V}^\mu\,=\,\partial_\n \left(Y^{(1)}\,F^{\n\m}\,-\,Y^{(int)}_{IJ}\,\partial^{[\nu} \chi^I\,\partial^{\mu]} \chi^J\right)\,+\,\partial_\n\left(Y^{(5)}\Box  F^{\m\n}\right)+\partial_\n\left(Y_{IJ}^{(7)}\left(F^{\m\n}\partial_\rho \chi^I \partial^\rho \chi^J\right)\right)+\mathcal{O}(\partial^5)\label{gg}
\end{equation}
Notice that from the definition \eqref{gg} we can immediately derive the identity
\begin{equation}
\partial_\m \mathcal{V}^\m\,=\,0\,
\label{A54}\end{equation}
by just using symmetry arguments.\\

We can calculate the field strength of the induced vector field $\mathcal{V}^\m$ as
\begin{equation}
\mathcal{F}_{\m\n}\,\equiv\,\partial_\m \mathcal{V}_\n\,-\,\partial_\n \mathcal{V}_\m\,=\,Y^{(1)}\,\Box F_{\m\n}-\Box\mathcal{K}_{\m\n}\,-Y^{(5)}\,\Box^2 F_{\m\n}-\,\Box \left(F_{\m\n}\,\Theta\right)\,+\,\mathcal{O}(\partial^6) \label{fs}
\end{equation}
where we defined the scalar
$$\Theta\equiv Y_{IJ}^{(7)} \partial_\m \chi^I \partial^\m \chi^J$$
 and neglected higher corrections.\\
We can now invert the expression \eqref{fs} assuming an ansatz of the type
\begin{equation}
F_{\m\n}\,=\,\Box^{-1}A_{\m\n}\,+\,B_{\m\n}\label{inv1}
\end{equation}
where $A,B$ are generic two forms. Combining \eqref{fs} and \eqref{inv1} and solving them in a perturbative expansion we obtain
\begin{equation}
\mathcal{F}_{\m\n}=Y^{(1)}A_{\m\n}-A_{\m\n}\Theta,\qquad Y^{(1)}B_{\m\n}-\mathcal{K}_{\m\n}-Y^{(5)}A_{\m\n}-B_{\m\n}\Theta=0\,.
\label{A55}\end{equation}
and consequently
\begin{equation}
A_{\m\n}=\frac{1}{Y^{(1)}}\,\mathcal{F}_{\m\n}\,+\,\frac{1}{{Y^{(1)}}^2}\,\mathcal{F}_{\m\n} \,\Theta\,+\,\mathcal{O}(\partial^6)
\label{A56}\end{equation}
\begin{equation}
B_{\m\n}\,=\,\frac{1}{Y^{(1)}}\mathcal{K}_{\m\n}\,+\,\frac{Y^{(5)}}{{Y^{(1)}}^2}\mathcal{F}_{\m\n}\,+\,\frac{\Theta \, (\mathcal{K}_{\m\n} Y^{(1)}+2 \,\mathcal{F}_{\m\n}\, Y^{(5)})}{{Y^{(1)}}^3}\,+\,\mathcal{O}(\partial^6)
\label{A57}\end{equation}
We use the redefinition in (\ref{def})
to write down the original field strength in terms of the induced one at leading order in derivatives
\begin{equation}
\bar{F}_{\m\n}\,=2\Box^{-1}\mathcal{F}_{\m\n}\,+\,2\mathcal{K}_{\m\n}\,+2\,\frac{Y^{(5)}}{Y^{(1)}}\,\mathcal{F}_{\m\n}\,+\, 2 \frac{\Theta}{Y^{(1)}}\Box^{-1}\mathcal{F}_{\mu\nu}+\,\mathcal{O}(\partial^4)
\label{A59}\end{equation}
and rewrite the original functional \eqref{W4} as

\begin{equation}
W(\mathcal{V}^\m,\chi^I)\,=\,\int d^4x\,\left[Y^{(0)}\,+\,\frac{Y^{(2)}_{IJ}}{2}\partial_\m \chi^I \partial^\m \chi^J-\,\frac{1}{4 Y^{(1)}}\left(\Box^{-1}\mathcal{F}_{\m\n}\Box^{-1}\mathcal{F}^{\m\n}\,+\mathcal{K}_{\mu\nu}\Box^{-1}\mathcal{F}^{\mu\nu}\right)\right.\,\label{W11}
\end{equation}
$$
\left.-\,\frac{1}{4Y^{(1)\,2}}\left(Y^{(5)}\,\mathcal{F}_{\mu\nu}\Box^{-1}\mathcal{F}^{\mu\nu}\,+\,\Theta\, \Box^{-1}\mathcal{F}_{\mu\nu}\Box^{-1}\mathcal{F}^{\mu\nu}\right) +\, \mathcal{O}(\partial^5)\right]
$$

\be
\Gamma(\mathcal{V}_\m,\chi_I,\mathbf{A}_\mu) =\int d^4 x\left[-\mathcal{V}^{\mu}(A_{\mu}-\mathbf{A}_\mu) \right] \, + \, W \, =
\label{A500}\ee

$$
=\,\int d^4x\,\left[Y^{(0)}\,+\,\frac{Y^{(2)}_{IJ}}{2}\partial_\m \chi^I \partial^\m \chi^J+\frac{1}{4Y^{(1)}}\left( \Box^{-1}\mathcal{F}_{\mu\nu}\Box^{-1}\mathcal{F}^{\mu\nu}+\mathcal{K}_{\mu\nu}\Box^{-1}\mathcal{F}^{\mu\nu}\right)\,+\right.
$$

$$
\left.+\frac{1}{4Y^{(1)\,2}}\left(Y^{(5)}\,\mathcal{F}_{\mu\nu}\Box^{-1}\mathcal{F}^{\mu\nu}\,+\,\Theta\, \Box^{-1}\mathcal{F}_{\mu\nu}\Box^{-1}\mathcal{F}^{\mu\nu}\right)+\, \mathcal{V}^{\mu}\mathbf{A}_{\mu} +\, \mathcal{O}(\partial^5)\right]
$$

The functional we obtain is definitely gauge-invariant with respect to the ``emergent'' U(1) symmetry but it is manifestly non-local.

{
We now resort to the same method we exploited for the single theory example in sec.\ref{singleNL}.
More specifically we perform the following further change of variables
\be
\mathcal{V}_{\m}={1\over 2}\e_{\m\n\r\s}\pa^{\n}\mathcal{B}^{\rho\sigma}={1\over 3!}\e_{\m\n\r\s}\mathcal{H}^{\n\r\s}\sp \mathcal{H}_{\m\n\r}\equiv \pa_{\m}\mathcal{B}_{\n\r}+\pa_{\n}\mathcal{B}_{\rho\m}+\pa_{\r}\mathcal{B}_{\m\n}
\label{A61}\ee
so that the condition
\be
\pa^{\mu}\mathcal{V}_{\m}=0
\label{A62}\ee
is identically satisfied. In other words we solve explicitly the constraint using symmetries.

We are now in the position of writing down the functional \eqref{A500} in terms of the new unconstrained variable

\begin{equation}\label{W222}
\Gamma(\mathcal{V}^\m,\chi^I, \mathbf{A}_\mu)\,=\,\int d^4x\,\left[Y^{(0)}+\frac{Y^{(2)}_{IJ}}{2}\partial_\m \chi^I \partial^\m \chi^J\, -\frac{1}{2 Y^{(1)}}\frac{1}{3!}\partial^{\mu}\left(\Box^{-1}\mathcal{K}_{\mu\nu}\right)\epsilon^{\nu}_{\kappa\lambda\sigma}\mathcal{H}^{\kappa\lambda\sigma}\right.
\end{equation}
$$
+{1\over 3!}\left(\e_{\m\kappa\lambda\s}\mathbf{A}^\mu-\frac{1}{2 Y^{(1)}}\partial^{\mu}\left(\Box^{-1}\mathcal{K}_{\mu\nu}\right)\epsilon^{\nu}_{\kappa\lambda\sigma}\right)\mathcal{H}^{\kappa\lambda\sigma}
$$

$$
\left.-\,\frac{1}{4 Y^{(1)}}\Big(\mathcal{H}_{\m\n\rho}\Box^{-1}\mathcal{H}^{\m\n\rho}\,+\,\,\frac{Y^{(5)}}{Y^{(1)}}\mathcal{H}_{\m\n\rho}\mathcal{H}^{\m\n\rho}+\,\frac{\Theta}{ Y^{(1)}}\mathcal{H}_{\m\n\rho}\Box^{-1}\mathcal{H}^{\m\n\rho}\Big)\,+\mathcal{O}(\partial^5)\right]
$$

The term in (\ref{W222}) that coupled the backround gauge field $\mathbf{A}$ to $\mathcal{H}$ can be written, after an integration by parts as
\be
{1\over 3!}\e_{\m\kappa\lambda\s}\mathbf{A}^\mu\mathcal{H}^{\kappa\lambda\sigma}\to -{1\over 4}F^{\kappa\mu}(\mathbf{A})\mathcal{B}^{\lambda\sigma}
\ee
and, interestingly, to the standard anomaly coupling between $\mathcal{B}$ and $\mathbf{A}$ for anomalous U(1)'s in four-dimensional compactifications of string theory, \cite{ABDK,review}.

By varying \eqref{W222} with respect to $\mathcal{H}$ we compute the equations of motion for  $\mathcal{H}$

\be \label{EOMH}
\pa_{\m}\left[\left(1+\frac{\Theta}{Y^{(1)}}\right)\Box^{-1}\,+\,\frac{Y^{(5)}}{Y^{(1)}}\right]\mathcal{H}^{\mu\nu\rho}\,+\,\frac{1}{3!}\pa_{\m}\left(
\partial^{\lambda}\left(\Box^{-1}\mathcal{K}_{\lambda\sigma}\right)-2 \mathbf{A}_{\nu}\right)\epsilon^{\sigma\mu\nu\rho}=0\,\;.
\ee
For the rest of the section, we set the source $\mathbf{A}_{\mu}=0$.

The last term above is CP-odd and provides a source term for the equations of ${\cal B}_{\m\n}$.
 Using \eqref{EOMH}  we can compute the equations of motion for the unconstrained two form $\mathcal{B}$
\begin{equation}
\left[\left(1+\frac{\Theta}{Y^{(1)}}\right)\Box^{-1}\,+\,\frac{Y^{(5)}}{Y^{(1)}}\right]\,\partial_\m \,\mathcal{H}^{\m\n\rho}\,+{\pa_{\m}\Theta\over Y^{(1)}}\Box^{-1}\mathcal{H}^{\m\n\rho}=
\label{A63}\end{equation}
$$
=-\frac{1}{3!}\pa_{\m}
\partial^{\lambda}\left(\Box^{-1}\mathcal{K}_{\lambda\sigma}\right)\epsilon^{\sigma\mu\nu\rho}= -{1\over 12}F_{\s\m}(Z)\epsilon^{\nu\rho\sigma\mu}
$$
with
\be
F_{\m\n}(Z)\equiv \pa_{\m}Z_{\n}-\pa_{\n}Z_{\m}\sp Z_{\m}=\pa^{\l}(\square^{-1}{\mathcal K}_{\l\m})
\ee
which is evidently non local but gauge-invariant.
We now consider the linearized part of this equation, namely
\begin{equation}
\left[\Box^{-1}\,+\,\frac{Y^{(5)}}{Y^{(1)}}\right]\,\partial_\m \,\mathcal{H}^{\m\nu\rho}= - {1\over 12}F_{\s\m}(Z)\epsilon^{\nu\rho\sigma\mu}\;.
\label{A63a}\end{equation}

Choosing an appropriate gauge, \textit{i.e.} the Lorentz gauge $\partial_\m \mathcal{B}^{\m\n}=0$, we can rewrite the equation above, in form notation,  as
\begin{equation}
\left[1\,+\,\frac{Y^{(5)}}{Y^{(1)}}\Box\right]\,\mathcal{B}_{\m\n}=-{1\over 3!}~(^*F(Z))_{\m\n}\,.\label{fifi}
\end{equation}
which is a similar massive equation as we found in the last subsection.
}

\section{Discussion\label{Dis}}

In the previous sections we have studied the emergence of a dynamical U(1) vector field from a global symmetry of a hidden theory. Here,  we would like to classify several distinct versions of this emergence and possible phenomenological applications.

We may envisage  the following  cases of emergence of a U(1) vector field that couples to the standard model:

\begin{enumerate}

 \item A hidden theory with a global U(1) symmetry and a current-current coupling to a SM global (non-anomalous) symmetry, An example could be $B-L$ in the SM.
In that case, the result will be that the hidden global symmetry will generate a (generically massive) vector boson that couples to the $B-L$ charges of the SM.  In such a case, the combined theory still has two independent U(1) symmetries, but one of them only is visible in the SM, (the hidden symmetry is not visible from the point of view of the SM).

 \item A hidden theory with a global U(1) symmetry and a coupling between a charged operator in the hidden theory to a charged operator of the SM under a SM  global (non-anomalous) symmetry. Such a coupling breaks the two U(1)'s into a single diagonal U(1). This leftover U(1)  couples to the emergent vector boson.

\item In the two cases above, the U(1) global symmetry of the SM may also be an anomalous global symmetry, like baryon or lepton number.   Although, some of the properties of the new vector interaction remain similar to what was described above, there are new features that are related to the anomaly of the global SM symmetry. In such a case, we expect to have similarities with the anomalous U(1) vector bosons of string theory.

\item   A hidden theory with a global U(1) symmetry and a current-current coupling to the (gauge-invariant) hypercharge current.
    In this case, both the hypercharge gauge field and the emergent vector couple to the hypercharge current. By a (generically non-local) rotation of the two vector fields, a linear combination will become the new hypercharge gauge field, while the other will couple to $|H|^2$ where $H$ is the SM Higgs.

\item  A hidden theory with a global U(1) symmetry, whose current is $J_{\m}$  and a coupling with the hypercharge field strength of the SM of the form
\be
S_{int}={1\over m^2}\int d^4 x F^{\m\n}(\pa_{\m}J_{\n}-\pa_{\n}J_{\m})
\ee
where the scale $M$ is of the same order as the messenger mass scale.
By an integration by parts this interaction is equivalent to the previous case, using the equations of motion for hypercharge.

\end{enumerate}

In all of the above,  we have an emergent U(1) vector boson that plays the role of a dark photon, and in this context the single most dangerous coupling to the standard model is the leading dark photon portal: a kinetic mixing with the hypercharge, \cite{ship}.
We can make estimates of such dangerous couplings using (weak coupling) field theory dynamics, but it is also interesting to make such estimates using dual string theory information at strong coupling. Such a study is underway, \cite{u1}.

\subsection{Emergent dark photons versus fundamental dark photons}

There are several studies so far of dark photons coupled to the standard model. Such studies involve fundamental dark photons and all experimental constraints have been parametrised in terms of the dark-photon coupling constant and its mass.
Extra parameters that affect phenomenological constraints may be minimal couplings to SM Fields and the mixing to hypercharge.
There may be also exotic couplings involving generalized (four-dimensional) Chern Simons terms between the hypercharge and the dark photon, if specific mixed anomalies exist, \cite{akt,CIK,ABDK,Anto}

Such constraints have been studied in detail, expecially in the last twenty years, and there are dedicated experiments for their detection, \cite{ship}.
The current constraints are exclusion plots on the effective coupling vs mass diagram for the dark photon, as shown on Figure 2.6, page 28 of \cite{ship}.

In the cases of emergent vectors studied here, as long as we look at such particles well below the compositeness scale, their dynamics to leading order may be indistinguishable    from fundamental dark photons.
 Unlike fundamental dark photons, the emergent photons described here, especially in the holographic context, have propagators that at intermediate energies behave differently from fundamental photons and therefore can have different phenomenology and different constraints from standard elementary dark photons. The same was found recently for , emergent axions in \cite{axion}

If such a compositeness scale is low and in such a case, non-local effects dominate and alter the phenomenological behavior of such vectors. This is
the case for emergent axions studied in \cite{axion}. The impact of this softer behavior depends on the particular experiment, and the energies at which it is sensitive. An phenomenological analysis is therefore necessarily energy and experiment dependent.

There is also another basic difference. With fundamental dark vectors, one
can more or less choose whatever he wants for the main parameters, coupling and
mass. In the emergent case,  one can vary at will the hidden theory
spanning a wide variety of theories. If however we want this same theory to provide emergent (observable) gravity coupled to the SM, as advocated in \cite{SMGRAV}, then this hidden theory must be a holographic theory.
  In that case, there are important changes
in the estimates of effective couplings from Effective Field Theory (EFT).

For example, it is found in \cite{u1}, that the effective couplings that lead to the mixing of the dark photon and the hypercahrge are smaller by extra factors of\footnote{The hidden theory number of colors.} N, if the dark photon is emergent from a hidden holographic theory.

Such phenomena are interesting to investigate, and we shall address them in a future publication.

\vskip 1cm

\section*{Aknowledgments}
\addcontentsline{toc}{section}{Aknowledgements\label{ack}}

\vskip 1cm

We would like to thank P. Anastasopoulos, C. Charmousis, M. Bianchi, G. Bossard, D. Consoli, B. Gouteraux, D. Luest, F. Nitti, A. Tolley, L. Witkowski for discussions.
We would also like to thank Matteo Baggioli for participating in early stages of this work.
\vskip 0.5cm

 This work was supported in part  by the Advanced ERC grant SM-grav, No 669288.

\newpage

%%%%%%%%%%%%%%%%%%
\appendix
\renewcommand{\theequation}{\thesection.\arabic{equation}}
\addcontentsline{toc}{section}{Appendix\label{app}}
\section*{Appendices}

\section{The example of free massive bosons and fermions}\label{Examples}

In this subsection we perform the direct computation of the 2-point current correlation function $\langle J^\mu (k) J^\nu (-k) \rangle$ for a free massive boson and a free complex massive fermion. For notational convenience we  use normal fonts in this subsection. Our goal is to verify the presence or absence of the contact term in \eqref{genpertuaj}.

\subsubsection{Free massive complex boson}

The current of the U(1) global symmetry is
\be
\label{genpertuba}
J^\mu = - \frac{i}{2} \left( \varphi^* \p^\mu \varphi - \varphi \p^\mu \varphi^* \right)
~.\ee
We now evaluate the 2-point function $G_{\mu\nu}(k) = i \langle J^\mu (k) J^\nu (-k) \rangle$ using standard Wick contractions and the propagator (following a mostly positive sign convention for the metric)
\be
\label{genpertubb}
\langle \varphi(x) \varphi^*(0) \rangle = \int \frac{d^d p}{(2\pi)^d} \frac{-i e^{-i p x}}{p^2 + m^2 - i \epsilon}
~.\ee
The Wick contractions lead to the expression
\be
\label{genpertubc}
i G_{\m \n}(k) = - \frac{1}{2} \int \frac{d^d p}{(2\pi)^d} \left[ \frac{(2 p_\mu + k_\mu)(2p_\nu + k_\nu)}{(p^2 + m^2)((p+k)^2 + m^2)} \, + \frac{2 \xi \eta_{\mu \nu}}{p^2 + m^2} \right]
~.
\ee
The first term is due to the usual Wick contractions and the second arises from normal ordering the operators that produce $\delta$-function contact terms in position space. From a diagrammatic point of view, the first term is coming from a loop diagram with two current insertions while the second is a contact tadpole term when the insertions are at the same point.
We have chosen an arbitrary coefficient $\xi$ for this term to eventually adjust imposing the invariance.

Using Feynman parameters, we obtain:
\bea
\label{genpertubd}
&&\int \frac{d^d p}{(2\pi)^d} \frac{(2 p_\mu + k_\mu)(2p_\nu + k_\nu)}{(p^2 + m^2)((p+k)^2 + m^2)}
\nonumber\\
&=&  \int_0^1 dx \int \frac{d^d p}{(2\pi)^d}  \frac{(2 p_\mu + k_\mu)(2p_\nu + k_\nu)} {( (p+k(1-x))^2 + x(1-x)k^2 + m^2) )^2}
\eea
and shifting $p^\mu \rightarrow p^\mu - k^\mu (1-x)$ to simplify the denominator we obtain
\be
\label{genpertubc2}
i G_{\mu\nu}(k) = - \frac{1}{2} \int \frac{d^d p}{(2\pi)^d} \, \int_0^1 d x \, \frac{2 \xi \eta_{\mu \nu} (p^2 + x^2 k^2 - m^2) \, + \, 4 p_\mu p_\nu \, + \, (2x-1)^2 k_\mu  k_\nu}{\left(p^2 + k^2 x(1-x) + m^2 \right)^2}
~,
\ee
where we kept only the rotationally invariant terms.
After employing standard dimensional regularization manipulations (introducing the scale $\mu$ and the parameter $\Delta = m^2 + k^2 x (1-x) \,$) we arrive at the result for the various parts of the correlator
\bea
\label{genpertube}
G_{\mu\nu}(k) &=&  \frac{ \mu^{4-d}}{(4\pi)^{d/2}} \left[ \eta_{\mu \nu}  \Gamma \left(1 - \frac{d}{2} \right)
 \int_0^1 dx \, \Delta^{d/2-1} - \frac{ k_\mu k_\nu}{2} \Gamma \left(2 - \frac{d}{2} \right)
 \int_0^1 dx  (2x-1)^2 \, \Delta^{d/2-2} \right] \, + \nn \\
 &+& \xi  \frac{ \mu^{4-d}}{(4\pi)^{d/2}}  \eta_{\mu \nu}   \left[\frac{d}{2} \Gamma \left(1 - \frac{d}{2} \right)
 \int_0^1 dx \, \Delta^{d/2-1} -   \Gamma \left(2 - \frac{d}{2} \right)
 \int_0^1 dx (k^2 x^2 + m^2) \, \Delta^{d/2-2} \right]
~. \nn \\
\eea
We can now expand in $4-d = \epsilon$ to find
\bea
\label{genpertube2}
G_{\mu\nu}(k) &=&  \frac{1}{(4\pi)^{2}} \left( \eta_{\mu \nu}
 \int_0^1 dx \,  \Delta \left[-\frac{2}{\epsilon} + \log \frac{c_1   \Delta}{\mu^2}  \right] - \frac{k_\mu k_\nu}{2}
 \int_0^1 dx  (2x-1)^2 \, \left[+\frac{2}{\epsilon} - \log \frac{c_1   \Delta}{\mu^2}  \right] \right) \, + \nn \\
 &+&  \xi  \frac{1}{(4\pi)^{2}}  \eta_{\mu \nu}  \left(
 \int_0^1 dx \,  \Delta \left[-\frac{4}{\epsilon} + 2 \log \frac{c_1   \Delta}{\mu^2}  \right] -
 \int_0^1 dx  (k^2 x^2 + m^2) \, \left[+\frac{2}{\epsilon} - \log \frac{c_1   \Delta}{\mu^2}  \right] \right) \nn \\
 &=&   \int_0^1 \frac{ dx }{(4\pi)^{2}} \left[\eta_{\mu \nu} m^2 (1+\xi) + \frac{(2x-1)^2}{2} k_\mu k_\nu + \eta_{\mu \nu} k^2 \left( x^2 (1+ 3 \xi) - x (1+ 2 \xi) \right) \right] \times \nn \\
 &\times& \left[-\frac{2}{\epsilon} + \log \frac{c_1   \Delta}{\mu^2}   \right]~.
 \eea
with $c_{1} = e^{\gamma}/\sqrt{2 \pi} $ a constant that depends on the Euler $\gamma$ that is to be subtracted in the MS-scheme or absorbed in the scale $\mu$. Notice also that we can add and subtract terms proportional to odd powers of $\sim 2x - 1$ since they integrate to zero. The divergences then can be subtracted and the result renormalised. Before doing so, we observe that the Ward identity should hold in the same way for both the divergent and the logarithmic terms, since they have the same tensor structure. In particular according to \eqref{genpertuaj}, we find a momentum independent contact term
\bea
k^\m G_{\mu\nu}(k) &=& \, -  \AA~k_\n \,   \, , \quad \AA =   \frac{1 }{ 8 \pi^{2}} \,  m^2 (1+\xi)   ~\, .
\eea
This is a quite generic feature. The correlator is transverse only for the scalar QED value $\xi = -1$ , else there is always a third longitudinal degree of freedom. This is precisely what the tadpole term in \eqref{genpertubc} affects, and hence can also be interpreted as a shift in the background due to this tadpole.

We conclude with the final renormalised result with respect to the scale $\mu_r$ (in the MS scheme), that reads for general $\xi$
\be
\label{genpertube3}
G^{ren}_{\mu\nu}(k) =   \int_0^1 \frac{dx }{(4\pi)^{2}} \times
 \ee
$$
\times \left[\eta_{\mu \nu} m^2 (1+\xi) + \frac{(2x-1)^2}{2} k_\mu k_\nu + \eta_{\mu \nu} k^2 \left( x^2 (1+ 3 \xi) - x (1+ 2 \xi) \right) \right] \, \log \frac{\Delta}{\mu_r^2} ~.
$$

\vskip 1.5cm
\subsubsection{The free massive fermion}

\vskip 1.5cm

As a second example consider the case of a free massive Dirac fermion
\be
\label{genpertuca}
S_{Dirac} = \int d^4 x \, \bar \psi ( i \gamma^\mu \p_\mu - m ) \psi
~.\ee
The current of the global U(1) symmetry is
\be
\label{genpertucb}
J^\mu = \bar\psi \gamma^\mu \psi
~.\ee
Wick contractions and the use of the Feynman propagator
\be
\label{genpertucc}
S_F (x-y) = \int \frac{d^4 p}{(2\pi)^4} \frac{ i ( \gamma^\mu p_\mu -  m)}{p^2 + m^2 - i \epsilon} e^{-i p(x-y)}
\ee
leads in Feynman parameters to the 2-point function ($\Delta = m^2 + k^2 x (1-x)$)
\be
\label{genpertucd}
i \langle J_\mu (k) J_\nu(-k) \rangle =
- 2  \int_0^1 d x \int \frac{d^d p}{(2 \pi)^d} \frac{2 p_\mu p_\nu - 2 x(1-x) k_\m k_\n - \eta_{\mu \nu} \left( p^2 - k^2 x(1-x) + m^2 \right)}{(p^2 + \Delta)^2}
~.\ee
No contact term contribution to the Ward identity is anticipated in this case and none is found by explicit computation.

After using dimensional regularisation, the result is explicitly transverse and similar to what was obtained for the free bosons
\be
\label{genpertubeferm}
G_{\mu\nu}(k) = - \frac{1}{4 \pi^2}
\left( k^2 \eta_{\mu\nu} -  k_\mu k_\nu \right) \int_0^1 dx (1-x) x \left[ \frac{2}{\epsilon} + \log \left(  \frac{ 4 \pi e^{- \gamma} \mu^2}{m^2 + k^2 x(1-x)} \right) \right]
~.\ee

\section{K\"allen-Lehmann spectral representation}\label{Spectralrepresentation}

One can write down the most general K\"allen-Lehmann spectral representation for the two-point function of currents as follows
\be\label{spectralposition}
 \langle J_\mu(x) J_\nu(0) \rangle = i G_{\m \n}(x) = \int_0^\infty d \mu^2 \int \frac{d^d k}{(2\pi)^d} \frac{ -i \, e^{- i k x}}{k^2 + \mu^2 - i \epsilon} \rho_{\m \n}(k, \, \mu^2)
\ee
where the spectral weight is split into longitudinal and transverse parts
\be
\rho_{\m \n}(k, \, \mu^2) = \left(\eta_{\m \n}   - \frac{ k_\m k_\n}{\mu^2}  \right) \rho^1( \mu^2) + k_\mu k_\nu \rho^0( \mu^2)\;.
\label{v2}\ee
The longitudinal part appears when the associated symmetry is broken and the current is not conserved. The scalar Goldstone pole in the case of spontaneous breaking is located in this part.

For a conserved current in a CFT we obtain
\be
\rho^1(\mu^2) \sim \mu^{d-2}, \quad \rho^0(\mu^2) = 0\;.
 \ee
Equation (\ref{v2}) is the common splitting employed in the literature that captures both vectors and axial vectors, see for example~\cite{Bernard:1975cd}.

Another possible splitting is the one performed in the main text \eqref{splitmain}, that corresponds to
\be
\rho_{\m \n}(k, \, \mu^2) =  \,  \eta_{\mu\nu} \, \AA(\mu^2)  +  \left( \eta_{\mu\nu}  - \frac{k_\mu k_\nu}{\mu^2}   \right) \BB(\mu^2) \, .
\ee
It is simple to find the relation between the two descriptions
\be
\BB(\mu) =  \rho^1( \mu^2) - \mu^2 \rho^0( \mu^2) \, , \quad \AA(\mu^2) = \mu^2 \rho^0( \mu^2) \, .
\ee
In order to write \eqref{genpertuak} in the main text we have exchanged the $k$ and $\mu$ integrations, so that
\be
\BB(k^2) = \int_0^\infty d \mu^2 \frac{\BB(\mu^2)}{k^2 + \mu^2 - i \epsilon} \, , \quad \AA(k^2) = \int_0^\infty d \mu^2 \frac{\AA(\mu^2)}{k^2 + \mu^2 - i \epsilon} \, .
\ee
One should be careful though, since to be able to perform this exchange, the integral over the spectral weight should be well defined, and frequently one needs to perform certain subtractions. In all cases though, the position space result \eqref{spectralposition} is well defined.

We may also rewrite (\ref{spectralposition})
as
\be
 G_{\m \n}(x) =(\pa_{\m}\pa_{\n}-\eta_{\m\n}\square)D_1(x)+\eta_{\m\n}\square D_0(x)
\label{b1}\ee

with
\be
D_1(x)=\int_0^\infty d \mu^2 {\cal B}(\mu^2) D_{d}(\mu) \sp
D_0(x)=\int_0^\infty d \mu^2 {\cal A}(\mu^2) D_{d}(\mu)
\ee
where
\be
D_d(\mu)\equiv - \int \frac{d^d k}{(2\pi)^d} \frac{e^{- i k x}}{k^2 + \mu^2 - i \epsilon}
\ee
is the scalar propagator.

Now we can study various spectral options for different theories such as having poles or a continuum of states. Near poles $k^2 \simeq - m_i^2$, the correlator has the behaviour corresponding to the exchange of a massive gauge boson. For a conserved current we find
\be\label{massivegaugeboson}
 G_{\m \n}(k)  \simeq - \frac{R(m_i)}{k^2 + m_i^2} \left(\eta_{\m \n} - \frac{k_\m k_\n}{m_i^2} \right) \, .
\ee
In this expression $R(m_i)>0$ is the positive definite spectral weight residue near such poles, coming solely from the $\rho^1(\mu^2)$ term of eqn. \eqref{spectralposition}\footnote{The other term $\rho^0(\mu^2)$ is relevant for \emph{Axial} currents.}. This two-point function leads to a repulsive force between two equal charges as shown in the next appendix~\ref{effectivestaticpotential}.

When the current is not conserved the longitudinal part of the correlator is non-zero. As this is proportional to $q^{\m}q^{\n}$ in momentum space, it gives rise to derivative interactions and does not mediate a static force. The same applies to spontaneous symmetry breaking.

\section{The effective static potential}\label{effectivestaticpotential}

In this appendix we shall investigate the nature of the static interaction mediated by the emergent vector as a function of different types of spectral densities in the vector two-point function.

We work under the assumption  that the hidden theory current two-point function is to be interpreted as the propagator for the emergent vector as shown in the main text. This implies  that we can directly compute the static potential between two external sources in real space by Fourier transforming the static part of the correlator (where $k^2 = - k_0^2 + |\vec{k}|^2$ in our conventions)
\be\label{staticpotential}
\Phi(r) =  \, \int_{-\infty}^\infty d t \,  G_{00}(t, \vec{r})  \, =  \, \int_{-\infty}^\infty   \frac{d^3 \vec{k}}{(2 \pi)^3}   G_{00}(k_0=0, \vec{k}) e^{i \vec{k} \vec{r}}
\ee
We first perform the angular integral to obtain ($y =  |\vec{k}|$)
\be\label{static1}
\Phi(r) = \int_0^\infty \frac{ d y}{2 \pi^2}  \, y  \, G_{00}(y) \, \frac{\sin r y }{r} \, .
\ee
Using the symmetry of the integrand under $y \rightarrow - y$ we can turn this integral into
\be\label{static3}
\Phi(r) = \frac{1}{4 \pi^2 r}  \int_{- \infty}^\infty d y \, y  \,  G_{00}(y)  \, e^{i r y} \, .
\ee

We now turn to the general spectral representation of the current-current correlator, presented in appendix~\ref{Spectralrepresentation}. Using \eqref{v2}, we find
\be\label{static7}
\Phi(r) = - \frac{1}{4 \pi^2 r} \int_{-\infty}^\infty d y \, y  \,  e^{i r y}   \int_0^\infty d \mu^2  \frac{ \rho^1(\mu^2)}{y^2 + \mu^2 - i \epsilon} \, .
\ee
Assuming now the presence of isolated poles at $m_i$ and a branch-cut continuum starting above a scale $M$ in the spectral weight, we can deform the contour in the complex $y$-plane, to one that encloses the upper-half imaginary axis (hairpin), resulting into
\be\label{static7a}
\Phi(r) = - \frac{1}{4 \pi^2 r} \int_{\mathcal{C}} d y \, y  \,  e^{i r y}   \int_0^\infty d \mu^2  \frac{\rho^1(\mu^2)}{y^2 + \mu^2 - i \epsilon} \, .
\ee
Taking advantage of the Sokhotski-Plemelj formula
\be\label{SP}
\frac{1}{x \pm i \epsilon} \, = \, \mathcal{P} \frac{1}{x} \, \mp \, i \pi \delta(x) \, ,
\ee
and noticing that only the imaginary part contributes to the hairpin contour (the principal values cancelling pairwise), we finally find
\be\label{static8}
\Phi(r) = \frac{1}{4 \pi r}  \int_{0}^\infty d u \, u \, e^{- r u} \left( \sum_i \rho^1 (m_i^2)  \delta (u^2 - m_i^2) + \theta ( u^2 > M^2) \rho^1 (u^2) \right) \, .
\ee
This is a general result, showing that in proper unitary theories, only the imaginary part of the correlator contributes to the static potential. We observe that each of the poles gives the potential of a massive vector boson
\be\label{static9}
\Phi(r) \sim \frac{1}{4 \pi r} \sum_i  e^{- m_i r} \, .
\ee
The force between two equal charges is then a repulsive force as expected from the exchange of a massive vector boson. The sign is fixed due to the positivity of the spectral weight. If the isolated pole is at $u^2 = 0$ we obtain a long range potential due to an exchange of a massless photon-like state
\be
\Phi(r) \sim \frac{1}{4 \pi r} \, .
\ee
{This case is not obviously excluded by the WW theorem, \cite{WW}, as one of the assumptions is that the massless pole should correspond to a charged state under the current whereas the states generated by a U(1) currets are chargeless.}

The branch-cut continuum generically results in a more intricate behaviour, two examples being the free bosonic and fermionic theories of appendix \ref{Examples}. The details depend on the precise continuum spectral weight of the hidden theory. Assuming a leading power law behaviour for the spectral weight near the mass gap of the form
\be\label{static10}
\rho^1(\mu^2) = \rho^1_{(0)} \, (\mu^2 - M^2)^a\left[1+O\left(\mu^2 - M^2\right)\right] \, ,
\ee
we find
\be\label{static11}
\Phi(r) \, \simeq \, \frac{1}{4 \pi r}  \int_{M} d u \, u \, e^{- r u}  (u^2 - M^2)^a  \rho^1_{(0)} \, \sim {e^{-Mr}\over (Mr)^{a+2}}\sp M r\to \infty
\ee
while for $a=0$ (logarithmic branch cut) we obtain
\be\label{static11a}
\Phi(r) \, \simeq \, \frac{1}{4 \pi r}  \int_{M} d u \, u \,  e^{- r u}  \log (u^2 - M^2)  \rho^1_{(0)} \, \sim \, {e^{-Mr}} \log\left[{Mr\over 2}e^{\gamma}\right]\sp M r\to \infty
\ee
In the limit where the mass gap of the continuous spectrum vanishes, ($M \rightarrow 0$),  we obtain instead
\be\label{static12}
\Phi(r) \, \sim\,\frac{1}{ r^{3+2 a}} \, ,
\ee
For a CFT $a=1$ and we recover the $\Phi(r) \sim 1/r^5$ scaling for the static potential
(the current-current  correlator behaves  as $1/|x|^6$).

Now we shall conclude with a detailed analysis for the cases of free bosons/fermions. In the case of free bosons \eqref{genpertube3} we find
\be\label{static2}
G^{f.b.}_{00}(y) =  \int_0^1 \frac{dx }{(4\pi)^{2}} \left[m^2 (1+\xi) -  y^2 \left( x^2 (1+ 3 \xi) - x (1+ 2 \xi) \right) \right] \log \frac{  \left(m^2 + y^2 x(1-x) \right)}{\tilde \mu^2} \, .
\ee
We can again deform the contour as before so that we enclose the cut of the function $G^{f.b.}_{00}(y)$. The cut starts at $y = 2 i m$.
We therefore write the integral as
\be\label{static4}
\Phi(r) = \frac{1}{4 \pi^2 r} \int_{2 m}^\infty d u \, u \,  e^{- r u} \Im  G^{f.b.}_{00}(u)
\ee
Taking the imaginary part directly we find (equivalently we could have used the Cutkosky rules in the original expression for the correlator)
\be\label{static5}
\Im  G^{f.b.}_{00}(u) = -\int_0^1 \frac{dx }{(4\pi)^{2}} \left[m^2 (1+\xi) +  u^2 \left( x^2 (1+ 3 \xi) - x (1+ 2 \xi) \right) \right] \pi \theta \left(u^2 x(1-x) - m^2 \right)
\ee
The roots giving the integral endpoints are $x_\pm = 1/2 \pm \sqrt{1/4 - m^2/u^2}$, so that we obtain
\be\label{static6}
\Im  G^{f.b.}_{00}(u) =  \frac{\sqrt{1- 4 m^2/u^2}}{ 6 \pi} \left[  u^2 - 4 m^2   \right]  \, ,
\ee
from which we observe that the contact term contribution ($\xi$ dependent) vanishes.
This results into
\be\label{potr}
\Phi(r) = \frac{ m^2}{8 \pi^3 r^3} K(2, 2mr) \sim \frac{e^{- 2 m r}}{r^{7/2}} \, , \quad r \rightarrow \infty
\ee
A similar manipulation of \eqref{genpertubeferm}, reveals that
\be\label{staticferm}
\Im  G^{f.f.}_{00}(u) =  \frac{\sqrt{1- 4 m^2/u^2}}{ 3 \pi} \left[  u^2 + 2 m^2   \right]  \, ,
\ee
so that the static potential again scales as
\be\label{potr1}
\Phi(r) \sim \frac{e^{- 2 m r}}{r^{7/2}} \, , \quad r \rightarrow \infty
\ee

\section{U(1) Effective action with multiple charged fields }\label{Effectiveactionexamples}\label{multiplefields}

In this appendix we analyse the effective action  in the presence
of multiple charge fields.

We therefore consider multiple charged fields and/or sources  $\Phi_i(x) \,$, $i = 1, ... N$ that all transform under a global $U(1)$ and have charges $q_i$. The Schwinger functional, valid for theories with a mass gap can be  now parametrized as
\be
W(A,\Phi_i)=\int d^4x\left(W_0(|\Phi_i|^2)+{W_1(|\Phi_i^2|)\over 4}F_A^2+\sum_{i}{W^2_{i}(|\Phi_i|^2)\over 2}|D\Phi_i|^2+{\cal O}(\pa^3)\right)
\label{AA24p}\ee
where $F_A=d A$, and $W_I(|\Phi|_i)$ is a shorthand for general potential functions that depend on all uncharged scalar monomials.
The covariant derivatives are
\be
D_{\m}\Phi_i=(\pa_{\m}+i q_i A_{\m})\Phi_i\sp D_{\m}\Phi_i^*=(\pa_{\m}-i q_i A_{\m})\Phi_i^*
\label{AA25p}\ee
and the ellipsis represents higher derivative terms, \textit{i.e.} terms with more than two derivatives. The functional in (\ref{AA24p}) is gauge-invariant under
\be
A_{\m}\to A_{\m}+ \pa_{\m}  \e(x) \sp \Phi_i\to \Phi_i~e^{-i q_i\e(x)}
\label{AA1}\ee
This is equivalent to the standard Ward identity for the total current
\be
\pa_{\m}{\delta W\over \delta A_{\mu}}+i\sum_{i}\left(\Phi_i{\delta W\over \delta \Phi_i}-\Phi_i^*{\delta W\over \delta \Phi_i^*}\right)=0 \, .
\label{AA20p}\ee

From \eqref{A24p} we can compute the current as (repeated indices are summed over)
\be
\tilde V_{\nu}\equiv {\delta W\over \delta A^{\n}}=-\partial^{\m}(W_1F^A_{\m\n})-{i\over 2}\sum_{i}\left[{W^2_i}(q_i \Phi_i^*\pa_{\n}\Phi_i- q_i \Phi_i\pa_{\n}\Phi_i^*)+W^2_i|q_i \Phi_i|^2A_{\n}\right]\,+{\cal O}(\pa^3)
\label{AA26p}\ee

We shall now invert the previous expression and compute $A_{\m}$ as a function of $\tilde V_{\m}$ in a derivative expansion. In particular we obtain an expansion of the form $A_\mu = \sum_{i=0}^\infty A_\mu^{(i)}$, where the $A_\mu^{(i)}$ contains $(i)$-derivatives.
The result for the first few terms is
\be
A_\n^{(0)} = \hat{V}_{\n} \, , \nn
\ee
\be
A_\n^{(1)} = \frac{i}{2 \sum_j W^2_j|q_j \Phi_j|^2} \sum_i W^2_i\left(q_i \Phi_i^*\pa_{\n}\Phi_i- q_i \Phi_i\pa_{\n}\Phi_i^* \right) \, ,  \nn
\ee
\be\label{AA31p}
A_\n^{(2)} = \frac{1}{ \sum_i W^2_i|q_i \Phi_i|^2}\,\partial^\m\left(W_1\,F^{\hat V}_{\m\n}\right) \, ,
\ee
and so on, where
\be
\hat{V}_\m\,\equiv\,{\tilde V_{\m}\over  \sum_i W^2_i|q_i \Phi_i|^2}
\sp F^{\hat{V}}_{\m\n}=\pa_{\m} \hat{V}_{\n}-\pa_{\n} \hat{V}_{\m}\;.
\label{AA6}\ee
 We truncate our expansion to two derivatives since our original functional~\ref{AA24p} was also valid up to two derivative terms.

Note, that from (\ref{AA31p},~\ref{AA6}),  $\hat{V}_{\m}$ is gauge invariant under the original gauge transformation~\ref{AA1} as it should.
We may rewrite the equations in (\ref{AA31p}) as
\be\label{AA21}
\frac{1}{ \sum_i W^2_i|q_i \Phi_i|^2}\partial^\m\left(W_1\,F^{\hat V}_{\m\n}\right)+  \, \hat V_{\n} + \,
\ee
$$
+  \frac{i}{2 \sum_j W^2_j|q_j \Phi_j|^2} \sum_i W^2_i\left(q_i \Phi_i^*\pa_{\n}\Phi_i- q_i \Phi_i\pa_{\n}\Phi_i^* \right) +{\cal O}(\pa^3)= {\bf A}_\nu
$$
where ${\bf A}_\nu$ is the background gauge field of the original theory.

Interestingly, equation (\ref{AA21}) is gauge-invariant under the following ``dual" gauge transformation:
\be
\hat V_{\m}\to \hat V_{\m}+\pa_{\m}\l \sp \Phi_i \to \Phi_i~e^{i q_i \l}  \, .
\label{gg2}\ee
but higher derivatives will not any more be invariant under this modified transforation.
We should also check whether this dual gauge invariance is present at the level of the effective action functional $\Gamma$.

 We therefore now move on to derive explicit expressions for the effective action functional in the derivative expansion.
Using the original functional~\ref{AA24p}, we compute the effective action $\Gamma$ as a Legendre transform with respect to the total current first in terms of $A_\mu$
\be
\Gamma(A_\mu,\Phi_i,\mathbf{A}_\mu) = \int d^4x\left(W_0(|\Phi_i|^2)+{W_1(|\Phi_i^2|)\over 4}F_A^2+{1\over 2}\sum_i W^2_i|D\Phi_i|^2\right) +
\ee
$$
+ \int d^4 x (A^{\n}-\mathbf{A}^\nu) \left(\partial^{\m}(W_1F^A_{\m\n})+\sum_iW_i^2\left[{iq_i\over 2}( \Phi_i^*\pa_{\n}\Phi_i- \Phi_i\pa_{\n}\Phi_i^*)-A_{\n}|q_i \Phi_i|^2\,\right]\right)+{\cal O}(\pa^3)
$$
Substituting  \ref{AA31p} and keeping terms up to two derivatives we finally find
\be
\label{AA22}
\Gamma(\hat{V}_\mu,\Phi_i ,\mathbf{A}_\mu) = \int d^4x\left[W_0 - {1\over 4}W_1 (F^{\hat V})^2+{1\over 2}\sum_{i}W^2_i\left(|\partial\Phi_i|^2  + |q_i \Phi_i|^2  \mathbf{A}^\mu  \hat{V}_\mu\right)  \right] -
\ee
$$
  -{1\over 2} \int d^4x\left(\sum_k W^2_k |q_k \Phi_k|^2\right)\left(\hat{V}_{\n} + {i\over 2}{\sum_i W^2_iq_i\left( \Phi_i^*\pa_{\n}\Phi_i- \Phi_i\pa_{\n}\Phi_i^* \right)\over  \sum_j W^2_j|q_j \Phi_j|^2} \right)^2   \, +{\cal O}(\pa^3)
$$
{As expected this functional is not gauge-invariant off-shell.
However, the equation that relates it to the Schwinger function is gauge-invariant. Therefore it is only gauge invariant on-shell to that order in the derivative expansion.

\section{The structure of higher derivative terms\label{high}}

We have seen in the main part of the paper, that in the leading orders in the derivative expansion, the equation of motion from the effective action of the vector $V_{\m}$ enjoys an emergent gauge invariance, distinct from the original gauge invariance of the Schwinger functional. It acts by a  standard U(1)  gauge transformation on $\hat V_{\m}$ and transforms the charged sources with the opposite charge, as in (\ref{dual}).

In this appendix we investigate the fate of the emergent gauge invariance acting on $V_{\m}$ and the charged fields found in the leading derivative expansion, by studying the higher order terms. We show that such an invariance does not exist, but it is an artifact of the first few terms of the expansion.

To indicate the problem we start with two such terms in the Schwinger functional that involve a charged source $\Phi$, a two derivative term and a four derivative term.
\be\label{k1}
W_2(A^\mu, \Phi) = \int d^4x\, \left[\frac{Z_2}{2} |D \Phi|^2 + \frac{Z_4}{4} |D \Phi|^4 +\cdots\right]
\ee
To have a consistent (gauge-invariant) derivative expansion we must count with the same weight $\pa_{\m}$ and $A_{\m}$ in this expansion.

We compute,
\be
V_{\mu}=\left(A_{\m}|\Phi^2|-{i\over 2}J_{\m}\right)\left(Z_2+Z_4D_{\m}\Phi D^{\m}\Phi^*\right)+\cdots
\label{k2}\ee
We rewrite this as
\be
\epsilon V_{\mu} = \e Z_2\left(A_{\m}|\Phi^2|-{i\over 2}J_{\m}\right)+\e^3 Z_4\left(A_{\m}|\Phi^2|-{i\over 2}J_{\m}\right)D_{\m}\Phi D^{\m}\Phi^*+{\cal O}(\e^4)
\label{dual2}\ee
where we introduces a small parameter $\e$ to count the order in $\pa_{\mu}$ and $A_{\m}$ and $V_{\mu}$.
We have also defined
\be
J_{\m}\equiv \Phi^*\pa_{\m}\Phi-\Phi\pa_{\m}\Phi^*
\label{kk2}\ee
from which we obtain
\be
\Phi\pa_{\m}\Phi^*={1\over 2}\left(\pa_{\m}|\Phi|^2-J_{\m}\right)
\sp
\Phi^*\pa_{\m}\Phi={1\over 2}\left(\pa_{\m}|\Phi|^2+J_{\m}\right)
\label{kk3}\ee
that will be useful further on.

Under a standard gauge transformation, the variables change as
\be
A_{\m}\to A_{\m}+\pa_{\m}\zeta\sp \Phi\to \Phi~e^{-i\zeta}\sp J_{\m}\to J_{\m}-2i|\Phi|^2\pa_{\m}\zeta
\ee

Inverting (\ref{dual2}) we obtain
\be
Z_2\left(A_{\m}|\Phi^2|-{i\over 2}J_{\m}\right)= V_{\mu}-\e^2 Z_4\left(A_{\m}|\Phi^2|-{i\over 2}J_{\m}\right)D_{\m}\Phi D^{\m}\Phi^*+{\cal O}(\e^3)=
\label{k3}\ee
$$
= V_{\mu} -\e^2 {Z_4\over Z_2}~V_{\mu}~D_{\m}\Phi D^{\m}\Phi^*+{\cal O}(\e^3)
$$
where in the last step we substituted the left hand side in the right side but not yet in
$D_{\m}\Phi D^{\m}\Phi^*$.
We now investigate this part
\be
D_{\m}\Phi D^{\m}\Phi^*=\pa_{\m}\Phi\pa^{\m}\Phi^*-iJ_{\m}A^{\m}+A_{\m}A^{\m}|\Phi|^2=
\label{k4}\ee
$$
=\pa_{\m}\Phi\pa^{\m}\Phi^*+{1\over 4|\Phi|^2}J_{\m}J^{\m}+{1\over |\Phi|^2}\left(A_{\m}|\Phi^2|-{i\over 2}J_{\m}\right)^2=
$$
$$
={\pa_{\m}|\Phi|^2\pa^{\m}|\Phi|^2\over 4|\Phi|^2}+{1\over |\Phi|^2}\left(A_{\m}|\Phi^2|-{i\over 2}J_{\m}\right)^2
$$
We now substitute (\ref{k3}) into (\ref{k4}) above to obtain
\be
D_{\m}\Phi D^{\m}\Phi^*={\pa_{\m}|\Phi|^2\pa^{\m}|\Phi|^2\over 4|\Phi|^2}+{V_{\m}V^{\m}\over Z_2^2 |\Phi|^2}+{\cal O}(\e)
\label{k5}\ee
and substituting this back into (\ref{k3}) we obtain
\be
Z_2\left(A_{\m}|\Phi^2|-{i\over 2}J_{\m}\right)= V_{\mu} - \e^2 {Z_4\over Z_2}~V_{\mu}~\left[{\pa_{\m}|\Phi|^2\pa^{\m}|\Phi|^2\over 4|\Phi|^2}+{V_{\m}V^{\m}\over Z_2^2 |\Phi|^2}+{\cal O}(\e)\right]+{\cal O}(\e^3)=
\label{k6}\ee
and finally rearranging
\be
A_{\m}={V_{\mu} \over Z_2|\Phi|^2 }+{i\over 2|\Phi|^2}J_{\m}- \e^2 {Z_4\over Z^2_2 |\Phi|^2}~V_{\mu}~\left[{\pa_{\m}|\Phi|^2\pa^{\m}|\Phi|^2\over 4|\Phi|^2}+{V_{\m}V^{\m}\over Z_2^2 |\Phi|^2}\right]+{\cal O}(\e^3)
\label{k7}\ee
We now redefine
\be
\hat V_{\m}={V_{\mu}\over Z_2|\Phi|^2}
\label{k8}\ee
to finally obtain
\be
A_{\m}=\hat V_{\mu} + {i\over 2}\pa_{\m}\log{\Phi\over \Phi^*}  - \e^2 {Z_4\over Z_2}~\hat V_{\mu}~\left[{\pa_{\m}|\Phi|\pa^{\m}|\Phi|}+ |\Phi|^2 \hat V_{\m} \hat V^{\m} \right]+{\cal O}(\e^3)
\label{k9}\ee
We observe that (\ref{k7}) has a consistent $\e$ expansion in which a derivative is ${\cal O}(\e)$ but $V_{\m}$ and $\hat V_{\m}$ is ${\cal O}(\e^0)$.
Still this equation is not gauge-invariant under the dual gauge invariance (\ref{dual}) beyond the leading ${\cal O}(1)$ piece.

We may now extend somewhat the Schwinger functional   by taking it to be
\be\label{k10}
W(A^\mu, \Phi) = \int d^4x\, f(|D \Phi|^2,|\Phi|^2)
\ee
and again compute
\be
V_{\m}=\e\left(-iJ_{\m}+2A_{\m}|\Phi|^2\right)f'(\e^2|D \Phi|^2,|\Phi|^2)=
\label{k11}\ee
where the prime above stands for the derivative of $f$ with respect to $|D \Phi|^2$ and we reintroduced the expansion parameter $\e$.
We now introduce
\be
d_{\m}\equiv A_{\m}-{i\over 2}{J_{\m}\over |\Phi|^2}=A_{\mu}-{i\over 2}\pa_{\mu}\log{\Phi\over \Phi^*}\sp F_{d}^{\m\n}\equiv \pa_{\m}d_{\n}-\pa_{\n}d_{\m} =F_A^{\m\n}
\label{k12}\ee
which is gauge-invariant under the orignal gauge transformations.

We use (\ref{k4}) to rewrite (\ref{k11}) as
\be
V_{\m}=2\e |\Phi|^2 d_{\m}f'\left(\e^2\left[{\pa_{\m}|\Phi|\pa^{\m}|\Phi|}+|\Phi|^2 d_{\m}d^{\m} \right],|\Phi|^2\right)
\label{k13}\ee
which we rewrite as
\be
d_{\m}={V_{\m}\over 2\e|\Phi|^2}~{1\over f'\left(\e^2\left[{(\pa|\Phi|)^2}+|\Phi|^2 d_{\m}d^{\m} \right],|\Phi|^2\right)}
\label{k14}\ee
and taking the square we obtain
\be
\e^2d^2={V^2\over 4|\Phi|^4}{1\over f'\left(\e^2\left[{(\pa|\Phi|)^2}+|\Phi|^2 d^2 \right],|\Phi|^2\right)^2}
\label{k15}\ee
It is clear that the solution to this equation for $d_{\m}$
wiill be of the form
\be
d_{\m}={V_{\m}\over 2\e|\Phi|^2}~g\left(V^2,\e^2(\pa|\Phi|)^2,|\Phi|^2\right)
\label{k16}\ee
with $g\left(V^2,\e^2(\pa|\Phi|)^2,|\Phi|^2\right)$ an appropriate function of its arguments.
Therefore, we observe that the solution has a form similar to what we found in the simpler example. It is a functional of $|\Phi|$ and $V_{\m}$ but the phase of $\Phi$ that here is a gauge degree of freedom,  has disappeared.
With a similar argument we can prove this statement for a general gauge-invariant Schwinger functional.

}

\section{The complete Legendre transform\label{complete}}

In this appendix we  proceed further and consider the effective action of the current as well as the other charged operators.
For this we must perform a complete Legendre transform that includes the charged fields/sources. To wit, we define the new functional
\be
\Gamma^T(\tilde{V}_\mu,\chi ,\mathbf{A}_\mu) = - \int d^4 x\left[\chi \Phi + \chi^* \Phi^* + \tilde V^{\mu}(A_{\mu}-\mathbf{A}_\mu) \right] + W(A,\Phi) \, .
\label{B22o}\ee
We also have the definitions
\be
 \tilde V_{\m}  \equiv {\delta W\over \delta A^{\mu}} \Bigg|_{A_\mu = \mathbf{A}_\mu} \, , \qquad \chi = \frac{\delta W}{\delta \Phi} \Bigg|_{A_\mu = \mathbf{A}_\mu} \, .
\label{B21o}\ee
The conservation law (\ref{A20p}) hence becomes
\be
\pa_{\mu}  \tilde V_{\m}  + i\left(\Phi \chi -\Phi^* \chi^* \right)  \Bigg|_{A_\mu = \mathbf{A}_\mu} = 0
\label{a3o}\ee
With these definitions we maintain gauge invariance explicitly since both the background and the gauge field transform in the same way, while $\tilde{V}$ is invariant.

The effective action has the defining property that it is extremal with respect to $\tilde V_{\m}$.
To show this we first obtain by direct functional differentiation
\be
{\delta \Gamma^T \over \delta \tilde V^{\m}} = (\mathbf{A}_\mu - A_\mu)   -{\delta A^{\n}\over \delta \tilde V^{\m}}\tilde V_{\n} -{\delta \Phi \over \delta \tilde V^{\m}} \chi -{\delta \Phi^* \over \delta \tilde V^{\m}} \chi^* +{\delta W\over \delta \tilde V^{\m}}
\label{a4o}\ee
and by using the chain rule
\be
{\delta W\over \delta \tilde V^{\m}} = \tilde V_{\n}{\delta A^{\n}\over  \delta \tilde V^{\m}} + {\delta \Phi \over \delta \tilde V^{\m}} \chi  + {\delta \Phi^* \over \delta \tilde V^{\m}} \chi^* \, ,
\label{a5o}\ee
the variation with respect to the emerging vector field $\tilde{V}_\mu$ is
\be
\frac{\delta \Gamma (\tilde V,\Phi,\mathbf{A}_\mu)}{\delta \tilde{V}_\mu} = (\mathbf{A}_\mu - A_\mu)  \, .
\ee
We therefore find that it is extremal on the background solution
\be
\frac{\delta \Gamma^T (\tilde V,\Phi,\mathbf{A}_\mu)}{\delta \tilde{V}_\mu} \Bigg|_{A_\mu = \mathbf{A}_\mu} = 0
\ee
In addition with this definition (and again using chain rule) we find
\be
\frac{\delta \Gamma^T (\tilde V,\chi ,\mathbf{A}_\mu)}{\delta \chi} = - \Phi \, , \qquad \frac{\delta \Gamma^T (\tilde V,\chi ,\mathbf{A}_\mu)}{\delta \chi^*} = - \Phi^*
\ee
Notice that gauge invariance is preserved once both $\Phi$ and $\chi$ transform together with opposite charges.

We now parametrize the IR effective action in a derivative expansion in the case of a single charged source for simplicity.
\be
W(A,\Phi)=\int d^4x\left(W_0(|\Phi|^2) + {W_1(|\Phi^2|)\over 4}F^2 + {W_2(|\Phi|^2)\over 2}|D\Phi|^2+\cdots\right)
\label{A24}\ee
where
\be
D_{\m}\Phi=(\pa_{\m}+iA_{\m})\Phi\sp D_{\m}\Phi^*=(\pa_{\m}-iA_{\m})\Phi^*
\label{A25}\ee
It is invariant under the gauge transformations
\be
A_{\m}\to A_{\m}+\pa_{\m}\e\sp \Phi\to \Phi~e^{i\e}
\label{BA1}\ee

We compute the first derivatives
\be
\tilde V_{\nu}\equiv {\Delta \Gamma\over \delta A_{\m}}=- \nabla^{\m}(W_1F_{\m\n}) - i{W_2\over 2}(\Phi^*\pa_{\n}\Phi-\Phi\pa_{\n}\Phi^*) + W_2A_{\n}|\Phi|^2
\label{A26}\ee
$$
=\tilde V_{\n}^0+ \tilde V_{\n}^1+\cdots
$$
with
\be
\tilde V^0_{\m}= W_2|\Phi|^2A_{\m}\sp \tilde V_{\m}^1= - i{W_2\over 2}(\Phi^*\pa_{\n}\Phi-\Phi\pa_{\n}\Phi^*)
\label{A27}\ee

\be
\chi=\left(W'_0(|\Phi|^2) + {W'_1(|\Phi^2|)\over 4}F^2 + {W'_2(|\Phi|^2)\over 2}|D\Phi|^2\right)\Phi^*- {1\over 2}\pa^{\m}(W_2\pa_{\m}\Phi^*) +
\label{A28}\ee
$$
+ {W_2\over 2}A_{\m}A^{\m}\Phi^* + iA^{\m}W_2\pa_{\m}\Phi^* + {i\over 2}\pa^{\m}(A_{\m}W_2)\Phi^*
$$
$$
=\chi_0+\chi_1+\chi_2+\cdots
$$
with
\be
\chi_0= W'_0 \Phi^* \, , \qquad \chi_1 = 0
\label{A29}\ee
\be
\chi_2=  \left({W_2\over 2}A_{\m}A^{\m} + {W_2'\over 2}A_{\m}A^{\m}|\Phi|^2\right)\Phi^* - i{W_2'\over 2}(\Phi^*\pa_{\m}\Phi + \Phi\pa_{\m}\Phi^*)\Phi^*A^{\m} +
\label{A30}\ee
$$
+ iW_2A^{\mu}\pa_{\m}\Phi^* + {i\over 2}\pa_{\m}(W_2A_{\m})\Phi^*\;.
$$
Notice that we treat $A_\mu$ on the same footing as $\partial_\mu$ in the derivative expansion for $\chi$, since they both couple in the same order to $\Phi$ and have the same dimension.

In order to facilitate the inversion procedure we pick $W_0 = |\Phi|^2, \, W_1, W_2 = \text{const}$.

To leading order, without derivative we can invert the relations as follows:
\be
\Phi = \chi^* \,  \sp \Phi^* = \chi \,  \sp A_{\mu}= + {\tilde V_{\m}\over W_2|\Phi|^2} = \hat{V}_\mu
\label{AA30}\ee
The fortunate thing is that if we wish to write down the the effective action up to quadratic order in derivatives, this is all we need since
\bea
\chi \Phi + \chi^* \Phi^* &=& \chi_0 \Phi + \chi_0^* \Phi^* + \chi_1 \Phi + \chi_1^* \Phi^* = 2 |\Phi|^2 + W_2 A^2 |\Phi|^2 + i W_2 A^\mu \left(\Phi \partial_\mu \Phi^* - \Phi^* \partial_\mu \Phi \right) \nn \\
&=& 2 |\Phi|^2 + W_2 A^2 |\Phi|^2 + i W_2 A^\mu \left(\Phi \partial_\mu \Phi^* - \Phi^* \partial_\mu \Phi \right)
\eea

 Considering in \eqref{A28},  $W_{0}=|\Phi|^2$ and that $W_{1}, W_{2}= const$, we obtain
\be
\chi=\Phi^{*}- {1\over 2}\pa^{\m}(W_2\pa_{\m}\Phi^*) +{W_2\over 2}A_{\m}A^{\m}\Phi^* + iA^{\m}W_2\pa_{\m}\Phi^* + {i\over 2}\pa^{\m}(A_{\m}W_2)\Phi^*
\label{A31}\ee
We then  expand
\be\label{A32}
\Phi^{*} = \Phi_{0}^{*}+\Phi_{1}^{*}+\Phi_{2}^{*}+\cdots
\ee
and
\be\label{A33}
A_{\mu} = A_{\mu}^{(1)}+A_{\mu}^{(2)} +A_{\mu}^{(3)}+\cdots
\ee
Notice that the expansion of the gauge field starts from first order because we treat the gauge field in the same footing as the space-time derivative.

We substitute the expansions above into \eqref{A31} and obtain the following relations between the dual fields,
\be\label{A34}
\Phi_{0}^{*} =\chi\,, \quad \Phi_{1}^{*} =0
\ee

\be \label{A35}
\Phi_{2}^{*} = \frac{W_2}{2}\left(\square\chi - A^{(1)}_\m  A^{(1) \, \m} \chi - 2 i\,A^{(1)}_{\mu}\pa^{\mu}\chi - i \chi\pa^{\mu}A_{\mu}^{(1)}\right)
\ee

Notice that $\Phi_{0}$ and $\Phi_{1}$ are fully determined, in order to determine $\Phi_{2}$, we need to solve the gauge field EOM's as well. To do this we substitute in the equation for the vector field $\tilde{V}_{\mu}$ \eqref{A26}, the expansions \eqref{A32}, \eqref{A33} and then we express the terms that depend on $\Phi$ in terms of $\chi$ using \eqref{A34}, \eqref{A35}. We obtain (up to three derivatives)
\bea\label{A36}
\tilde{V_{\nu}}= -i\, \frac{W_{2}}{2}\left(\chi\pa_{\nu}\chi^{*}-\chi^{*}\pa_{\nu}\chi\right)+W_2 A^{(1)}_{\nu}\mid\chi\mid^{2}+W_2 A^{(2)}_{\nu}\mid\chi\mid^{2} +W_2 A^{(3)}_{\nu}\mid\chi\mid^{2} \nn \\
+ W_2 A^{(1)}_{\nu} (\chi^* \Phi_2^* + \Phi_2 \chi) -i\, \frac{W_{2}}{2}\left(\Phi_2^* \pa_{\nu}\chi^{*} + \chi\pa_{\nu}\Phi_2 - \chi^{*}\pa_{\nu}\Phi_2^* - \Phi_2 \pa_{\nu}\chi\right) -W_1\pa^{\mu}F_{\mu\nu}^{A^{(1)}} \nn \\
\eea
and
\be \label{A37}
A^{(1)}_{\mu}=\hat{V}_{\nu}+\frac{i}{2}\pa_{\nu}\log\frac{\chi^{*}}{\chi} \, , \quad A_{\nu}^{(2)} = 0
\ee
where
\be\label{A39}
\hat{V}_{\mu}\equiv\frac{\tilde{V}_{\mu}}{W_{2}\mid\chi\mid^2} \, .
\ee
We may then rewrite (\ref{A35}) as
\be\label{A40}
\Phi_{2}^{*}=  \frac{W_2}{2}\left[\pa^2\chi - \left(\hat{V}_{\nu}+\frac{i}{2}\pa_{\nu}\log\frac{\chi^{*}}{\chi}\right)^{2}\chi - 2 i\left(\hat{V}_{\nu}+\frac{i}{2}\pa_{\nu}\log\frac{\chi^{*}}{\chi}\right)\pa^{\mu}\chi -\right.
\ee
$$
\left.-
 i \chi\pa^{\mu}\left(\hat{V}_{\nu}+\frac{i}{2}\pa_{\nu}\log\frac{\chi^{*}}{\chi}\right)\right]
$$
Finally
\be\label{A38}
A_{\nu}^{(3)}=\frac{W_1}{W_{2}\mid\chi\mid^{2}}\pa^{\mu}F^{\hat{V}}_{\mu\nu} + \cdots
\ee
We observe that the inversion (as an expansion in powers of derivatives)
\be
A_\m = \hat{V}_{\nu}+\frac{i}{2}\pa_{\nu}\log\frac{\chi^{*}}{\chi} + \frac{W_1}{W_{2}\mid\chi\mid^{2}}\pa^{\mu}F^{\hat{V}}_{\mu\nu} +  \, O(\partial^{3+})
\ee
admits the following dual gauge invariance
\be
 \hat V_{\m}\to \hat V_{\m}+\pa_{\m} \epsilon \sp \chi\to \chi~e^{-i\e} \, .
\ee
but as in the case of the partial Legendre transform, it is an artifact of the leading orders of the derivative expansion.

We have expressed up to second order $A_{\mu}, \Phi$ in terms of $\hat{V}_{\mu}, \chi$ and we can substitute them in the expression for the functional \eqref{B22o} in order to have the effective action in terms of the classical vev fields
\be\label{A41}
\Gamma\left(\hat{V}_{\mu}, \chi, \mathbf{A}_\mu \right)=\int\, d^{4}x\, \left[ - |\chi|^2 +\frac{W_{2}}{2}\mid\pa\chi\mid^2 + W_{2}\mid\chi\mid^2\hat{V}^{\mu} \mathbf{A}_{\mu}  \right.
\ee
$$
\left. -{W_2\over 2}|\chi|^2\left(\hat{V}_{\m} +\frac{i}{2}\pa_{\m}\log{\chi^*\over \chi}\right)^2\right] + {\cal O}(\pa^2)\;.
$$

\section{Massless bulk gauge field propagators\label{gau12}}

We consider a Maxwell gauge field in a Euclidean asymptotically AdS$_{d+1}$ bulk, in conformal coordinates
\be
ds^2=e^{2A(z)}(dz^2+dx_{\m}dx^{\m})
\label{i1}
\ee
with action
\be
S=Z{M_p^{d-1}\over 4}\int dz~d^{d} x\sqrt{g}F_{mn}F^{mn}\eqc
\label{i2}\ee
where $m,n$  are (d+1)-dimensional  indices while we reserve here $\m,\n$ for the boundary indices.
The equations to be solved are the Maxwell equations
\be
\nabla^{m}F_{mn}=0~~~\to ~~~\pa_{m}\left(\sqrt{g}g^{mr}g^{ns}F_{rs}\right)=0\eqp
\label{i3}\ee
We shall pick a gauge that fixes all gauge degrees of freedom, namely the generalized Coulomb gauge, which  amounts to imposing the conditions
\be
A_z=0\sp \pa^{\m}A_{\m}=0\eqp
\label{i4}\ee
where the indices above are raised with the Minkowski (boundary) metric.

Translating the equations (\ref{i3}) in the metric (\ref{i1}), the gauge fields satisfy
\be
\left(\pa_{z}^2+(d-3)A'\pa_z +\square_{d}\right)A_{\m}(z,x)=0
\label{i5}\ee
together, with the Coulomb gauge condition in (\ref{i4}).
A prime above is a derivative with respect to $z$ and $\square_d$ is the flat boundary Laplacian.
Transforming to momentum space along the boundary directions we obtain
  \be
\left(\pa_{z}^2+(d-3)A'\pa_z -p^2\right)a_{\m}(z,p)=0\sp p^{\m}a_{\m}=0
\label{i6}\ee

We  first study this equation in $AdS_{d+1}$ with $e^A={L\over z}$.
Then (\ref{i6}) becomes
 \be
\left(\pa_{z}^2-{(d-3)\over z}\pa_z -p^2\right)a_{\m}(z,p)=0\sp p^{\m}a_{\m}=0
\label{i7}\ee
with linearly independent solutions $u^{d-2\over 2}K_{d-2\over 2}(pz)$ and $z^{d-2\over 2}I_{d-2\over 2}(pz)$ with $p\equiv \sqrt{p^2}$.
The regular solution in the IR ($z\to \infty$) is the one proportional to the $K$ function. We can now construct a bulk-to-bulk propagator as
\be
G_{\mu\nu}(z,x;z',x')=-{(zz')^{d-2\over 2}\over L^{d-3}} \int {d^{d} p\over (2\pi)^{d}}e^{-ip\cdot (x-x')} \, a_{\m\n}(p)
\left[\left\{ \begin{array}{lll}
\displaystyle I_{d-2\over 2}(pz)K_{d-2\over 2}(pz'),&\phantom{aa} &z<z'\\ \\
\displaystyle I_{d-2\over 2}(pz')K_{d-2\over 2}(pz),&\phantom{aa}&z>z'.
\end{array}\right\}+\right.
 \label{i8}
\ee
$$
\phantom{\left\{ \begin{array}{l}
\displaystyle I_{d-2\over 2}\\ \\
\displaystyle I_{d-2\over 2}.
\end{array}\right.}\left.+C~K_{d-2\over 2}(pz)\right]
$$
where  $a_{\m\nu}$ is a symmetric dimensionless transverse projector, $p^{\m}~ a_{\m\n}=0$.
The arbitrary constant $C$ reflects the  freedom of imposing boundary conditions near the AdS boundary. For Neumann boundary conditions $C=0$, and this is what we entertain in the rest.

{Another equivalent formula for the bulk to bulk propagator is derived using the spectral representation for the Dirac delta function}
\be
\delta(z-z') = \frac{\sqrt{z z'}}{2} \int_0^\infty d k^2  J_{\nu}(k z) J_{\nu}(k z') \, ,
\ee
in order to obtain the mixed propagator (position space in the radial - momentum space in the transverse directions)
\be
G_{\mu\nu}(z,z'; p)= -{(zz')^{d-2\over 2}\over L^{d-3}}a_{\m\n}(p) \int_0^\infty {d k^2 \over (2\pi)^{d}} \frac{J_{(d-2)/2}(k z) J_{(d-2)/2}(k z')}{k^2 + p^2 + i \epsilon} \, .
\ee
Upon performing the integral, we obtain the integrand of eqn. \eqref{i8}.

We now set the source and the observation point in the  radial direction, to be the same point, $z=z'$, and study the momentum dependence of the bulk-to-bulk propagator, dropping the tensor structure
\be
G_{0}(z',z',p)=-{(z')^{d-2}\over L^{d-3}}~ I_{d-2\over 2}(pz')K_{d-2\over 2}(pz')
 \label{i9}
\ee
The large and small momentum dependence of $G_0$ can be obtained from the properties of the Bessel functions. For large momenta, we obtain
\be
 G_{0}(z',z',p)=-{(z')^{d-3}\over 2L^{d-3}}\left[{1\over p}+{\cal O}\left(p^{-3}\right)\right]
 \label{i10} \ee
 which is the leading behavior of the bulk-to-bulk propagator in flat space, as expected.

At the position $z=z'$ the induced metric is $\gamma_{\m\n}={L^2\over u'^2}\eta_{\m\n}$. Performing a rescaling of coordinates so that the metric at that point becomes $\eta_{\m\n}$ we obtain the rescalled bulk to bulk propagator.
\be
\bar G_{\mu\nu}(z,x;z',x')=-z'~\left({z\over z'}\right)^{d-2\over 2} \int {d^{d} p\over (2\pi)^{d}}e^{-ip\cdot (x-x')}
a_{\m\n}(p) \left\{ \begin{array}{ll}
\displaystyle I_{d-2\over 2}(pz)K_{d-2\over 2}(pz'), &z<z'\\ \\
\displaystyle I_{d-2\over 2}(pz')K_{d-2\over 2}(pz),&z>z'.
\end{array}\right.
 \label{i88}
\ee
Then (\ref{i9}) and (\ref{i10}) become
\be
\bar G_{0}(z',z',p)=-z'~ I_{d-2\over 2}(pz')K_{d-2\over 2}(pz')
 \label{i9a}
\ee
\be
 \bar G_{0}(z',z',p)=-{1\over 2}\left[{1\over p}+{\cal O}\left(p^{-3}\right)\right]
 \label{i10a} \ee

For small momenta, the propagator has a regular expansion in power of $p^2$, with the first non-analyticity appearing at order $(p^2)^{(d-2)\over 2}\log p^2$ for even $d$.

Moving beyond AdS, we can state that the leading behavior of the propagator both in the UV and IR remains the same as in AdS$_{d+1}$ above. The details of the holographic RG flow enter at subleading orders and modify the coefficients of the IR expansion and provide novel non-analyticities in the $IR$ ($p\to 0$).
However, as such non-analyticities are driven by irrelevant operators in the IR, their powers are subleading and we can write, in general
\be
 \bar G_{0}(z',z',p)={1\over m}\left[d_0+(d_2+d_2'~\log {p^2\over m^2}){p^2\over m^2}+{\cal O}(p^4)\right]
 \label{i11} \ee
 for all RG flows that end in AdS$_{5}$ in the IR.
 $m$ is the characteristic scale of the QFT and $d_i$ are dimensionless coefficients that can computed from the background data.

 The situation with flows that are ending in an IR singularity,  that is ``good" in the Gubser classification,  do not change  the expansion in (\ref{i11}) as was already argued in \cite{self}.

The large distance properties of the static potential mediated by  (\ref{i11}) are as follows, \cite{kkt}
\be
V(r)={1\over 2\pi^2 r}\int_0^{\infty}pdp \sin(pr)  \bar G_{0}(z',z',p)\simeq {3d_2'\over 2\pi m^3 r^5}\left[1+{\cal O}(r^{-2})\right]
\label{i22}\ee
where the integrals are defined by introducing a convergence factor like $e^{-ap}$ and taking the limit $a\to 0$ in the end.

\newpage

%%%%%%%%%%%%%%%%%%%%%%%%%%%%%%%%%%%%%%%%%%%%%%%%%%%%%%%%%%%%%%%%%%%%%%%%%%%%%


\begin{thebibliography}{99}


\bibitem{nielsen}
H. B. Nielsen, {\em Do we need fundamental laws of nature?}, Gamma 36 and 37 (1979);\\
  D.~Forster, H.~B.~Nielsen and M.~Ninomiya,
  {\em ``Dynamical Stability of Local Gauge Symmetry: Creation of Light from Chaos,''}
  \href{http://www.doi.org/10.1016/0370-2693(80)90842-4}{Phys.\ Lett.\  {\bf 94B} (1980) 135};\\
  D.~L.~Bennett, N.~Brene and H.~B.~Nielsen,
  {\em ``Random Dynamics,''}
  \href{http://www.doi.org/10.1088/0031-8949/1987/T15/022}{Phys.\ Scripta T {\bf 15} (1987) 158}.








\bibitem{CG}
S.~R.~Coleman and S.~L.~Glashow,
{\em ``High-energy tests of Lorentz invariance,''}
\hrj{10.1103/PhysRevD.59.116008}{Phys. Rev. D \textbf{59} (1999), 116008};
\hre{hep-ph}{9812418}.
%1204 citations counted in INSPIRE as of 15 May 2020

\bibitem{Matt}
D.~Mattingly,
{\em ``Modern tests of Lorentz invariance,''}
\hrj{10.12942/lrr-2005-5}{Living Rev. Rel. \textbf{8} (2005), 5};
\hre{gr-qc}{0502097}
%685 citations counted in INSPIRE as of 15 May 2020

\bibitem{Horava}
P.~Horava,
{\em ``Quantum Gravity at a Lifshitz Point,''}
\hrj{10.1103/PhysRevD.79.084008}{Phys. Rev. D \textbf{79} (2009), 084008};
\hri{0901.3775}{[hep-th]}.
%1899 citations counted in INSPIRE as of 19 May 2020

\bibitem{Pospelov}
M.~Pospelov and Y.~Shang,
{\em ``On Lorentz violation in Horava-Lifshitz type theories,''}
\hrj{10.1103/PhysRevD.85.105001}{Phys. Rev. D \textbf{85} (2012), 105001};
\hri{1010.5249}{[hep-th]}.
%102 citations counted in INSPIRE as of 15 May 2020

\bibitem{lorentz}
  E.~Kiritsis,
  {\em ``Lorentz violation, Gravity, Dissipation and Holography,''}
 \hrj{10.1007/JHEP01(2013)030}{ JHEP {\bf 1301} (2013) 030};
\hri{1207.2325}{[hep-th]}.
  %%CITATION = doi:10.1007/JHEP01(2013)030;%%
  %39 citations counted in INSPIRE as of 22 Jul 2018

 \bibitem{RVB}
  Pauling Linus Carl,
  \em{``A resonating-valence-bond theory of metals and intermetallic compounds",}
  196, Proc. R. Soc. Lond. A. \\
  Anderson, P. W.,
  \em{``Resonating valence bonds: A new kind of insulator?"},
  Materials Research Bulletin, 8(2), 153-160. (1973).


\bibitem{wen}
  M.~A.~Levin and X.~G.~Wen,
  {\em ``Colloquium: Photons and electrons as emergent phenomena,''}
 \hrj{10.1103/RevModPhys.77.871}{ Rev.\ Mod.\ Phys.\  {\bf 77} (2005) 871};
  \hre{cond-mat}{0407140}].
  %%CITATION = doi:10.1103/RevModPhys.77.871;%%
  %63 citations counted in INSPIRE as of 18 Feb 2019

\bibitem{wen2}
  Z.~C.~Gu and X.~G.~Wen,
  {\em ``A Lattice bosonic model as a quantum theory of gravity,''}
  \hre{gr-qc}{0606100}.
  %%CITATION = GR-QC/0606100;%%
  %28 citations counted in INSPIRE as of 18 Feb 2019

  \bibitem{frac} A. Y. Kitaev, Ann. Phys. (N.Y.) 303, 2 (2003);\\
 X.-G. Wen, Phys. Rev. Lett. 90, 016803 (2003);\\
 R. Moessner and S. L. Sondhi, Phys. Rev. Lett. 86, 1881 (2001);\\
 X.-G. Wen, Phys. Rev. Lett. 88, 11602 (2002);\\
 X.-G. Wen, Phys. Rev. B 68, 115413 (2003);\\
 O. I. Motrunich and T. Senthil, Phys. Rev. Lett. 89, 277004 (2002); O. I. Motrunich and T. Senthil, Phys. Rev. B 71, 125102 (2005).



  \bibitem{Senthil}
  Senthil, T., Vishwanath, A., Balents, L., Sachdev, S., Fisher, M. P.,
  {\em ``Deconfined quantum critical points",}
  Science, 303(5663), 1490-1494.(2004).




%\cite{Lee:2010fy}
\bibitem{Lee1}
  S.~S.~Lee,
  {\em ``TASI Lectures on Emergence of Supersymmetry, Gauge Theory and String in Condensed Matter Systems,''}
  \hri{1009.5127}{[hep-th]}.
  %%CITATION = ARXIV:1009.5127;%%
  %25 citations counted in INSPIRE as of 18 Feb 2020

%\cite{Lee:2018dqw}
\bibitem{Lee2}
  S.~S.~Lee,
  {\em ``Emergent gravity from relatively local Hamiltonians and a possible resolution of the black hole information puzzle,''}
\hrj{10.1007/JHEP10(2018)043}{  JHEP {\bf 1810} (2018) 043};
  \hri{1803.00556}{[hep-th]]}.
  %%CITATION = doi:10.1007/JHEP10(2018)043;%%
  %2 citations counted in INSPIRE as of 18 Feb 2020


  \bibitem{NJL}
Y.~Nambu and G.~Jona-Lasinio,
{\em ``Dynamical Model of Elementary Particles Based on an Analogy with Superconductivity. 1.,''}
\href{https://doi.org/10.1103/PhysRev.122.345}{Phys. Rev. \textbf{122} (1961), 345-358}.


\bibitem{Bjorken1}
  J.~D.~Bjorken,
  {\em ``A Dynamical origin for the electromagnetic field,''}
  \href{https://www.sciencedirect.com/science/article/pii/0003491663900691?via\%3Dihub}{Annals Phys.\  {\bf 24} (1963) 174.}
  %%CITATION = doi:10.1016/0003-4916(63)90069-1;%%
  %334 citations counted in INSPIRE as of 04 Oct 2018



\bibitem{Bjorken2}
  J.~Bjorken,
  {\em ``Emergent gauge bosons,''}
  \hre{hep-th}{0111196}.
  %%CITATION = HEP-TH/0111196;%%
  %85 citations counted in INSPIRE as of 04 Oct 2018





  \bibitem{BB}
  I.~Bialynicki-Birula,
  {\em ``Quantum Electrodynamics without Electromagnetic Field,''}
  \href{https://doi.org/10.1103/PhysRev.130.465}{Phys.\ Rev.\  {\bf 130} (1963) 465}.
  %%CITATION = doi:10.1103/PhysRev.130.465;%%
  %88 citations counted in INSPIRE as of 04 Oct 2018

\bibitem{BZ}T.~Banks and A.~Zaks,
{\em ``Composite Gauge Bosons in Four Fermi Theories (Or Honey and Condensed Vectors),''}
\hrj{10.1016/0550-3213(81)90220-0}{Nucl. Phys. B \textbf{184} (1981), 303-322}
%19 citations counted in INSPIRE as of 03 Nov 2020

\bibitem{cpn}
A.~D'Adda, M.~Luscher and P.~Di Vecchia,
{\em ``A 1/N Expandable Series of Nonlinear Sigma Models with Instantons,''}
\href{www.doi.org/10.1016/0550-3213(78)90432-7}{Nucl. Phys. B \textbf{146} (1978), 63-76}.

%696 citations counted in INSPIRE as of 15 May 2020

\bibitem{wcpn}
E.~Witten,
{\em ``Instantons, the Quark Model, and the 1/N Expansion,''}
\href{www.doi.org/10.1016/0550-3213(79)90243-8}{Nucl. Phys. B \textbf{149} (1979), 285-320}.


  \bibitem{Polyakov}
A.~M.~Polyakov,
{\em ``Gauge Fields and Strings,''}
Contemp. Concepts Phys. \textbf{3} (1987), 1-301
%167 citations counted in INSPIRE as of 15 May 2020

\bibitem{seiberg}
  N.~Seiberg,
  {\em ``Exact results on the space of vacua of four-dimensional SUSY gauge theories,''}
\hrj{10.1103/PhysRevD.49.6857}{  Phys.\ Rev.\ D {\bf 49} (1994) 6857};
  \hre{hep-th}{9402044}.
  %%CITATION = doi:10.1103/PhysRevD.49.6857;%%
  %732 citations counted in INSPIRE as of 19 Feb 2019


\bibitem{Komar}
  Z.~Komargodski,
  {\em ``Vector Mesons and an Interpretation of Seiberg Duality,''}
\hrj{10.1007/JHEP02(2011)019}{  JHEP {\bf 1102} (2011) 019};
  \hri{1010.4105}{[hep-th]}.
  %%CITATION = doi:10.1007/JHEP02(2011)019;%%
  %36 citations counted in INSPIRE as of 18 Feb 2019




\bibitem{malda}
  J.~M.~Maldacena,
  {\em ``The Large N limit of superconformal field theories and supergravity,''}
\hrj{10.1023/A:1026654312961}{  Int.\ J.\ Theor.\ Phys.\  {\bf 38} (1999) 1113};
   [Adv.\ Theor.\ Math.\ Phys.\  {\bf 2} (1998) 231];
  \hre{hep-th}{9711200}.
  %%CITATION = doi:10.1023/A:1026654312961, 10.4310/ATMP.1998.v2.n2.a1;%%
  %13692 citations counted in INSPIRE as of 21 May 2018



%\cite{Harlow:2015lma}
\bibitem{Harlow}
  D.~Harlow,
  {\em ``Wormholes, Emergent Gauge Fields, and the Weak Gravity Conjecture,''}
\hrj{10.1007/JHEP01(2016)122}{  JHEP {\bf 1601} (2016) 122};
  \hri{1510.07911}{[hep-th]]}.
  %%CITATION = doi:10.1007/JHEP01(2016)122;%%
  %104 citations counted in INSPIRE as of 18 Feb 2020








\bibitem{SMGRAV}
  E.~Kiritsis,
  {\em ``Gravity and axions from a random UV QFT,''}s
  \hrj{10.1007/JHEP10(2019)113}{EPJ Web Conf.\  {\bf 71} (2014) 00068};
  \hri{1408.3541}{[hep-ph]}.
  %%CITATION = doi:10.1051/epjconf/20147100068;%%
  %2 citations counted in INSPIRE as of 25 Apr 2018


\bibitem{KT}
P.~Kraus and E.~T.~Tomboulis,
{\em ``Photons and gravitons as Goldstone bosons, and the cosmological constant,''}
\hrj{10.1103/PhysRevD.66.045015}{Phys. Rev. D \textbf{66} (2002), 045015};
\hre{hep-th}{0203221}.
%107 citations counted in INSPIRE as of 30 Sep 2020


\bibitem{axion}
  P.~Anastasopoulos, P.~Betzios, M.~Bianchi, D.~Consoli and E.~Kiritsis,
  {\em ``Emergent/Composite axions,''}
 \hrj{10.1007/JHEP10(2019)113}{ JHEP {\bf 1910} (2019) 113};
 \hri{1811.05940}{[hep-ph]}s.
  %%CITATION = doi:10.1007/JHEP10(2019)113;%%
  %5 citations counted in INSPIRE as of 27 Mar 2020



     \bibitem{WW}
     S.~Weinberg and E.~Witten,
{\em ``Limits on Massless Particles,''}
\hrj{10.1016/0370-2693(80)90212-9}{Phys. Lett. B \textbf{96} (1980), 59-62}.
%527 citations counted in INSPIRE as of 30 Sep 2020


\bibitem{hidden}
M.~Bando, T.~Kugo, S.~Uehara, K.~Yamawaki and T.~Yanagida,
{\em ``Is teh $\rho$-Meson a Dynamical Gauge Boson of Hidden Local Symmetry?,''}
\href{www.doi.org/10.1103/PhysRevLett.54.1215}{Phys. Rev. Lett. \textbf{54} (1985), 1215};
%887 citations counted in INSPIRE as of 25 May 2020
H.~Georgi,
{\em ``Vector Realization of Chiral Symmetry,''}
\href{www.doi.org/10.1016/0550-3213(90)90210-5}{Nucl. Phys. B \textbf{331} (1990), 311-330}.

%157 citations counted in INSPIRE as of 25 May 2020

\bibitem{hh}
T.~Sakai and S.~Sugimoto,
{\em ``Low energy hadron physics in holographic QCD,''}
\hrj{10.1143/PTP.113.843}{Prog. Theor. Phys. \textbf{113} (2005), 843-882};
\hre{hep-th}{0412141};\\
%1287 citations counted in INSPIRE as of 25 May 2020
D.~K.~Hong, M.~Rho, H.~U.~Yee and P.~Yi,
{\em ``Nucleon form-factors and hidden symmetry in holographic QCD,''}
\hrj{10.1103/PhysRevD.77.014030}{Phys. Rev. D \textbf{77} (2008), 014030};
\hri{0710.4615}{[hep-ph]}.
%90 citations counted in INSPIRE as of 25 May 2020



\bibitem{u1}
P.~Anastasopoulos, M.~Bianchi, D.~Consoli and E.~Kiritsis,
{\em ``String (gravi)photons, ''dark brane photons'', holography and the hypercharge portal,''}
\hri{2010.07320}{ [hep-ph]}.





    \bibitem{csaki}
  C.~Csaki, C.~Grojean, L.~Pilo and J.~Terning,
  {\em ``Towards a realistic model of Higgsless electroweak symmetry breaking,''}
\hrj{10.1103/PhysRevLett.92.101802}{  Phys.\ Rev.\ Lett.\  {\bf 92} (2004) 101802};
  \hre{hep-ph}{0308038}.
  %%CITATION = doi:10.1103/PhysRevLett.92.101802;%%
  %500 citations counted in INSPIRE as of 21 Feb 2019

\bibitem{grav}
P.~Betzios, E.~Kiritsis and V.~Niarchos,
{\em ``Emergent gravity from hidden sectors and TT deformations,''}
\hri{2010.04729}{[hep-th]}.


 %\cite{Alekhin:2015byh}
\bibitem{ship}
S.~Alekhin et al.
{\em ``A facility to Search for Hidden Particles at the CERN SPS: the SHiP physics case,''}
\hrj{10.1088/0034-4885/79/12/124201}{Rept.\ Prog.\ Phys.\  \textbf{79} (2016) no.12, 124201};
\hri{1504.04855}{ [hep-ph]}.
%471 citations counted in INSPIRE as of 31 Mar 2020



\bibitem{review}
E.~Kiritsis,
{\em ``D-branes in standard model building, gravity and cosmology,''}
\hrj{10.1016/j.physrep.2005.09.001}{Phys. Rept. \textbf{421} (2005), 105-190};
\hri{hep-th/0310001}{[hep-th]};\\
\href{https://arxiv.org/abs/hep-th/0310001v1}{Fortsch. Phys. \textbf{52} (2004) no.2-3, 200-263},  doi:10.1002/prop.200310120
%171 citations counted in INSPIRE as of 06 May 2020


\bibitem{KA}
E.~Kiritsis and P.~Anastasopoulos,
{\em ``The Anomalous magnetic moment of the muon in the D-brane realization of the standard model,''}
\hrj{10.1088/1126-6708/2002/05/054}{JHEP \textbf{05} (2002), 054};
\hre{hep-ph}{0201295}.
%63 citations counted in INSPIRE as of 06 May 2020

\bibitem{AKR}
I.~Antoniadis, E.~Kiritsis and J.~Rizos,
{\em ``Anomalous U(1)s in type 1 superstring vacua,''}
\hrj{10.1016/S0550-3213(02)00458-3}{Nucl. Phys. B \textbf{637} (2002), 92-118};
\hre{hep-th}{0204153};\\
%\cite{Antoniadis:2001np}
I.~Antoniadis, E.~Kiritsis and T.~Tomaras,
{\em ``D-brane standard model,''}
Fortsch. Phys. \textbf{49} (2001), 573-580;
\hre{hep-th}{0111269};\\
%61 citations counted in INSPIRE as of 01 Jun 2020
I.~Antoniadis, E.~Kiritsis, J.~Rizos and T.~Tomaras,
{\em `D-branes and the standard model,''}
\hrj{10.1016/S0550-3213(03)00256-6}{Nucl. Phys. B \textbf{660} (2003), 81-115};
\hre{hep-th}{0210263}.
%144 citations counted in INSPIRE as of 06 May 2020


\bibitem{akt}
I.~Antoniadis, E.~Kiritsis and T.~Tomaras,
{\em ``A D-brane alternative to unification,''}
\hrj{10.1016/S0370-2693(00)00733-4}{Phys. Lett. B \textbf{486} (2000), 186-193};
\hri{hep-ph}{0004214}.


\bibitem{ADKS}P. ~Anastasopoulos, T.~Dijkstra, E.~Kiritsis and A.~Schellekens,
{\em ``Orientifolds, hypercharge embeddings and the Standard Model,''}
\hrj{10.1016/j.nuclphysb.2006.10.013}{Nucl. Phys. B \textbf{759} (2006), 83-146};
\hre{hep-th}{0605226}.
%169 citations counted in INSPIRE as of 06 May 2020


\bibitem{CIK}
C.~Coriano, N.~Irges and E.~Kiritsis,
{\em ``On the effective theory of low scale orientifold string vacua,''}
\hrj{10.1016/j.nuclphysb.2006.04.009}{Nucl. Phys. B \textbf{746} (2006), 77-135};
\hre{hep-ph}{0510332}.



\bibitem{self}
  C.~Charmousis, E.~Kiritsis and F.~Nitti,
  {\em ``Holographic self-tuning of the cosmological constant,''}
\hrj{10.1007/JHEP09(2017)031}{  JHEP {\bf 1709} (2017) 031};
  \hri{1704.05075}{[hep-th]};\\
  %%CITATION = doi:10.1007/JHEP09(2017)031;%%
  %13 citations counted in INSPIRE as of 28 Mar 2020
Y.~Hamada, E.~Kiritsis, F.~Nitti and L.~T.~Witkowski,
{\em ``The self-tuning of the cosmological constant and the holographic relaxion,''}
\hri{2001.05510}{[hep-th]}.


\bibitem{stras}
M.~J.~Strassler and K.~M.~Zurek,
{\em ``Echoes of a hidden valley at hadron colliders,''}
\hrj{10.1016/j.physletb.2007.06.055}{Phys. Lett. B \textbf{651} (2007), 374-379};
\hre{hep-ph}{0604261}.
%661 citations counted in INSPIRE as of 25 May 2020

\bibitem{Coleman}
S.~Coleman,
\href{https://doi.org/10.1017/CBO9780511565045}{\em ``Aspects of Symmetry''},  Cambridge University Press, 1985, ISBN: 9780511565045.
%47 citations counted in INSPIRE as of 05 May 2020


\bibitem{gk}
A.~Giveon and E.~Kiritsis,
{\em ``Axial vector duality as a gauge symmetry and topology change in string theory,''}
\hrj{10.1016/0550-3213(94)90460-X}{Nucl. Phys. B \textbf{411} (1994), 487-508};
\hre{hep-th}{9303016}.
%122 citations counted in INSPIRE as of 01 Jun 2020


\bibitem{sf}
K.~Sfetsos, K.~Siampos and D.~C.~Thompson,
{\em ``Generalised integrable $\lambda$ - and $\eta$-deformations and their relation,''}
\hrj{10.1016/j.nuclphysb.2015.08.015}{Nucl. Phys. B \textbf{899} (2015), 489-512};
\hri{1506.05784}{[hep-th]}.
%84 citations counted in INSPIRE as of 01 Jun 2020





    %\cite{Bianchi:2001kw}
\bibitem{Bianchi:2001kw}
  M.~Bianchi, D.~Z.~Freedman and K.~Skenderis,
  {\em ``Holographic renormalization,''}
\hrj{10.1016/S0550-3213(02)00179-7}{  Nucl.\ Phys.\ B {\bf 631} (2002) 159};
  \href{hep-th}{0112119}.
  %%CITATION = doi:10.1016/S0550-3213(02)00179-7;%%
  %454 citations counted in INSPIRE as of 03 Nov 2018

%\cite{Bianchi:2001de}
\bibitem{Bianchi:2001de}
  M.~Bianchi, D.~Z.~Freedman and K.~Skenderis,
  {\em ``How to go with an RG flow,''}
\hrj{10.1088/1126-6708/2001/08/041}{  JHEP {\bf 0108} (2001) 041};
  \hre{hep-th}{0105276}.
  %%CITATION = doi:10.1088/1126-6708/2001/08/041;%%
  %281 citations counted in INSPIRE as of 03 Nov 2018





  %\cite{Dvali:2000hr}
\bibitem{DGP}
G.~Dvali, G.~Gabadadze and M.~Porrati,
{\em ``4-D gravity on a brane in 5-D Minkowski space,''}
\hrj{10.1016/S0370-2693(00)00669-9}{Phys.\ Lett.\ B \textbf{485} (2000), 208-214};
\hre{hep-th}{0005016}.
%2812 citations counted in INSPIRE as of 31 Mar 2020


\bibitem{Nu}
C.~Nunez, I.~Papadimitriou and M.~Piai,
{\em ``Walking Dynamics from String Duals,''}
\hrj{10.1142/S0217751X10049189}{Int. J. Mod. Phys. A \textbf{25} (2010), 2837-2865};
\hri{0812.3655}{[hep-th]}.
%85 citations counted in INSPIRE as of 06 May 2020

\bibitem{P}
S.~Kumar, D.~Mateos, A.~Paredes and M.~Piai,
{\em ``Towards holographic walking from N=4 super Yang-Mills,''}
\hrj{10.1007/JHEP05(2011)008}{JHEP \textbf{05} (2011), 008}
\hri{1012.4678}{[hep-th]}.
%20 citations counted in INSPIRE as of 06 May 2020

\bibitem{A}
L.~Anguelova, P.~Suranyi and L.~Wijewardhana,
{\em ``Holographic Walking Technicolor from D-branes,''}
\hrj{10.1016/j.nuclphysb.2011.06.010}{Nucl. Phys. B \textbf{852} (2011), 39-60}
\hri{1105.4185}{[hep-th]}.
%34 citations counted in INSPIRE as of 06 May 2020

\bibitem{KJ}
M.~Jarvinen and E.~Kiritsis,
{\em ``Holographic Models for QCD in the Veneziano Limit,''}
\hrj{10.1007/JHEP03(2012)002}{JHEP \textbf{03} (2012), 002};
\hri{1112.1261}{[hep-ph]};\\
D.~Arean, I.~Iatrakis, M.~Jarvinen and E.~Kiritsis,
{\em ``V-QCD: Spectra, the dilaton and the S-parameter,''}
\hrj{10.1016/j.physletb.2013.01.070}{Phys. Lett. B \textbf{720} (2013), 219-223};
\hri{1211.6125}{[hep-ph]};\\
{\em ``The discontinuities of conformal transitions and mass spectra of V-QCD,''}
\hrj{10.1007/JHEP11(2013)068}{JHEP \textbf{11} (2013), 068};
\hri{1309.2286}{[hep-ph]}.
%45 citations counted in INSPIRE as of 06 May 2020
%40 citations counted in INSPIRE as of 06 May 2020
%127 citations counted in INSPIRE as of 06 May 2020





\bibitem{IS}
K.~A.~Intriligator and N.~Seiberg,
{\em ``Lectures on supersymmetric gauge theories and electric-magnetic duality,''}
\hrj{10.1016/0920-5632(95)00626-5}{Subnucl. Ser. \textbf{34} (1997), 237-299};
\hre{hep-th}{9509066}.
%756 citations counted in INSPIRE as of 25 May 2020


\bibitem{ABDK}
P.~Anastasopoulos, M.~Bianchi, E.~Dudas and E.~Kiritsis,
{\em ``Anomalies, anomalous U(1)'s and generalized Chern-Simons terms,''}
\hrj{10.1088/1126-6708/2006/11/057}{JHEP \textbf{11} (2006), 057};
\hre{hep-th}{0605225}.
%161 citations counted in INSPIRE as of 29 May 2020




\bibitem{QT}
F.~Quevedo and C.~A.~Trugenberger,
{\em ``Phases of antisymmetric tensor field theories,''}
\hrj{10.1016/S0550-3213(97)00337-4}{Nucl. Phys. B \textbf{501} (1997), 143-172};
\hre{hep-th}{9604196}.
%123 citations counted in INSPIRE as of 05 May 2020


\bibitem{JT}
B.~Julia and G.~Toulouse,
{\em ``The Many Defect Problem: Gauge Like Variables for Ordered Media Containing Defects,''}
\hrj{10.1051/jphyslet:019790040016039500}{J. Phys. Lett. \textbf{40} (1979), 396}.
%21 citations counted in INSPIRE as of 30 Sep 2020









\bibitem{Bernard:1975cd}
C.~W.~Bernard, A.~Duncan, J.~LoSecco and S.~Weinberg,
{\em ``Exact Spectral Function Sum Rules,''}
\href{www.doi.org/10.1103/PhysRevD.12.792}{Phys. Rev. D \textbf{12} (1975), 792}.






\bibitem{kkt}
E.~Kiritsis, N.~Tetradis and T.~Tomaras,
{\em ``Induced gravity on RS branes,''}
\hrj{10.1088/1126-6708/2002/03/019}{JHEP \textbf{03} (2002), 019};
\hre{hep-th}{0202037}.
%101 citations counted in INSPIRE as of 01 Jun 2020





\bibitem{Anto}
I.~Antoniadis, A.~Boyarsky, S.~Espahbodi, O.~Ruchayskiy and J.~D.~Wells,
{\em ``Anomaly driven signatures of new invisible physics at the Large Hadron Collider,''}
\hrj{10.1016/j.nuclphysb.2009.09.009}{Nucl. Phys. B \textbf{824} (2010), 296-313};
\hri{0901.0639}{[hep-ph]}.

\end{thebibliography}
\end{document}